\documentclass[pre,aps,twocolumn,superscriptaddress,showpacs,floatfix,amssymb,amsmath]{revtex4-1}
\usepackage{epsfig}
\usepackage{graphics}
\usepackage{subcaption} 
\usepackage{caption}

\usepackage{float} 
\usepackage{array}
\usepackage{multirow}

\usepackage{yfonts}

\usepackage{graphicx}
\usepackage{amssymb}
\usepackage{mathrsfs}
\usepackage{bbm}
\usepackage{epsfig}
\usepackage{makecell}

\usepackage[usenames,dvipsnames]{color}

\usepackage{bbm}
\usepackage{color}

\definecolor{MyWhite}{RGB}{255,250,240} 
\definecolor{light-blue}{RGB}{240,248,255}
\definecolor{light-gray}{gray}{0.95}

\newcommand{\beq}{\begin{equation}}
\newcommand{\eeq}{\end{equation}}

\begin{document}

\title{\Large Multiple scales and phases in discrete chains  \\
 with application to  folded proteins 
}

\vskip 5.0cm
\author{A. Sinelnikova}
\email{Anna.Sinelnikova@physics.uu.se}
\affiliation{Department of Physics and Astronomy, Uppsala University,
P.O. Box 516, S-75120, Uppsala, Sweden}
\author{A. J. Niemi}
\email{Antti.Niemi@physics.uu.se}
\homepage{http://www.folding-protein.org}
\affiliation{Nordita, Stockholm University, Roslagstullsbacken 23, SE-106 91 Stockholm, Sweden}
\affiliation{Department of Physics and Astronomy, Uppsala University,
P.O. Box 516, S-75120, Uppsala, Sweden}
\affiliation{Laboratory of Physics of Living Matter, School of Biomedicine, Far Eastern Federal University, Vladivostok, Russia}
\affiliation{Department of Physics, Beijing Institute of Technology,
Haidian District, Beijing 100081, P. R. China}
\author{Johan Nilsson}
\email{Johan.Nilsson@physics.uu.se}
\affiliation{Department of Physics and Astronomy, Uppsala University,
P.O. Box 516, S-75120, Uppsala, Sweden}
\author{M. Ulybyshev}
\email{Maksim.Ulybyshev@physik.uni-regensburg.de}
\affiliation{Institute of Theoretical Physics, University of Regensburg,
Universit\"atsstra\ss e 31, Regensburg,  D-93053 Germany}

%

\begin{abstract}
\noindent
Chiral heteropolymers such as larger globular proteins can simultaneously support multiple length scales. The interplay 
between different scales brings about conformational diversity, and governs the structure of the energy landscape. 
Multiple scales  produces also  complex dynamics, which in the case of proteins sustains live matter. 
However, thus far no clear understanding exist,
how to distinguish the various scales that determine the structure and dynamics of a complex protein. 
Here we propose a systematic method to identify the scales in chiral heteropolymers such as a protein. For this 
we introduce a novel order parameter,  that not only reveals the scales but also probes the phase structure. In particular, we argue 
that a chiral heteropolymer can simultaneously display traits of 
several different phases, contingent on the length scale at which it is scrutinized. 
%
Our approach builds on a variant of Kadanoff's block-spin transformation that we employ to coarse grain 
piecewise linear chains such as the C$\alpha$ backbone of a protein.  We derive analytically and then 
verify numerically a number of  properties that the order parameter can display.
We demonstrate how, in the case of crystallographic protein structures in Protein Data Bank, 
the order parameter reveals the presence of different length scales, and we propose that a relation 
must exist between the scales, phases, and the complexity of folding pathways. 
 \end{abstract}


\maketitle

\section{Introduction}
A linearly conjugated polymer is conventionally viewed as 
a piecewise linear polygonal chain, it connects
a sequence of vertices that coincide with the locations of its skeletal atoms
\cite{huggins,flory1,flory2,flory3,degen,degen1,desc,edw1,degen2,edw2}.
For example, in the case of a protein the vertices  
coincide with the positions of the
C$\alpha$ atoms, and the connecting line segments concur with the  diagonals of the peptide planes.
Conventionally, a  phase is then assigned to the polymer by inspecting the fractal geometry of the chain.  
For this let ($\mathbf{x}_0, \dots  ,\mathbf{x}_{N} $) denote the $N+1$ 
vertices of a given discrete chain  $\Gamma$.   
When $N$ becomes large the radius of gyration 
\begin{equation}
R_{gyr} \ = \ \sqrt{ \, \frac{1}{2(N+1)^2}  \sum\limits_{i,j = 0}^{N} ( \mathbf{x}_i  - \mathbf{x}_j )^2\, }   
\label{Rg}
\end{equation}
admits an asymptotic expansion of the form \cite{degen2,schafer,naka,sokal}
\begin{equation}
R^2_{gyr} \ \approx \ L_0^2 N^{2\nu} ( 1 + R_1 N^{-\Delta_1} +  ... )  \buildrel{N \ {\rm large}}\over{\longrightarrow} 
L_0^2 N^{2\nu} 
\label{R}
\end{equation}
The pre-factor $L_0$ is an effective segment (Kuhn) length. In the case of a polymer its value depends on the atomic 
level details of the chain and the environment, it is not a universal quantity. 
The scaling exponent  $\nu$ coincides with the inverse Hausdorff dimension of $\Gamma$. It is 
a universal quantity \cite{degen2,edw2,grosberg-1994,naka,sokal,schafer} with 
a numerical value that does not depend on the atomic level details of the 
chain. It acts as an order parameter that 
can detect the
phase of the chain. 
The exponents $\Delta_1, \, \Delta_2, \, ...$ 
that characterise the finite size corrections are similarly  universal  \cite{sokal}.

In the case of a discrete chain with a homogeneous structure, 
the scaling exponent $\nu$ is commonly presumed to have only four possible
values, corresponding to  the four different phases of a homopolymer.
At the level of classical mean field theory \cite{huggins,flory1,flory2,degen2,edw2,grosberg-1994,naka,sokal,schafer}
\begin{equation}
\nu \ =  \ \left\{  \ \, \begin{matrix} 1/3  \vspace{0.1cm}  \\ 1/2  \vspace{0.1cm} \\  3/5  \vspace{0.1cm} \\ 1 \end{matrix} \right. \ \ \ \ \  \begin{matrix}
\ \,  \hspace{-4.6cm} {\rm Collapsed}   \vspace{0.1cm} \\  
 {\rm \hspace{-2.8cm} Random ~walk ~(RW)}   \vspace{0.1cm} \\  {\rm Self-avoiding ~random ~ walk ~(SARW)}   \vspace{0.1cm} \\  
 \hspace{-4.0cm} {\rm Straight ~ rod}  \end{matrix} 
\label{nuval}
\end{equation}
The 
value of $\nu$ is determined using (\ref{R}), 
by successively increasing the number of vertices $\mathbf x_i$  and by observing how the
radius of gyration scales when $N$ becomes large. This procedure can work in the case of a homopolymer, when
the number of vertices can be increased in an unambiguous manner.
Unfortunately,  it does not work  in the case  of a heteropolymer such as a protein, 
where the amino acid  assignment is fixed: There is no unambiguous way 
to extend the length of a protein, to determine the scaling of its radius of gyration when $N$ grows,
the number of vertices 
can not be  systematically increased as required by (\ref{R}).
Instead one can try and deduce  the value of $\nu$ statistically, by comparing the radius of gyration of a given protein
to a statistical pool of different lengths but similar kind of protein structures such as those classified as $\alpha$-helical or $\beta$-stranded   
in Protein Data Bank (PDB)  \cite{pdb}.  This procedure has some merits, but it lacks rigor.  Moreover, it brings about
intriguing but difficult-to-verify proposals, including  a suggestion that  since {\it e.g.} $\alpha$-helical 
and $\beta$-stranded proteins have different $\nu$-values,  the ensuing chains reside
in different phases \cite{dewey,hong,huang,jcp}.  To clarify all these issues, there is need to introduce another
order parameter, one  that directly and unambiguously probes the phase of a given heteropolymer, with no need to extend its length.
 
Here we propose such a new order parameter.  It builds on the properties of a new observable we introduce, 
under  a variant
of Kadanoff's block-spin transformation \cite{kada,wilson,fisher,golden}  that we design to 
coarse grain a fixed chain in an effective manner. 
We show how the observable detects different 
scales, as the coarse graining proceeds. 
For a homopolymer, we confirm the universal 
phase structure in line with (\ref{nuval}). However, in the case of a 
heterogeneous chain such as a folded protein we find that our observable is  {\it variable}.
Its value depends on the scale and oscillates, apparently between different phases, as the coarse graining proceeds.
We interpret this variable character of the observable 
in terms of a multiphase structure: Depending on the distance scale at which a heteropolymer chain is inspected, 
it can  display different phase properties.

We recall the following: In the case of $N$ point-like,
chemically independent components  there are {\it a priori}  $N$ different dimensionful parameters. 
The Gibbs phase rule states that in the presence of $F$ intensive thermodynamical 
variables the number $P$  of co-existing phases is limited by  
\begin{equation}
P \leqslant  N - F +2 
\label{phase-rule}
\end{equation}
When the elemental constituents are chain-like, the rule can change.
If a relation between the
number of dimensionful parameters and  the number of 
thermodynamical phases persists,
even a single chain might exhibit 
different phase characteristics when we  inspect it at different length scales. 
For example, consider a crystallographic globular protein in a collapsed phase. 
At the same time, at distance scales that are short in comparison to the radius of gyration, its  
structure can be dominated  {\it e.g.} by $\alpha$-helices or $\beta$-strands which are both
in the phase of a straight rod. Thus there is an intermediate length scale, at which the 
character of the protein structure transits from the straight rod phase to the collapsed phase.  
The scaling exponent (\ref{nuval}) is unsuitable for detecting how such a transition from
a regime dominated by a straight rod phase  to a collapsed phase regime takes place. But our new
observable can detect the presence of a scale that could produce a transition.

We  start by describing the generic theoretical properties of our observable.  
We then proceed to develop it into a tool that detects a phase.
For this we device a variant of the Kadanoff block-spin transformation, specially tailored 
to inspect chain-like objects. We analyse the transformation properties of the observable numerically,
in terms of Monte Carlo simulations of a homopolymer model.  We show that 
the phase structure is in line with the classification (\ref{nuval}).   
We then continue to crystallographic protein structures.  We reveal how
a complex globular protein can display apparently 
different phase properties at different length scales. 
We propose that the presence of a multiple phase structure 
can have profound effects on the folding and unfolding 
transitions, and to other dynamical and structural properties of proteins.

\section{Observables and phase diagrams}

\subsection{New observable}  

Let  $\mathbf{t}_i$ denote the segment from the vertex $\mathbf x_{i-1}$
to the subsequent vertex $\mathbf x_{i}$ along a discrete linear chain $\Gamma$,  with a total of $N+1$ 
vertices  ($\mathbf{x}_0, \dots  ,\mathbf{x}_{N} $)
\beq
\mathbf{t}_i = \mathbf{x}_i  - \mathbf{x}_{i-1}
\label{ti}
\eeq
We introduce the following observable
\begin{eqnarray}
{\mathcal P_\Gamma}(N)  &=&   
\sum_{1\leqslant i<j\leqslant {N}}
\frac{\mathbf{t}_i\cdot\mathbf{t}_j}{\vert\mathbf{t}_i\vert\vert\mathbf{t}_j\vert} 
\ \equiv  \hspace{-.2cm} 
\sum_{1\leqslant i<j\leqslant {N}}  \hspace{-.2cm}
\cos \kappa_{ij}
\label{eq_cos}
\\
&\approx& \ \mathcal P_\Gamma N^{\sigma} (1 +  \ {\mathcal Q_1} N^{-\delta_1}  + \dots )
\label{eq_B}
\end{eqnarray}
Here both the $N$-independent factor $\mathcal P_\Gamma$ and the scaling exponent $\sigma$ 
are the quantities that are of interest to us in the sequel. We shall also consider the finite size corrections specified by
$\mathcal Q_1, \ \delta_1 \ \ etc. $ 
The quantity (\ref{eq_cos}), (\ref{eq_B}) bears resemblance to the radius of gyration (\ref{Rg}), except
that (\ref{eq_cos}), (\ref{eq_B}) is dimensionless. We also note that (\ref{eq_cos}), (\ref{eq_B}) relates to, but is quite different from,  
the concept of folding angle introduced in \cite{stam}.

In the sequel we shall introduce a chain-specific coarse graining transformation of (\ref{eq_cos}), (\ref{eq_B})  
akin the Kadanoff block-spin transformation \cite{kada,wilson,fisher}
of renormalisation group equations \cite{grosberg-1994,schafer}.
We follow how {\it both}  $\mathcal P_\Gamma$ and $\sigma$  evolve during the ensuing flow, 
and deduce the phase properties of $\Gamma$.
Even though the numerical value of $\mathcal P_\Gamma$ apparently lacks universality,  
the sign of $\mathcal P_\Gamma$  and the numerical value of  $\sigma$ 
are both specific to the phase where the chain resides. 

As an example, consider the  straight rod phase where  $\nu = 1$. Take a chain that has a 
linear structure, such that all the vertices lie in the vicinity of a given straight line. Then, in the limit of large
number of vertices
\beq
{\mathcal P_\Gamma}(N)  \ \buildrel{N \gg 1}\over{\longrightarrow} \  C\, \frac{N(N-1)}{2} \ = \  \frac{C}{2} ( N^2 -N)
\label{str-rod}
\eeq
Here  $C$ characterises the average value of cosine of the angle between two vectors $\mathbf t_i$ 
and $\mathbf t_j$. When the  chain becomes straight so that the vectors $\mathbf{t}_k $ are close to parallel, 
we have $C \to 1$. This example makes it clear that (\ref{eq_cos}), (\ref{eq_B})  can never 
grow faster than $N^2$. We also note the finite size correction which is proportional to $N$ in (\ref{str-rod}), it coincides
with the number of nearest neighbour segments along the chain. Finally, in the case of regular protein structures
the bond angle $\kappa_{i,i+1}$ between two neighbouring C$\alpha$ atoms along a $\beta$-strand has
the value $\kappa_{i,i+1} \approx 1$~(rad)   while along $\alpha$-helices the value is $\kappa_{i,i+1} \approx \pi/2$~(rad)
\cite{frenet}.  
Thus, in the case of $\beta$-stranded proteins we expect a positive valued finite size correction $\mathcal O(N)$
due to nearest neighbour  vertices, while in the case of $\alpha$-helical proteins the 
finite size correction due to nearest neighbours should be tiny.

\subsection{Statistical Ensembles}  

We proceed to develop (\ref{eq_cos}), (\ref{eq_B}) 
into an order parameter that can probe the phase structure of chains.  For this we
analyse the statistical ensemble average
\beq
\langle {\mathcal P_\Gamma(N)} \rangle \ = \ {\rm Tr}\left\{ {\mathcal P_\Gamma(N)} \rho(\Gamma))  \right\}
\label{density-m}
\eeq 
of the observable ({\ref{eq_B}) in the different phases  (\ref{nuval}).
Here $\rho(\Gamma) $ is a density matrix that determines the thermodynamical ensemble. In our numerical simulations we
assume that the system is in a thermodynamical equilibrium state, with
$\rho(\Gamma) $  admitting the Gibbsian form
\begin{equation}
\rho(\Gamma)  \ \propto  \ e^{-\beta H}
\label{gibbs}
\end{equation}
with $\beta$ the temperature factor and $H$ the Hamiltonian of the chain.

\subsubsection{The random walk}

In a random walk the vertices along the chain are mutually independent.  
The density matrix has the form \cite{degen2,grosberg-1994}
\beq
\rho_0(\Gamma) 
= \delta(\mathbf{x}_0) \prod_{i=1}^{N-1} g(\mathbf{x}_{i-1}-\mathbf{x}_i),
\label{rho-0}
\eeq 
where $g(\mathbf{x} -\mathbf{x}')$ is a probability distribution with Gaussian 
pairwise probabilities,
\beq
g(\mathbf{x}-\mathbf{x}') = \left(\frac{1}{2\pi a^{2}}\right)^{3/2}\exp\left[-\frac{1}{2a^2}(\mathbf{x}-\mathbf{x}')^2\right] .
\eeq
We fix the initial point $\mathbf{x}_0$ to the origin, in order to eliminate the  space
volume as an (infinite) overall normalisation factor.
We find 
\[
\langle {\mathbf{t}_i\cdot\mathbf{t}_j} \rangle \ = \ {\rm Tr}\left\{ (\mathbf{t}_i\cdot\mathbf{t}_j) \rho_0(\Gamma)  \right\} 
\]
\beq
= \int \! d ^{ \tiny N+1}  \mathbf{x} \
 ( \mathbf{t}_{i} \cdot \mathbf{t}_{j} )
\delta(\mathbf{x}_0)\prod_{k=1}^{N-1}g(\mathbf{x}_{k}-\mathbf{x}_{k-1}) =  0
\ \  (i\not=j)
\label{vanish}
\eeq 
Thus the ensemble average of the observable (\ref{eq_cos}), (\ref{eq_B}) vanishes in the random walk phase.

\subsubsection{The hard sphere  repulsion and SARW}

In the self-avoiding random walk (SARW)  phase there are  repulsive interactions between the vertices. These interactions
can have a varying range in terms of the spatial separation, from the short distance Pauli (steric) repulsion to the extensive reach of
Coulomb interaction. 
But the effect is always that of a long range interaction, when we measure distance along the chain.
In a weak coupling limit we can try to handle these interactions perturbatively, using a virial expansion 
around a random walk chain.   For this
we assume a homogeneous chain with $N$ vertices,  with  an interaction potential 
of the form 
\beq
E = \hspace{-0.3cm} \sum_{1\leqslant i<j\leqslant N-1}  \hspace{-0.2cm}  U(\mathbf{x}_i -\mathbf{x}_j ) 
\label{eq_E}
\eeq
between the vertices; the summation 
extends over all vertex pairs. The density matrix has the Gibbsian form (\ref{gibbs}),
\beq
\rho(\Gamma)  
= \hspace{-0.4cm} \prod_{1\leqslant i<j \leqslant N-1}\hspace{-0.4cm} e^{-\beta U(\mathbf{x}_i -\mathbf{x}_j )} \prod_{i=1}^{N-1} 
g(\mathbf{x}_{i-1} -\mathbf{x}_i )
\label{eq_rho}
\eeq
We proceed to an explicit calculation of (\ref{eq_cos}), (\ref{eq_B})  in the  simplified case of
excluded volume {\it i.e.} we assume there is only a hard sphere steric  repulsion between the vertices:
\beq
U(\mathbf{x}_i -\mathbf{x}_j ) \  = \ 
\begin{cases} \infty & \mbox{if } \vert\mathbf{x}_i -\mathbf{x}_j \vert\leqslant \Delta 
\\ 0 & \mbox{if } \vert\mathbf{x}_i -\mathbf{x}_j \vert > \Delta \end{cases}
\label{eq_exVolume}
\eeq
Note that in this hard sphere limit  the temperature dependence becomes absent. Thus there are no (temperature dependent) phase
transitions. A single phase  prevails and, in the absence of any other interaction, the chain resides in the SARW phase
by construction.   
 
We follow \cite{grosberg-1994} and introduce  the Mayer function 
\beq
f(\mathbf{x}_i-\mathbf{x}_j) =  e^{-\beta U(\mathbf{x}_i-\mathbf{x}_j)}-1
= \begin{cases} -1, & \mbox{if } \vert\mathbf{x}_i-\mathbf{x}_j \vert\leqslant \Delta \\ \hspace{0.3cm}  0, & \mbox{if } \vert\mathbf{x}_i-\mathbf{x}_j \vert > \Delta \end{cases}
\eeq
and we consider the limit where the hard sphere radius $\Delta$ at the vertex is very small so that
\beq
f(\mathbf{x}_i-\mathbf{x}_j)=-\frac{4}{3}\pi \Delta^3 \delta (\mathbf{x}_i-\mathbf{x}_j )\equiv -B \delta (\mathbf{x}_i-\mathbf{x}_j),
\eeq
Here $B$ specifies the excluded volume around a vertex.
The virial expansion is 
\begin{eqnarray}
e^{-\beta E}&=& \prod_{1\leqslant i<j\leqslant N-1} \hspace{-0.2cm} [1+f(\mathbf{x}_i-\mathbf{x}_j) ]= \nonumber \\
&=&1+\hspace{-0.3cm} \sum_{1\leqslant i<j\leqslant N-1} \hspace{-0.2cm} f(\mathbf{x}_i-\mathbf{x}_j)+\\
&+&\sum_{i,j,k,l}f(\mathbf{x}_i-\mathbf{x}_j)f(\mathbf{x}_k-\mathbf{x}_l)+\ldots \nonumber
\label{virial}
\end{eqnarray}
The  term which is linear in the Mayer function describes collisions between a pair of vertices; 
note that  the linear term engages interactions that have
a long range {\it along the chain} despite being short range {\it in space}.   The 
bilinear  term describes triple collisions,  and so on. In the limit of a dilute chain only pair collisions 
can be relevant. Thus, in this limit we obtain for our observable the second order virial approximation
\beq
\begin{gathered}
\langle {\mathcal P_\Gamma}(N)\rangle \approx 
\int \left\{ d\mathbf{x}_0 \cdots d \mathbf{x}_{N-1} \prod_{i=1}^{N-1} g(\mathbf{x}_{i}-\mathbf{x}_{i-1})\right.
\times\\
\times\left.\left(1-B\hspace{-0.4cm}\sum\limits_{1\leqslant i< j\leqslant N-1}\hspace{-0.4cm}
\delta (\mathbf{x}_i-\mathbf{x}_j )\right) {\mathcal P}_{\Gamma}(\mathbf{x}_0 ,\ldots , \mathbf{x}_{N-1} )\delta(\mathbf{x}_0)\right\}
\end{gathered}
\label{eq_obs}
\eeq
We  substitute 
(\ref{eq_B}) in (\ref{eq_obs}). The integrals are elemental and in the limit where the vertex size is very
small in comparison to the segment length $B \ll a^3$, the result  is
\beq
\langle {\mathcal P}_\Gamma)(N) \rangle =  
\left(\frac{3}{2\pi}\right)^{\frac{3}{2}}\!\!
\frac{B}{2 a^3}\! 
 \hspace{-0.1cm} \sum_{1\leqslant i<j\leqslant N-1}\!\!  
\hspace{-0.cm} \frac{1}{\sqrt{j-i}}
+ \mathcal O(\frac{B}{a^3})^2\!  
\label{eq_B_fin}
\eeq

\subsubsection{The large-N limit in SARW}

For large  $N$ 
we can  estimate the sum in (\ref{eq_B_fin}) using an integral approximation
\beq
\begin{gathered}
\sum_{1\leqslant i<j\leqslant N-1}\left(j-i\right)^{-\frac{1}{2}}\buildrel{N\gg1}\over{\longrightarrow} \int_0^Ndx\int_0^xdy
\frac{1}{\sqrt{x-y}} \sim N^{3/2}
\end{gathered}
\label{eq_integral}
\eeq
We then get for the large-$N$ limit 
\begin{equation}
\langle {\mathcal P}_\Gamma(N) \rangle  \buildrel{N \gg 1}\over{\longrightarrow} 
D\, \frac{B}{a^3}  \, N^{3/2}  \equiv \mathcal P_\Gamma  \, N^{3/2} \ > \ 0 \ \ \ {\rm (SARW)}
%
\label{sarw-as}
\end{equation}
with some chain specific positive constant $D$.
We note 
that the observable (\ref{sarw-as}) is proportional to $N^{3/2}$ while the number of terms 
that contribute in  (\ref{eq_cos}), (\ref{eq_B}) increases like $N^2$, according to (\ref{str-rod}). 
Thus, there must be cancellations of order $N^2$: We conclude that {\it in the leading
order}  there is an equal  contribution from terms with positive and negative values of $\cos \kappa_{ij}$, and the 
result (\ref{sarw-as}) follows due to sub-leading predominance of positively valued $\cos \kappa_{ij}$.
Moreover, since the  $x\to y$ singularity in (\ref{eq_integral}) is integrable, in the large-$N$ limit
the contribution from small separation values of $|i-j|$ becomes insignificant in comparison to the 
contribution from large separation values $|i-j|$. Thus the  fine
details of the interaction potential become increasingly irrelevant. Accordingly, we argue that the dominant
scaling exponent $\sigma = 3/2$  in (\ref{sarw-as}) is {\it universal},  for discrete chains 
in the SARW phase when $N$ is very large.

Finally,  the {\it positive} sign of (\ref{sarw-as}) can be understood as follows: In the SARW phase both
the radius of gyration and the end-to-end distance  increase 
faster in $N$,  than in the RW phase. This implies that there is a tendency in SARW phase, for the 
different vectors $ \mathbf t_i$, $ \mathbf t_j$ to be more parallel to each
other than in the RW phase. In the RW phase the ensemble average of the angle $\kappa_{ij}$ between any two vectors 
$ \mathbf t_i$ and $\mathbf t_j$ is $\pi/2$. Thus,  in the SARW phase
there is an inclination towards $\kappa_{ij} < \pi/2$ implying that (\ref{sarw-as}) is positive.

\subsubsection{Corrections to large-$N$ limit in SARW phase}
\label{subsec:corrections}

Since the result (\ref{sarw-as}) reflects a $\mathcal O(N^2)$ balance 
between positive and negative values of $\cos \kappa_{ij}$,
we can expect that when $N$ is not very large  
the higher order correction terms in (\ref{eq_B}) are notable. To analyse these, we observe that 
oftentimes there is a steric repulsion that enforces a minimum distance between {\it any} two vertices.  
For example,  in PDB protein structures the minimal distance between any two C$\alpha$ atoms 
that do not share a  peptide plane is (practically) always  larger than the diagonal size $\sim$3.8 \AA ~of a peptide plane. 
Thus the angle $\kappa_{i, i+1}$ between  any two neighboring vectors
$\mathbf t_i$ and $\mathbf t_{i+1}$ is always less than $2\pi/3$. In fact, for most proteins we can confirm that 
\cite{frenet}
\[
\kappa_{i,i+1} \leqslant \kappa_{max} \approx \pi/2 \hspace{0.8cm} {\rm (for ~ PDB ~ proteins)}
\]
In the general case, the value of $\kappa_{max}< \pi$ is determined by the ratio 
between the {\it effective} hard-sphere radius (\ref{eq_exVolume}) and the segment length. 
Accordingly, there must be a finite-$N$ correction to (\ref{sarw-as}) which reflects the local details of
steric repulsion. We estimate the correction  in the hard sphere limit, by separating out
the effect of very short distance interactions {\it i.e.} contribution from small values of $k = |i-j|$. For this we simply subtract the 
effect of the ensuing interactions by replacing (\ref{eq_integral}) with
\[
\sum_{1\leqslant i<j\leqslant N-1} \left(j-i\right)^{-\frac{1}{2}} \ \buildrel{ \frac{k}{N} \ll 1}\over{ \longrightarrow } \ \sum\limits_{\substack{i=0 \\ j=i+k} }^{N} (j-i)^{-\frac{1}{2}}
\]
\beq 
\longrightarrow
\ \int\limits_k^N dx \int\limits_0^{x-k}  dy 
\ \frac{1}{\sqrt{x-y}} \ \sim \ \frac{2}{3} N^{3/2} - \sqrt{k} N
\label{eq_integral-2}
\eeq
In lieu of (\ref{sarw-as}) we then have an estimate that excludes the short distance effects of steric repulsion:
\beq
\mathcal P_\Gamma (N) \ \sim \ \mathcal P_\Gamma  \, ( N^{3/2} - \frac{3}{2} \sqrt{k} N )  \ \ \ \ \left( \frac{k}{N} \ll 1\right)
\label{sarw-as2}
\eeq
Note that this result is derived using the limit, where the radius of the hard sphere repulsion becomes small. 
In the case of most proteins we have noted that mostly $\kappa_{i,i+1} \leqslant \pi/2 $ (rad)   thus we expect  that  
the $\mathcal O(N)$ contribution from {\it nearest} neighbour vertices is often non-negative.
Accordingly, in practical scenarios the estimate (\ref{sarw-as2}) should apply when $k$ 
is not very small.

\subsubsection{The collapsed phase}

In the collapsed phase we can not
estimate (\ref{eq_B}) using a perturbation (virial) expansion around an ideal RW chain. 
In the collapsed phase both repulsive and attractive long distance interactions along chain are present, 
the chain properties are reigned by non-perturbative effects. 

According to  (\ref{nuval}) when $N$ increases, for a chain in the collapsed phase the radius of gyration  
grows slower in $N$  than in the RW phase. Thus, in a statistical ensemble 
the angle $\kappa_{i,j}$ between any two vectors $\mathbf t_i$ and $\mathbf t_j$
should have a statistical inclinations towards values that are larger than $\pi/2$.  Otherwise,
a collapsed chain does not curl upon itself at a rate which fills the space faster than in the RW phase. 
Since the primary contribution to  (\ref{eq_B})  derives from large values of $k=|i-j|$,
in accordance with  (\ref{sarw-as2}) we expect that in the collapsed phase and with small values of $k$
\beq
\mathcal P_\Gamma (N) \ \buildrel{N \gg 1}\over{\longrightarrow} \ \mathcal P^{\rm coll}_\Gamma  \left(N^{\sigma} + f(k) N\right) \ < \ 0
\label{coll-as}
\eeq
where the pre-factor $\mathcal P^{\rm coll}_\Gamma<0$. {\it In particular}, the exponent $\sigma> 1$. This is because 
the chain collapse is due to interactions that have a long distance 
along the chain, and the number of possible vertex pairs increases faster in $N$ than the number of  
nearest neighbour vertex pairs. Note the small-$k$ near-neighbour correction term that we have included in (\ref{coll-as}):
As in (\ref{sarw-as2}) there should be  such a  term, it includes 
short distance repulsion between those vertices that are {\it very} close to each other
along the chain {\it e.g.} nearest neighbours. 
The $f(k)$ is some function of the short distance 
cut-off value $k$, in general it is model specific. 

\subsection{Renormalisation group flow}

When the number of vertices $N$ is very large we may coarse grain the chain by repeating a 
Kadanoff block-spin transformation of  the vectors that determine the segments, as shown in Figure~\ref{fig-1}.
%
%
%
%
%
\begin{figure}
\centering
\includegraphics[width=0.4\textwidth]{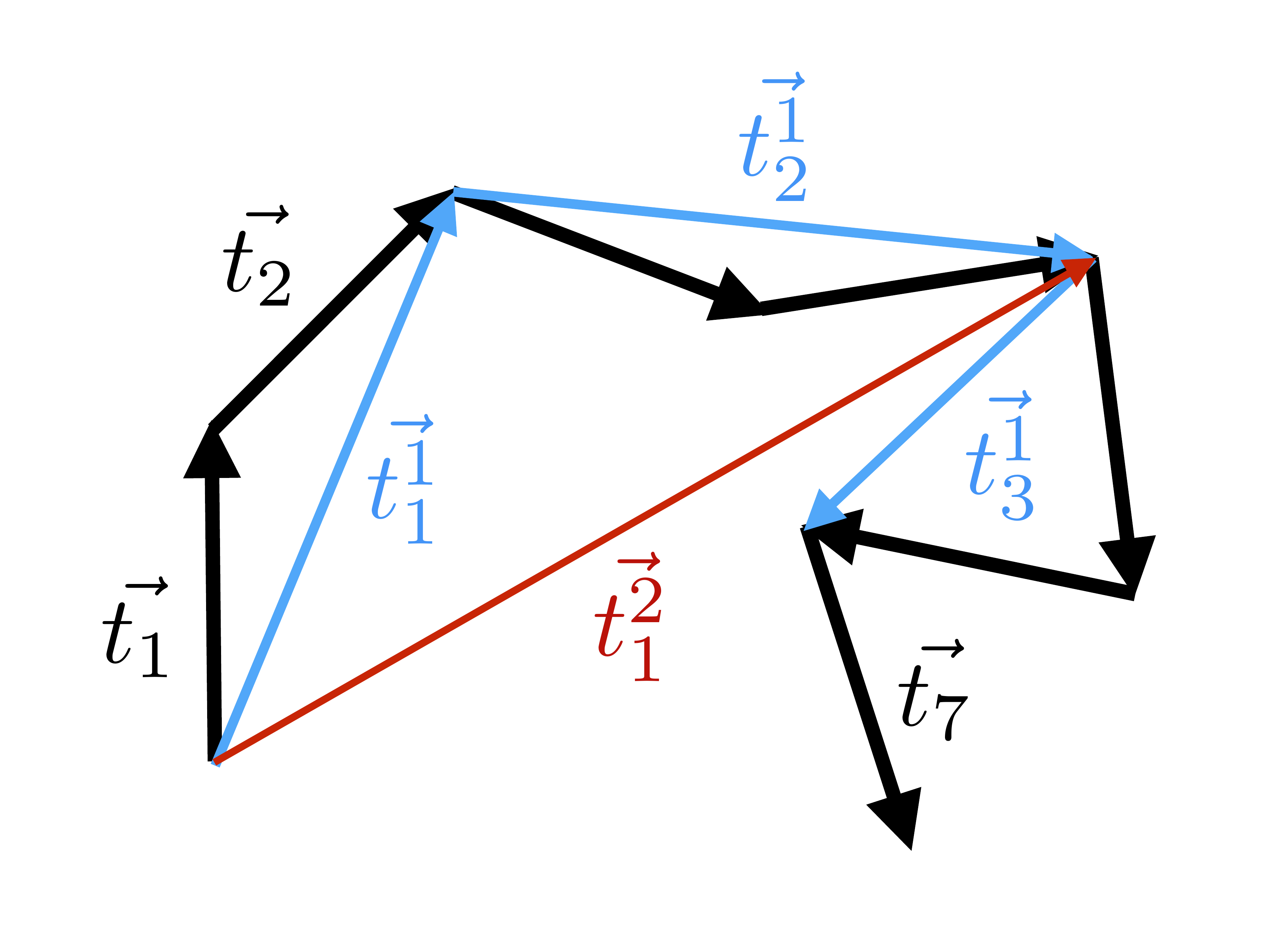}
\caption{(Color online) A scaling  akin to Kadanoff's block-spin transformation that 
successively combines two preceding segments into one following 
segment, and eliminates the middle vertex. The segments in the initial chain are bold black (vectors $\mathbf{t}_1, \dots
\mathbf{t}_7$), those in the first level of iteration are blue (vectors $\mathbf{t}^1_1, \dots
\mathbf{t}^1_3$), and the thinnest red segment (vector $\mathbf{t}_1^2$) is obtained at the second level of iteration. } 
\label{fig-1}
\end{figure}
This gives rise to a renormalisation group (RG) evolution of $\mathcal P_\Gamma(N)$.
The Figure~\ref{fig-2} shows the phase diagram that we expect to find in the case of a homopolymer, for very large values of $N$;
we deduce the phase diagram from our preceding analysis of (\ref{eq_B})  - see also  
\cite{degen2,grosberg-1994}.  For the moment we overlook the straight rod phase. 
%
%
%
%
%
%
%
\begin{figure}
\includegraphics[width=0.45\textwidth]{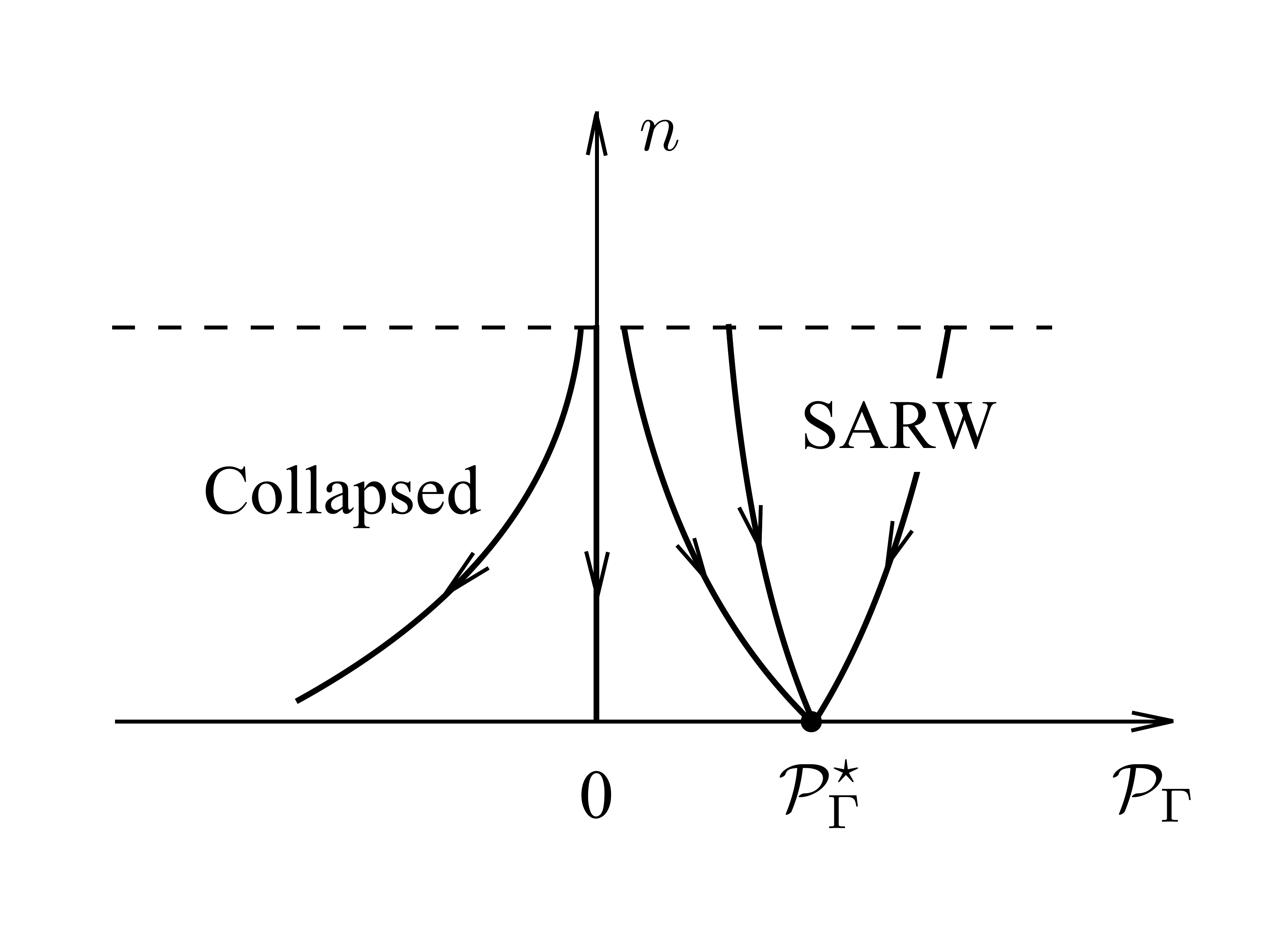}
\caption{Expected RG flow of the coefficient $\mathcal P_\Gamma$ in (\ref{eq_B}) in different phases, with  
$\mathcal P_\Gamma = 0$ the Gaussian fixed  
point and $\mathcal P^\star_\Gamma$  the fixed point of SARW. } 
 \label{fig-2}
\end{figure}
As shown in the Figure, we have found that in the SARW phase the strength of repulsive
interaction between vertex pairs (second virial  coefficient) evolves
towards a non-trivial positive fixed point value \cite{degen2,grosberg-1994}. 
Consequently, by  repeated action of
the block-spin transformation shown in Figure~\ref{fig-1} we expect the observable (\ref{density-m}) 
to flow towards a  fixed positive value, in the SARW phase. 
We expect this value to be universal, quite 
independently of the underlying energy function (\ref{eq_E}). 
On the other hand, since (\ref{density-m}) vanishes in the RW phase the ensuing RG evolution
defines a vertical basin of attraction towards the Gaussian fixed point, as shown in Figure~\ref{fig-2}. This 
flow separates the SARW phase from  the  collapsed phase, where the flow is towards a negative value
of $\mathcal P_\Gamma$. The collapsed phase is commonly assumed to correspond to the space filling fixed point 
value (\ref{nuval}) of the scaling exponent $\nu$, in the case of a homopolymer.

However, we note that there are {\it numerous} examples of discrete space chains 
with geometrically nontrivial attractors. A generic, deterministic and  chaotic 3D  flow 
approaches an attractor that can have {\it a priori} an arbitrary fractal Hausdorff dimension. This opens the
possibility for a more complex phase structure, also in the case of discrete chains that deserves to be addressed.
We conclude that {\it at this point} we expect the following correspondences between the phase 
of a chain, the sign of (\ref{density-m}) and the numerical value of $\sigma$: 
\begin{table}[tbh]
\caption{Phases in terms of the new observable.}
\vspace{10mm}
\begin{tabular}{|c|c|c|}
\hline 
~~~{\rm phase} ~~~ & ~~~$\langle {\mathcal P_\Gamma} \rangle$~~~ &  ~~~ $\sigma$  ~~~      \\

\hline\hline
\vspace{-0.3cm}
& & \\
Rod & $>\ 0$ & 2 \\
SARW & $>\ 0$ & $\approx$ 3/2 \\
RW & = 0 & $- \hspace{-0.1cm}-$ \\
Collapsed & $< \ 0 $ & $ > \ 1 $  \\
\hline
\end{tabular}
\label{table-1}
\end{table}

\section{Coarse graining chains}

The scaling transformation shown in Figure~\ref{fig-1}  is a direct adaptation of Kadanoff's
block-spin transformation. It decreases the number of segments at an exponential rate. 
Thus  a chain becomes very rapidly coarse grained. For example, a typical protein backbone 
with a couple of hundreds C$\alpha$  atoms can support  only a handful of block-spin transformations. 
This is hardly sufficient to  define a smooth RG flow, not to mention the identification of distinct
length scales  that
govern the  chain properties at intermediate distance scales. 

\subsection{New scaling procedure for chains}

We proceed to develop a chain specific variant of the block-spin transformation. One that can be iterated a large number of times, 
comparable
to the number of vertices in the chain.
With the help of our new coarse graining transformation  we then hope to detect and identify
the different length scales that characterise a given chain, even when there are only 
a relatively few vertices such as in the case of a generic protein backbone. 

We start by introducing   a scaling parameter $s$. We define it to be the number of old segments which are connected by the new one,
during a coarse graining process.
In the case of the conventional (Kadanoff) block-spin transformation $s$ is always an integer.
For example, in Figure~(\ref{fig-1}) we have $s=2$. For $s=3$ we connect every third vertex, while for $s=1$ we 
simply repeat the original chain. 

Canonically, in the case of a spin system the parameter $s$ can only have integer values. But in the case of a chain, it turns out that 
we can promote $s$ into an {\it a priori} arbitrary number and here we are particularly interested in  values 
$s\in (1,2]$. For this we introduce a new coarse graining procedure, and in Figure~\ref{fig-3} we show how it 
proceeds when $s=4/3$:
%
%
%
%
%
%
%
%
\begin{figure}
\includegraphics[ width=0.37\textwidth]{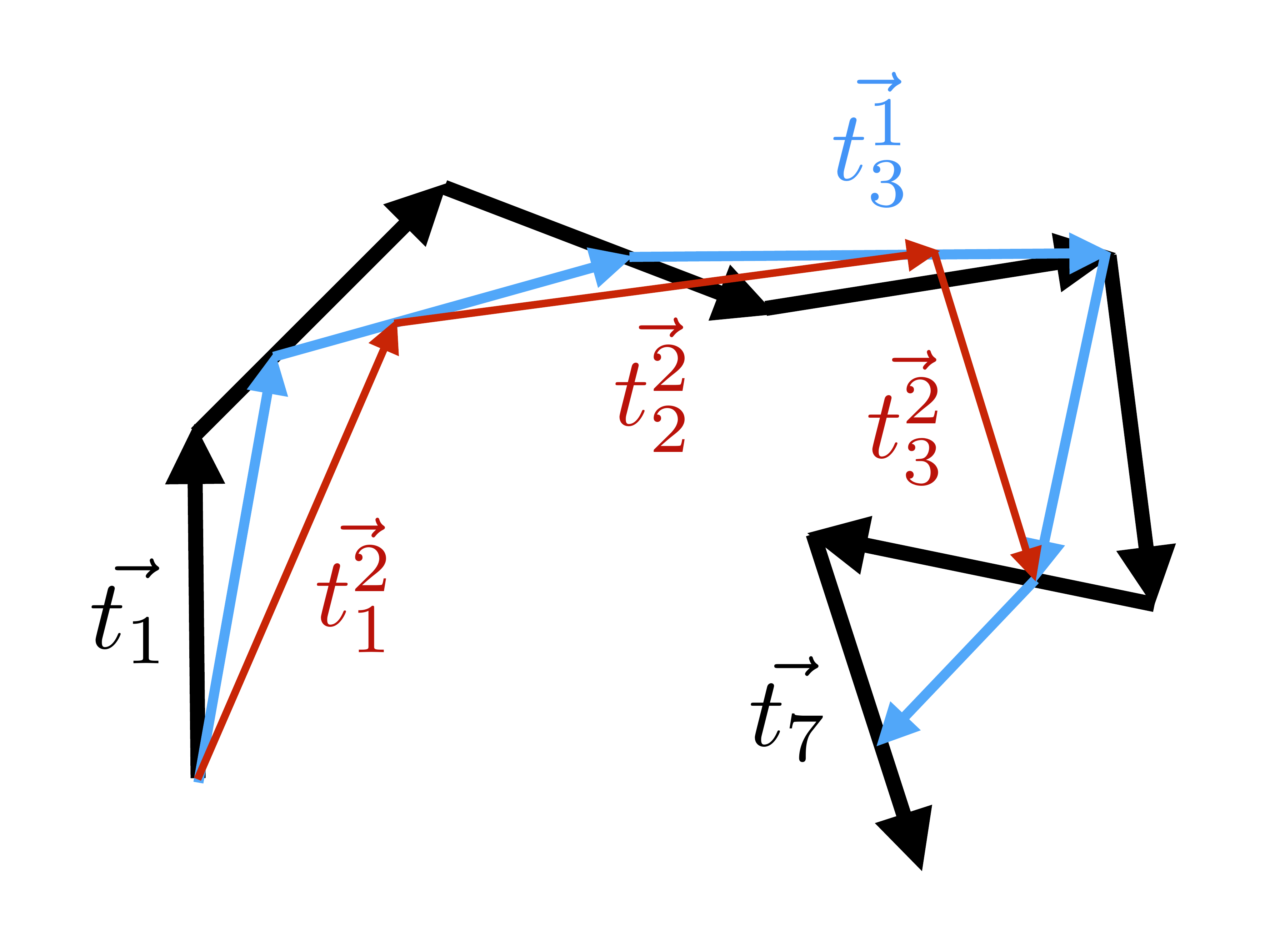}
\caption{(Color online) Coarse graining procedure for scaling parameter 
$s=4/3$. The initial chain is bold black (vectors $\mathbf{t}_1, \dots
\mathbf{t}_7$), the first step of the coarse graining 
procedure is thin blue arrows (vectors $\mathbf{t}^1_1, \dots
\mathbf{t}^1_5$) and the second step is the thinnest  red lines (vectors $\mathbf{t}^2_1, \dots
\mathbf{t}^2_3$).}
\label{fig-3}
\end{figure}
We initiate the coarse graining with the vectors  $\mathbf{t}_i$ that determine 
the segments of the chain, at the current iteration level.  
We then define the vector $\mathbf{t}_1^{new}$  
which determines the  first segment of the following iteration step by 
\[
\mathbf{t}_1^{new} = \mathbf{t_1} + \frac{1}{3}\mathbf{t_2}
\]
To construct  the second segment $\mathbf{t}_2^{new}$ in the chain of the following iteration step,  
we  add the remaining two thirds of $\mathbf{t_2}$ together with
two thirds of $\mathbf{t_3}$,
\[
\mathbf{t}_2^{new} = \frac{2}{3}\mathbf{t_2} + \frac{2}{3}\mathbf{t_3}
\]
Finally, for $\mathbf{t}_3^{new}$ we add one third of $\mathbf{t_3}$ and $\mathbf{t_4}$, so that 
\[
\mathbf{t}_3^{new} = \frac{1}{3}\mathbf{t_3} +\mathbf{t_4}
\] 
The third vertex of the following iteration step then coincides with the 4$^{th}$ vertex of the preceding iteration step. 
The process  is  repeated with $\mathbf{t_5}$ and so forth, until the entire chain becomes covered.  
Note that as shown in the Figure~\ref{fig-3},  the last vertex of the preceding chain is  not necessarily 
reached by the last vector of the following chain. The Figure shows  this in the case when the 
preceding chain has seven vertices and we have chosen $s=4/3$. By repeating the coarse 
graining,  at the end of the second iteration (red line in the Figure)  we again 
miss part of the end in the preceding  chain. This loss of structure at the end of the chain can be avoided by choosing 
the scaling parameter $s_p$ at iteration step $p$ so that 
\beq
N_p = s_p m 
\label{inZ}
\eeq
where $N_p$ is the number of vertices at the iteration level $p$ and $m$ is some integer.
Thus, the smallest value we can choose for $s_p$ is
\beq
s^{opt}_p = \frac{N_p}{N_p -1} \ = \ 1 + \frac{1}{N_p -1}
\label{sopt}
\eeq
Now the end points of the chain do not move, but the scaling parameter varies with the iteration step. However, this variation is quite small.
As an example, for a chain with  300 
vertices which is quite typical in the case of a protein backbone,  we 
estimate that after  $\sim$200  iteration steps  the optimal value $s^{opt}_p $ becomes changed  
by less than 0.7\% as shown in Figure~\ref{fig-4}.
%
%
%
%
%
%
%
%
\begin{figure}
 \includegraphics[width=0.45\textwidth]{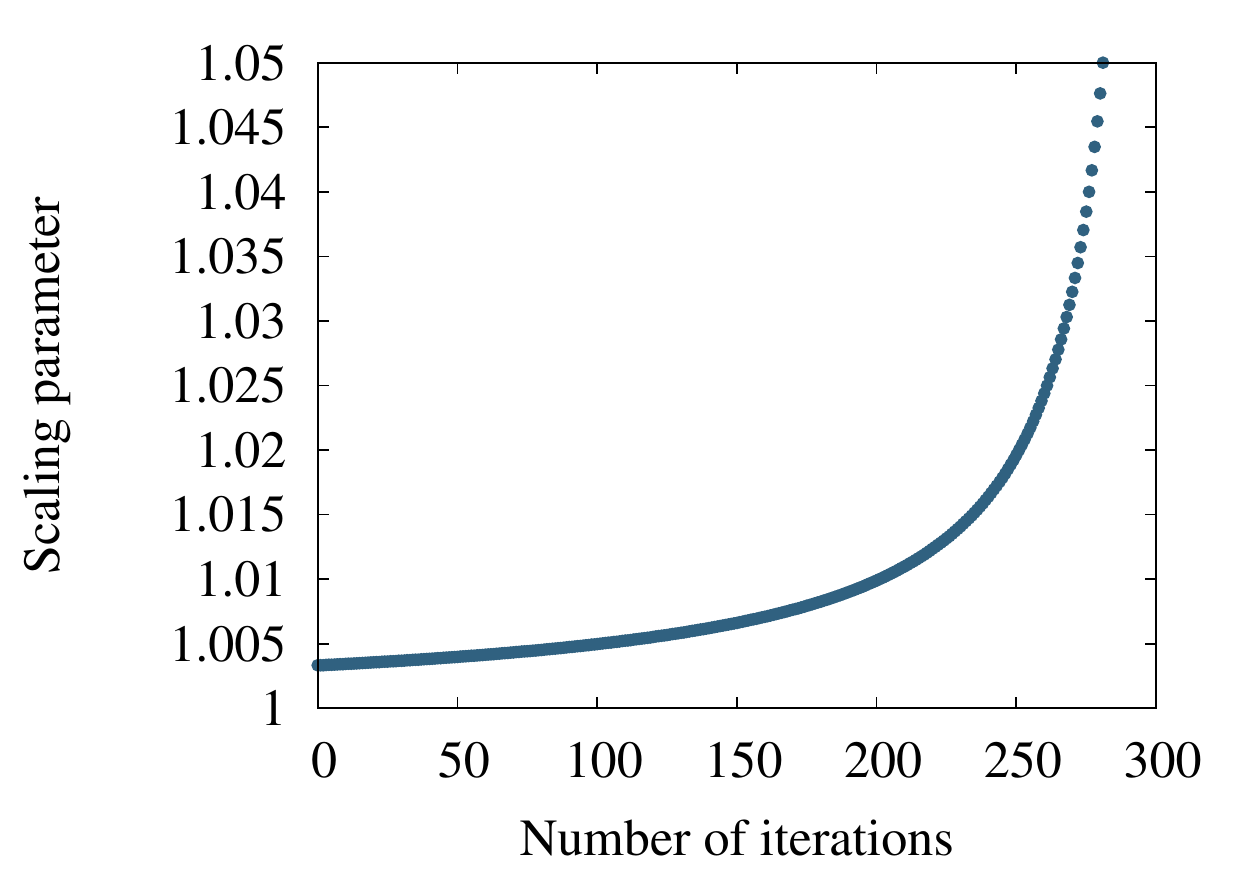}
 \vskip 0cm 
 \caption{(Color online) Dependence of the optimal scaling parameter 
 $s^{opt}_p$ value on the number of iteration steps $p$, for a chain 
with 300 initial vertices.} 
 \label{fig-4}
\end{figure}

Figure~\ref{fig-5} shows the effect of coarse graining on the chain geometry.  
The effect is to suppress any abrupt short wave-length oscillation in the geometry; 
those sections of the chain with many twists and turns become more regular as shown in the Figure:
A chain becomes visibly smoother while preserving its overall shape, 
as the coarse graining advances. 
%
%
%
%
%
%
%
%
\begin{figure}
  	\includegraphics[width=0.45\textwidth]{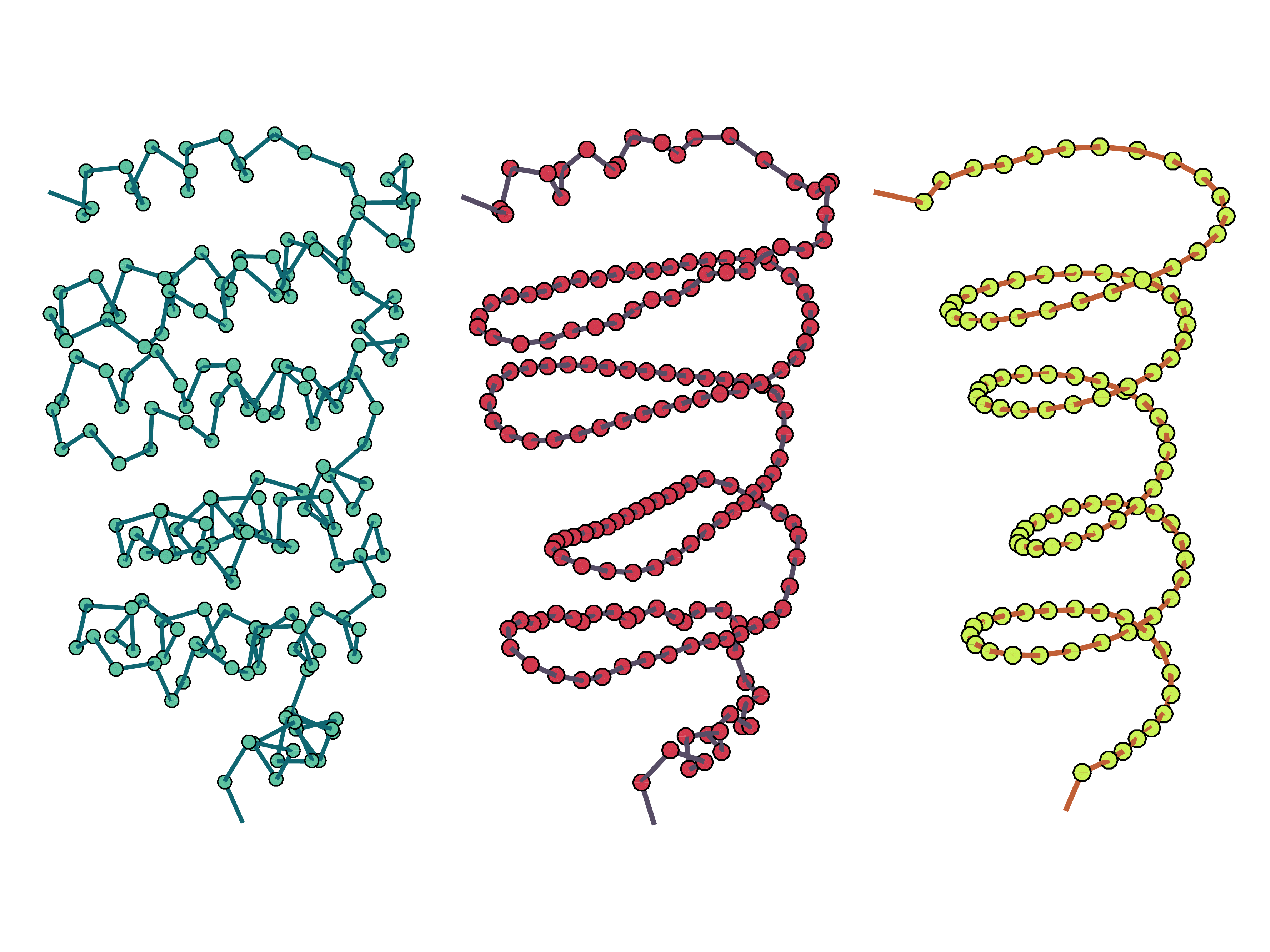}
	\caption{(Color online) Illustration of the effect of the coarse graining procedure for a helical chain; the example is from the C$\alpha$ backbone
	of PDB structure 5DN7. On the left is the PDB structure, in the middle the structure after 10 coarse graining steps, and on the right after 80 coarse graining steps.}
\label{fig-5}
\end{figure}
\section{Homopolymer model}

We shall employ (\ref{eq_B_fin}) in combination with our coarse graining procedure to investigate 
the homopolymer phase structure numerically, using a universal energy function.

\subsection{Frenet frames}

To fully describe chain geometry, we need to introduce a framing. 
For this we consider four generic consecutive vertices $\mathbf{x}_{i-1}, \mathbf{x}_{i},\mathbf{x}_{i+1}, \mathbf{x}_{i+2}$ along 
a piecewise linear string-like chain. Let  $\mathbf{t}_i, \mathbf{t}_{i+1},\mathbf{t}_{i+2}$ be the three segments that connect these
four vertices.
For each vertex we evaluate the ensuing bond ($\kappa$) and torsion ($\tau$) angle as follows: 
The bond angle is obtained directly in terms of the segments,
\beq
\kappa_{i} \ \equiv \ \kappa_{i+1 , i} \ = \ \arccos \left( \frac{\mathbf{t}_{i+1} \cdot \mathbf{t}_i}{\vert\mathbf{t}_{i+1} | | \mathbf{t}_i\vert }\right).
\label{bond}
\eeq
For the torsion angles we first introduce the normal vector 
\beq
\mathbf{b}_i = \mathbf{t}_{i-1}\times\mathbf{t}_i
\eeq
of the $(\mathbf x_{i-2}, \mathbf x_{i-1}, \mathbf x_i)$ plane.  The torsion angle $\tau_i$ is then
\beq
\tau_{i} \ \equiv \ \tau_{i+1,i} \ = \ {\rm sign} [(\mathbf{b}_{i-1}\times \mathbf{b}_i)\cdot\mathbf{t}_i] 
\times\arccos\left(  {\mathbf b}_{i+1} \cdot {\mathbf b}_i \right) 
\eeq
Note that a torsion angle can be introduced 
whenever  $\mathbf{t}_{i-1}$ and $\mathbf{t}_{i}$ are linearly independent.
We also note that  ($\kappa_i, \tau_i)$ yield spherical coordinates around $\mathbf x_i$, and that
the three vectors ($\mathbf t_i, \ \mathbf b_i, \ \mathbf n_i = \mathbf b_i \times \mathbf t_i$) constitute an 
orthogonal frame at this vertex.

Inversely, once the bond and torsion angles and in addition the segment lengths are known, we recover the chain as follows:
From the angles we  first compute the frames ($\mathbf t_i, \mathbf b_i, \mathbf n_i$) 
using the discrete Frenet equation, as described in \cite{frenet}. The entire chain is then given by the solution of
the discrete Frenet equation
\[ 
\mathbf x_{i} \ = \ \sum\limits_{k=1}^{i}  \mathbf t_k 
\] 

\subsection{Landau free energy}

We deduce the  Landau free energy of a chain using a symmetry principle \cite{oma-old,ulf}: 
The energy function of a structureless,  piecewise linear discrete chain should not depend on the
way the chain is framed. 
Thus the energy function must remain intact under frame rotations around the segment vectors $\mathbf t_i$. Let
($\mathbf e^1_i, \mathbf e^2_i$) denote  two orthogonal unit vectors that relate to ($\mathbf b_i, \mathbf n_i$)  by a generic SO(2)
rotation around $\mathbf t_i$.   The 
orthogonal basis ($\mathbf t_i, \mathbf e^1_i, \mathbf e^2_i$) could then be used instead of the Frenet basis 
($\mathbf t_i, \mathbf b_i, \mathbf n_i$), to construct the energy function.
Mathematically, this determines  a local SO(2) gauge structure. 

The C$\alpha$ backbone of a protein is 
akin our piecewise linear discrete chain, with an average $\sim$3.8 \AA ~ distance between vertices; from this perspective,
the only influence of side-chains is to introduce a heterogeneous interaction between the C$\alpha$ atoms.
Therefore, the C$\alpha$ backbone must employ a SO(2) invariant energy function of the (virtual) 
backbone bond and torsion angles.  
The bond angles $\kappa_i$ transform like a two-component SO(2) scalar field 
and the torsion angles $\tau_i$ transform like a SO(2) gauge 
field under a local frame rotation \cite{oma-old,ulf}. {\it  Universality } then implies that
the leading order C$\alpha$ energy function for a protein backbone with $N$ residues (vertices) must relate to
the lattice Abelian  Higgs Model (AHM)  Hamiltonian. This follows directly because the AHM Hamiltonian 
is the most general SO(2) gauge invariant Hamiltonian there is. 
In  the unitary gauge AHM Hamiltonian coincides with 
the following discrete nonlinear Schr\"odinger (DNLS) Hamiltonian  with a spontaneously broken symmetry 
~\cite{oma-old,ulf,hu-13,theo-2014,theo-2015,ivan}
\[
 E(\kappa,\tau)  = 
\sum\limits_{i=1}^{N-1}  (\kappa_{i+1}-\kappa_i)^2 
+   \sum\limits_{i=1}^N \left\{ \lambda\, (\kappa_i^2 - m^2)^2
+ \frac{d}{2} \,  \kappa_i^2  \tau_i^2 \right.
\]
\begin{equation}
\left.  - \ b \kappa_i^2 \tau_i  - a  \tau_i + \frac{c}{2} \tau_i^2\right\}  
\ + \ \sum\limits_{i\not= j} U(\mathbf x_i - \mathbf x_j)
\label{eq:A_energy}
\end{equation}
Here ($\lambda,m,a,b,c,d$) are parameters, they are specific to a given amino acid sequence in the case of a protein. 
The terms in the first row coincide with 
a {\it naive} discretisation of the continuum nonlinear Schr\"odinger
equation. On the second row, the  first term ($b$) is the conserved momentum in the DNLS model,  the second ($a$) 
is the Chern-Simons term, and the third ($c$) is the Proca mass term; see \cite{hu-13,theo-2014,theo-2015,ivan} for detailed analysis. 
We note that both momentum and Chern-Simons terms are chiral.  

Besides the terms that we have 
displayed explicitly in (\ref{eq:A_energy}) there are also two-body interactions (\ref{eq_E}) 
that have a long range along the chain, and are governed by the last term in (\ref{eq:A_energy}).
These interactions include Pauli exclusion,  electromagnetic,  van der Waals {\it etc.} 
interactions between the various
atoms.  Here we consider a simple homogeneous variant of $U(\mathbf x_i - \mathbf x_j)$
that in addition of the hard sphere 
(Pauli) repulsion (\ref{eq_exVolume}) has a spatially short range 
attractive component, 
\begin{equation}
U(r) \ = \left\{ { { \ + \infty \ \ \  \ \ \ \ \ \ \ \ \ \ \ \ \ \ \ \   \quad   0 < r < R_0 } \atop
 {\ U_0 \! \left\{ \tanh(r-R_0)-1\right\}  \quad  R_0  < r < +\infty } } \right.
\label{U}
\end{equation}
For $r<R_0$ there is a hard-core  
repulsion but for $r>R_0 $ there is a spatially short range attractive interaction with 
strength determined by the parameter $U_0$. In the case of proteins we choose $  R_0 \sim
\Delta = 3.8 \ {\rm \AA} $
which is the distance between two neighboring C$\alpha$ atoms.  We refer to \cite{Ann} for  a detailed analysis of 
the effects of long range (along chain) interactions in (\ref{eq:A_energy}).

\subsubsection{Cooperativity and first order phase transition}

The free energy (\ref{eq:A_energy})  can be validated by verifying its compatibility with Privalov's criterion \cite{priva-1,priva-2,priva-3}. 
It states that protein folding is a cooperative process which in the case of a short two-stage folding protein resembles 
a first order phase transition.

For (\ref{eq:A_energy}) cooperativity is  
due to solitons that are supported by the DNLS equation \cite{cherno}, solitons are the paradigm cooperative organisers 
in numerous physical scenarios. Here a soliton 
emerges when we first eliminate the torsion angles using their equation of motion,
\begin{equation}
\tau_i[\kappa] \ = \ \frac{a+b\kappa_i^2}{c+d\kappa_i^2}
\label{tau}
\end{equation} 
For bond angles we then  obtain
\begin{equation}
\kappa_{i+1} = 2\kappa_i - \kappa_{i-1} + \frac{dV[\kappa]}{d\kappa_i^2}  
 \label{kappaeq}
\end{equation}
where
\[
V[\kappa] = - \left( \frac{bc-ad}{d}\right) \, \frac{1}{c+d\kappa^2}  - \left( \frac{b^2 + 8 \lambda m^2}{2b}\right) \kappa^2 + \lambda \kappa^4
\]
The difference equation (\ref{kappaeq}) can be solved iteratively using the algorithm developed in  \cite{nora}. 
A soliton solution
models a super-secondary protein structure such as a helix-loop-helix motif, with the loop corresponding to 
the soliton proper. 

To identify the putative first order transition character we observe that in the 
case of a protein, the  bond angles are  rigid and slowly varying while the torsion angles are highly flexible. 
Thus, over sufficiently large distance scales we may try and proceed self-consistently in a Born-Oppenheimer 
approximation,  using a mean field $\kappa_i \sim \kappa$ and then solving for $\kappa$ in terms of torsion 
angles.  From (\ref{eq:A_energy})  
\begin{equation}
\frac{\delta E}{\delta \kappa} = 0 \ \Rightarrow \ \kappa^2 \ = \ m^2 + \frac{b}{2\lambda} \tau - \frac{d}{4\lambda} \tau^2 \ 
\label{mftk}
\end{equation}
In those cases that are of interest to us, this equation always has a solution:  Both $\kappa$ and $\tau$ are multivalued angular variables, 
and for proteins the parameters $b$ and $d$ are small in comparison with $m^2$ and $\lambda$.
We substitute the solution into (\ref{eq:A_energy}) which gives for the energy
\begin{equation}
 -  \frac{d^2}{16 \lambda} \tau^4 + \frac{bd}{4\lambda} \tau^3 - \left( \! \frac{b^2}{\lambda} - 
2dm^2 - 2c \right) \tau^2 + \left( a+bm^2 \right) \tau 
\label{mftF}
\end{equation}
We identify here the canonical form of the de Gennes free energy 
of a first order phase transition ~\cite{DeGennes-book}. This completes our
qualitative validation of (\ref{eq:A_energy}) in line with Privalov's criterion 
~\cite{priva-1,priva-2,priva-3}, at the level  of mean field theory. 

We conclude with the following comment:  
Despite the suggestive analogy between (\ref{mftF}) and the de Gennes free energy of a first order transition,
a chain collapse from the SARW phase to a space filling phase proceeds 
through an intermediate that includes the random walk phase; see Figure~\ref{fig-2}. 
The intermediate can be either a tricritical $\theta$-point \cite{huggins,flory1,flory2,flory3,degen2} in which case we encounter the characteristics of a first order phase transition in line with Privalov's criterion, or it can be an extended $\theta$-regime 
possibly with
its own internal structure, possibly including molten globule folding intermediates \cite{ptitsyn-1,ptitsyn-2}:  
An analysis at the level of a Landau-Ginsburg theory is suggestive, 
but not sufficient,  in determining the character of a phase transition. Entropic corrections
are important for chain collapse, and  
accounted for in  the usual manner  of Landau-Ginsburg-Wilson theory \cite{golden}. 


\subsubsection{Radius of gyration vs. temperature and two-stages of collapse}

To scrutinize the details of chain collapse in the homopolymer model, we 
investigate the temperature dependence of the radius of gyration using numerical simulations.
We employ the heat bath algorithm that has been
detailed in \cite{Ann}. Our parameter values for  (\ref{eq:A_energy}) are 
shown in Table \ref{table-2}
\begin{table}[tbh]
\caption{The parameter values  in (\ref{eq:A_energy})  during our simulations.}
\vspace{10mm}
\begin{tabular}{|c|c|c|c|c|c|c|}
\hline 
~~~$\lambda$~~~ & ~~~$m$~~~ & ~ $a$ ~ & ~ $b$ ~ & ~ $c$ ~ & ~$d$~  & ~ $U_0$ ~  \\
\hline
 $3.5$ & $1.5$ & $10^{-4}$ & $0$ & $10^{-4}$  & $10^{-4}$ & $0.5$ \\
\hline
\end{tabular}
\label{table-2}
\end{table}
where the numerical value of  $m$ corresponds 
to $\alpha$-helical protein structures; for $\beta$-stranded chains we choose $m=1$.  The parameters that  relate to
torsion angles are relatively small, in comparison to those that relate to bond angles only.
This is in line with proteins where bond angles are known to be  quite rigid while torsion angles 
are often found to be highly flexible; see the analysis in connection of Eqs. (\ref{mftk})-(\ref{mftF}).

The Figure~\ref{fig-6} shows how the value of radius of gyration (\ref{Rg}) increases with increasing temperature factor, 
in the case of a homopolymer
chain with $N=300$ 
vertices. Here $T=1/\beta$ is the inverse of the Gibbs temperature factor in (\ref{gibbs}). We find that
at low temperatures $\log_{10}T< 0$ the chain is in the collapsed phase, where the radius of gyration is temperature independent. 
%
%
%
%
%
%
%
%
\begin{figure}
\includegraphics[
width=0.45\textwidth]{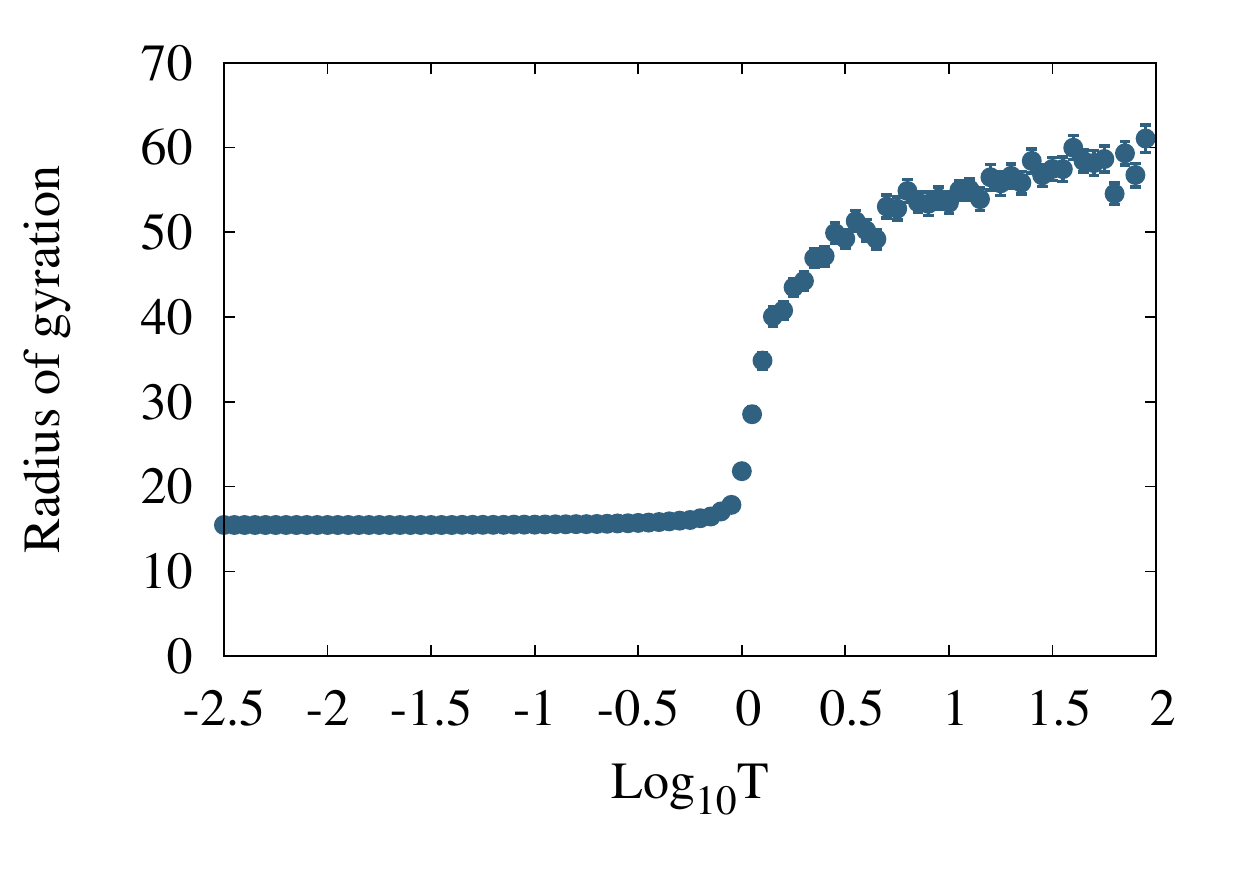}
\caption{(Color online) The evolution of the radius of gyration as a function of temperature factor $T = 1/\beta$, in the homopolymer model
with $N=300$ vertices.
}
\label{fig-6}
\end{figure}
When temperature is (roughly) in the range  $0 <\log_{10}T< 0.5 $ 
the chain is in the transient $\theta$-region where the radius of gyration 
rapidly increases as a function of the temperature; the RW phase is located in this
$\theta$-region.  
For $\log_{10}T> 0.5$ the chain enters the SARW phase where the 
radius of gyration value eventually stabilises  into a temperature independent value. The apparent 
two-state character with lack of structure in the $\theta$ regime is in line with the two-stage folding 
nature (Privalov's criterion) of the Landau free energy function, in the case of a homopolymer 
chain. We refer to \cite{Ann} for additional 
details of the chiral homopolymer phase structure.

\section{Random Chain simulations }

From Figure~\ref{fig-6} we confirm that the RW phase appears in the phase diagram of the homopolymer model  (\ref{eq:A_energy})
in the $\theta$-regime, between the high temperature
SARW phase and the low temperature collapsed
phase \cite{Ann}. The width of the $\theta$-regime 
relates to finite size effects, in this regime the radius of gyration  
is very sensitive to temperature variations. 
%
%
%
%
%
%
%
%
\begin{figure}
\includegraphics[width=0.45\textwidth]{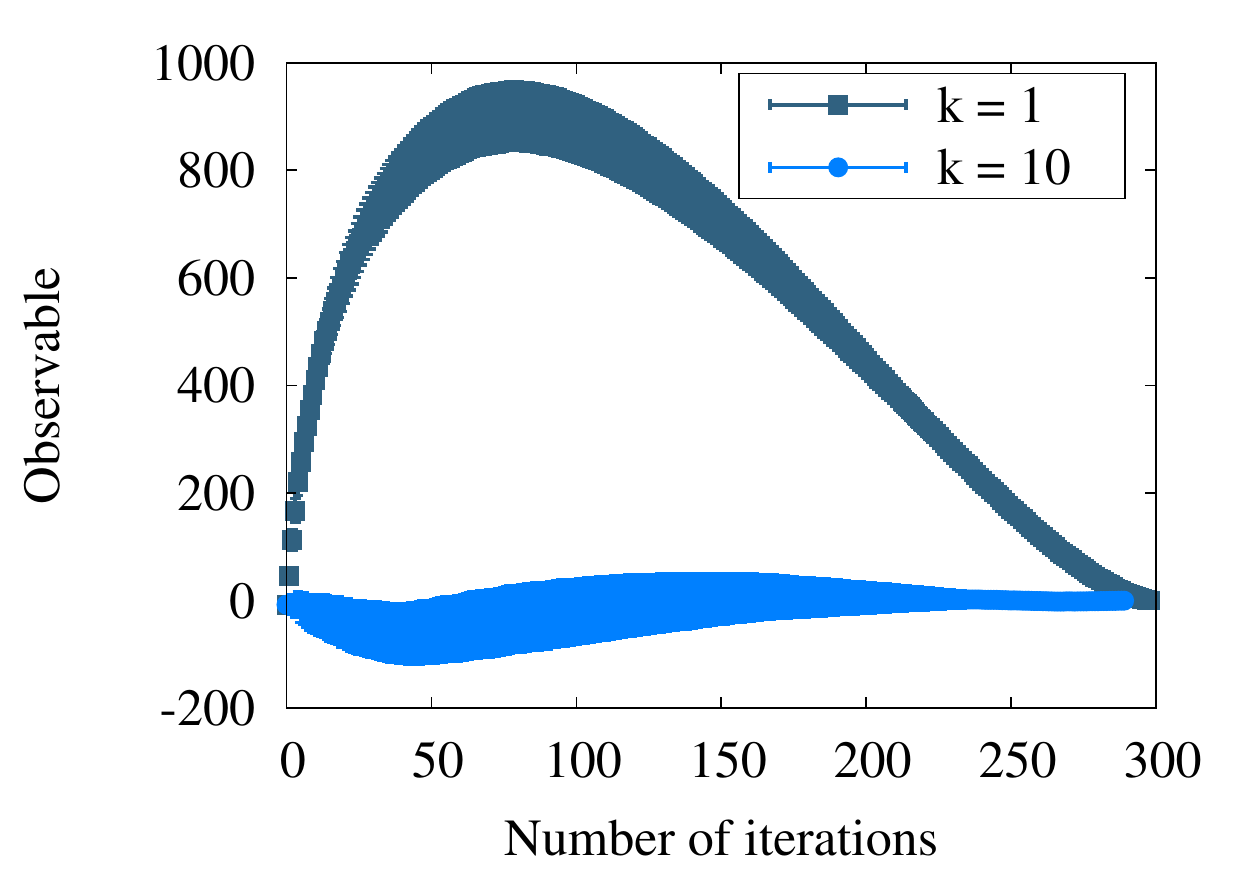}
\caption{(Color online) The evolution of the observable (\ref{density-m}) for $k=|i-j|\geqslant 1 $ and $k=|i-j|\geqslant 10 $
in (\ref{eq_cos_cutoff}). The figure shows the ensemble average of 128 simulations together with the one $\sigma$ deviation 
distance from the average value, in each case.}
\label{fig-7}
\end{figure}
Accordingly we find it delicate to try and describe a statistical ensemble of RW phase homopolymer chains  with  energy function 
(\ref{eq:A_energy}), for the exact ideal value $\nu = 1/2$ of the scaling exponent in (\ref{R}). 
Instead we proceed to simulate the RW phase directly. For this we
promote $\kappa_i\in [0,2\pi)$ and $\tau_i\in [0,2\pi)$ into independent random variables. We fix the 
segment length to a constant value, {\it e.g.} 3.8 (\AA ). In particular, 
we ignore 
all the effects of the energy function (\ref{eq:A_energy}) in the Gibbsian, including the short distance Pauli repulsion.

The Figure~\ref{fig-7} shows the evolution of the observable  (\ref{density-m}) in the RW model, as a function of 
coarse graining steps and in the case of a chain with $N=300$ initial vertices. The lateral axis depicts the progress of
the iterative coarse graining procedure.
In our simulations 
we use the value $s=s^{opt}$ that we determine from (\ref{sopt}) for the scaling parameter. This enables us to iterate
the coarse graining $n=300$ times. 

In the sequel we investigate the sensitivity of the observable to short distance corrections (see discussion below and in Sec.~\ref{subsec:corrections}) by modifying (\ref{eq_cos}) using a short-distance cutoff $k$ as follows:
\begin{equation}
{\mathcal P_\Gamma}(N)  \to 
\sum_{i=1}^N \sum_{j=i+k}^N
\langle \cos \kappa_{ij} \rangle,
\label{eq_cos_cutoff}
\end{equation}

The Figure~\ref{fig-7} shows  the evolution of (\ref{density-m}) both
when we account for all values of the segment separation {\it i.e.} when we have  
$k=|i-j|\geqslant 1 $ in (\ref{eq_cos_cutoff}), and when we eliminate the short segment distance effects {\it i.e.} we only account for 
pairs with $k=|i-j|\geqslant 10 $ in (\ref{eq_cos_cutoff}). 

In the case when we include all values $k\geqslant 1$ in Figure~\ref{fig-7},  the observable initially vanishes in line with (\ref{vanish}). 
When we proceed to coarse grain,  
the value of the observable starts rapidly increasing. It then decreases, approaching a vanishing value towards the  
end of the coarse graining process.
The intermediate  increase in the value of the observable can be understood as follows: 
Consider the blue segments (arrows) in  Figure~\ref{fig-3} that show the outcome of the first coarse graining 
step.  The first blue segment connects the first vertex of the initial chain to the second (black) segment 
of the initial chain. The second blue segment then connects the 
second segment  of the initial (black) chain to the second vertex of  the coarse grained chain, 
located on the third segment of the initial chain.  The fact that both coarse grained
segments engage the same (second)   segment of the initial chain introduces a correlation between the 
(blue) coarse grained segments; the nearest neighbour segments along the coarse grained chain are 
not mutually fully independent. This interdependence, caused by the coarse graining process, implies that 
the observable (\ref{eq_cos}) does not vanish during the flow. Instead, after initially increasing, the observable
decreases towards a vanishing value when the number of coarse graining
steps becomes very large. At the end of the flow, when there are only three
vertices and two connecting segments left, the observable vanishes:  The angle between the two final segments 
is randomly distributed.

The Figure~\ref{fig-7}  shows also the evolution of the observable, once
we remove the contribution of the first 10 nearest neighbour pairs, those with $k=|i-j| \leqslant 10 $. This removal of 
short distance correlations, caused by the coarse graining procedure, yields an 
observable that is in line with the RW phase behaviour, one that vanishes with one standard deviation precision. 
Thus the  result shown in the Figure~\ref{fig-7}  with $k=|i-j| \leqslant 10 $ suggests that the correlations 
introduced by the coarse graining procedure have a short range, in terms of segments along the chain.

In Figure~\ref{fig-8} (a)
%
%
%
%
%
%
%
%
\begin{figure}
	\begin{subfigure}{0.45\textwidth}
		\includegraphics[width=0.99\textwidth]{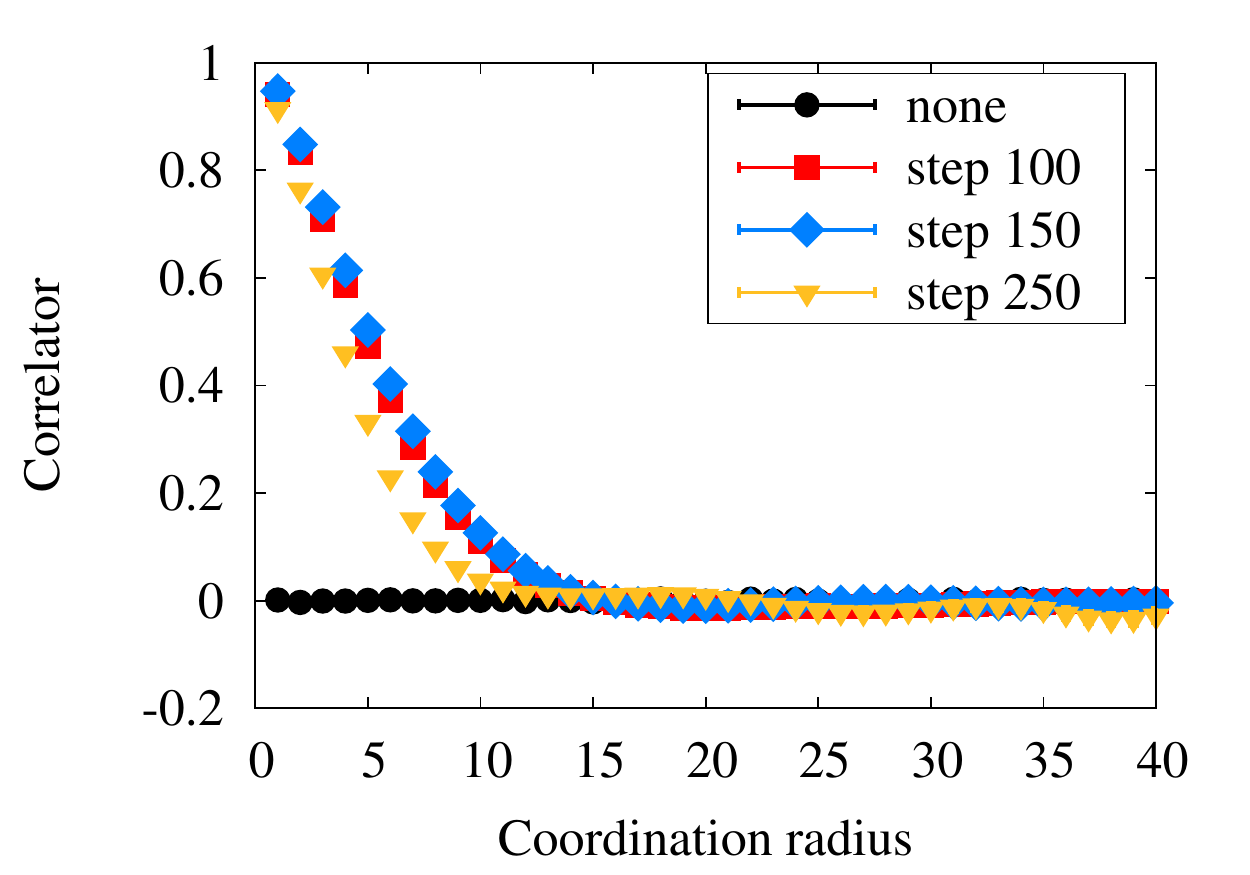}
 		\label{fig-8a}
 		\vskip -0.5cm 
 		\caption{(Color online) The dependence of the correlator (\ref{corr-theta}) on the coordination radius $k=|i-j|$ for different numbers of coarse graining steps.}
	\end{subfigure}
  \hfill
  	\begin{subfigure}{0.45\textwidth}
  		\includegraphics[width=0.99\textwidth]{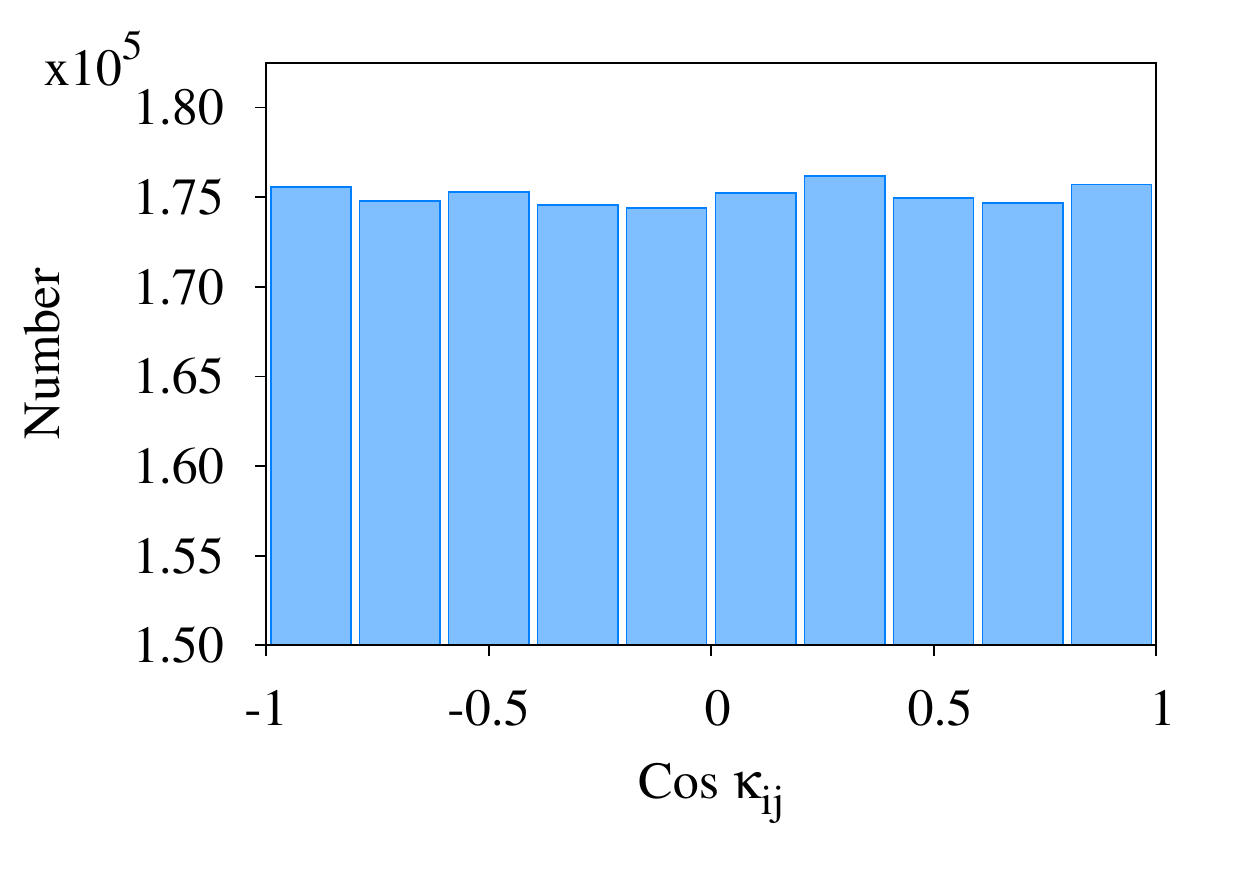}
 		\label{fig-8b}
 		\vskip -0.7cm 
 		\caption{(Color online) Histogram for the values of $\cos\kappa_{ij}$ after $n=150$ iteration steps, $k=10$ nearest neighbour subtraction.}
	\end{subfigure}
	\caption{(Color online) The RW phase.}
\label{fig-8}
\end{figure}
we show how the correlation length of the cosines in (\ref{eq_cos})
\begin{equation}
G(k) \ = \ < \cos \kappa_{ij}>
\label{corr-theta}
\end{equation}
depends on the segment distance $k=|i-j|$ and on the number $n$
of coarse graining steps. We find that quite independently of the number of 
coarse graining steps, the quantity (\ref{corr-theta}) decays at an (apparently) 
exponential rate in $k$, so that after around $k \sim 10$
these correlations are vanishingly small; in the Figure~\ref{fig-8} (a)  we display the correlation
length of (\ref{corr-theta}) after  $n=100, \ 150$ and  $250$ coarse graining steps.  

We conclude that in the RW phase
there are finite size effects due to short distance correlations between the coarse grained segments. But these  correlations
have a short range and become vanishingly small beyond $k = |i-j| \sim$10, in the RW phase.
The results shown in Figure~\ref{fig-8} (b) confirm this: In this Figure  we display 
the distribution of $\cos\kappa_{ij}$ for $k=10$ and
after $n=150$ coarse graining steps.

We note that the histogram in Figure \ref{fig-8} is in line with what we can expect in 
RW phase: In RW phase, since the value of the observable vanishes,  
we expect the distribution of $\cos \kappa_{ij}$ to  
be symmetrical with respect to $\cos \kappa = 0$, 
Our simulations confirm that  the histogram tends to a uniform flat distribution for $k=|i-j| \gg 1$, as expected.

\section{Homopolymer simulations}

We proceed to investigate (\ref{density-m}) in combination with our coarse graining, 
in the SARW and collapsed phases of the homopolymer model (\ref{eq:A_energy}). We use the 
parameter values shown in Table \ref{table-2}.
 In our simulations we
employ the heat bath algorithm that has been detailed in \cite{Ann}. 
We study chains with $N=300$, $N=700$ and $N=1000$ initial segments. 
We control the thermodynamical phase by adjusting the ambient temperature in the heat bath algorithm 
\cite{Ann}. We coarse grain the chains using the optimal scaling parameter  (\ref{sopt}).  
The number of vertices then decreases slowly, and the number of coarse grain iterations supported by the chains
becomes comparable to the number of initial vertices.

\subsection{Scaling effects on radius of gyration}

We first 
analyse how the radius of gyration (\ref{Rg}) evolves under coarse graining, in the SARW and collapsed phases.
The Figure~\ref{fig-9}  
%
%
%
%
%
%
%
\begin{figure}
\includegraphics[
width=0.45\textwidth]{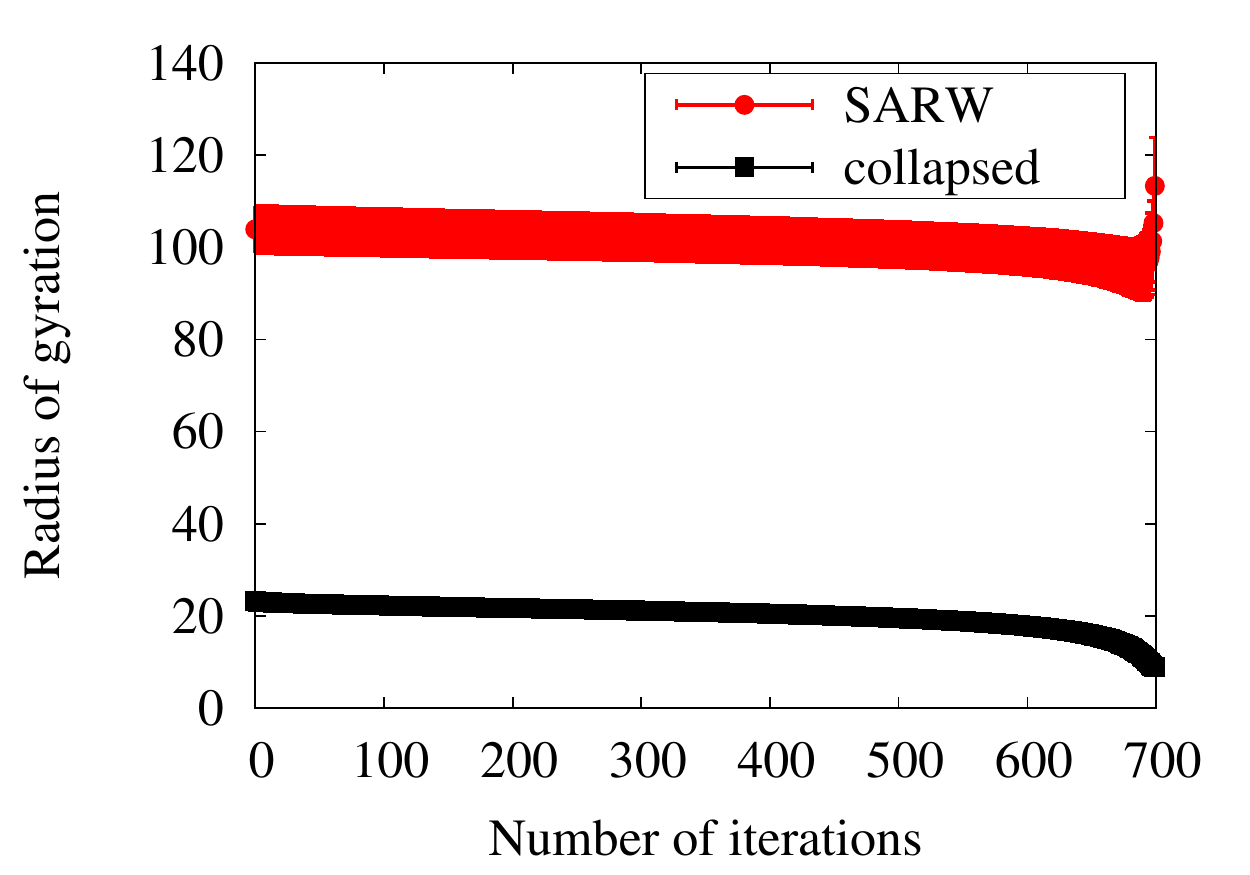}
\caption{(Color online) Variation of radius of gyration vs. number of coarse graining steps, for two different phases. 
The initial chain has 700 vertices, the error-bars are for one standard deviation.}
\label{fig-9}
\end{figure}
shows the result for a chain with $N=700$ initial segments. The stability of the radius of gyration 
during coarse graining proposes that the 
chain preserves its overall geometry as the coarse graining proceeds. Note that  for a renormalisation group
flow which builds on our coarse graining procedure, the radius of gyration would appear to be akin a renormalisation
group invariant quantity. 

The Figure~\ref{fig-10}
%
%
%
%
%
%
%
\begin{figure}
\includegraphics[
width=0.45\textwidth]{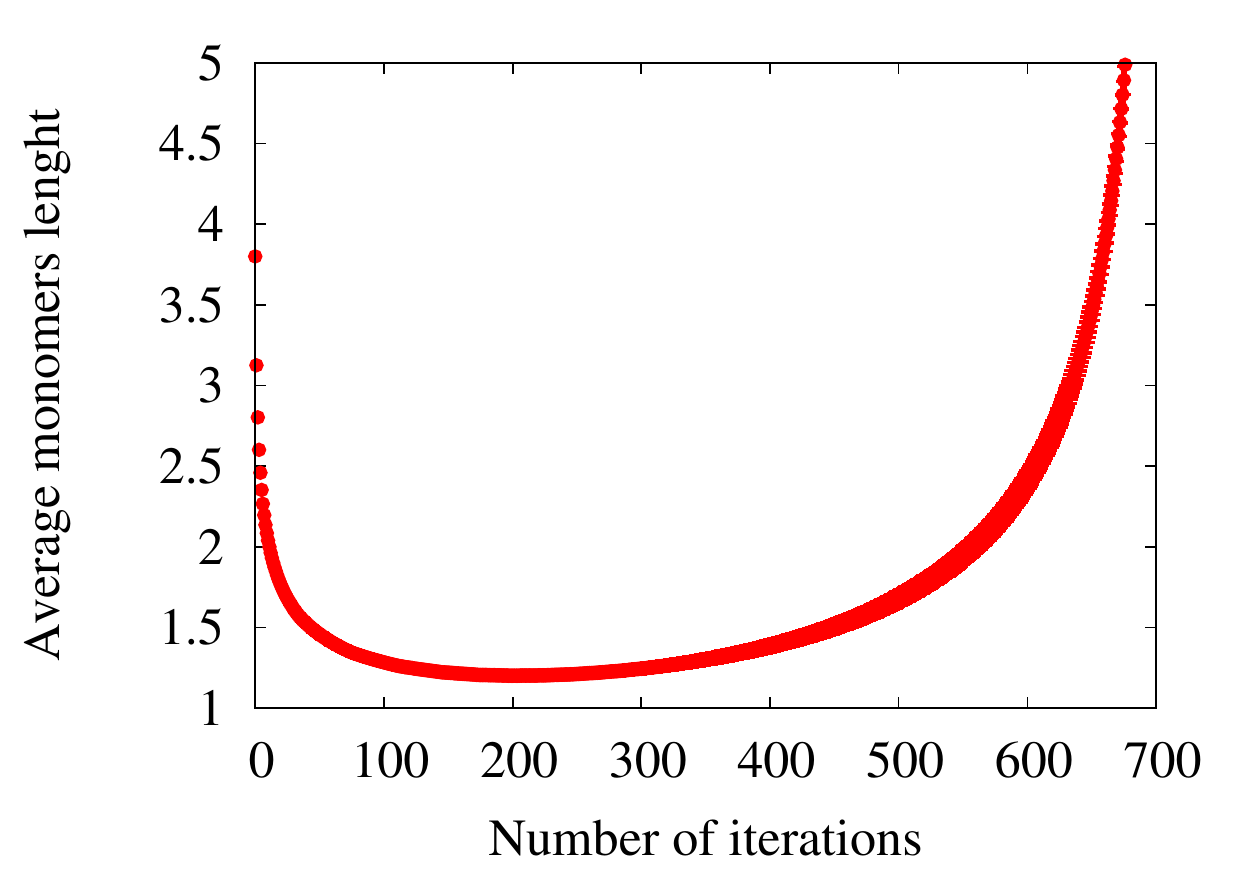}
\caption{(Color online)  Dependence of the average segment length on scaling step $n$ for $N=700$ initial segments 
in the SARW phase. 
}
\label{fig-10}
\end{figure}
shows how the effective segment (Kuhn)  length varies during the coarse graining process for a chain in the SARW phase, with 
an initial segment length of 3.8 (\AA ) and $N=700$ initial segments. 
We observe that,  with the parameter values in Table \ref{table-2}, initially the effective segment decreases and 
reaches a a minimum value $\sim$1.9 (\AA ) after 
around 200 coarse graining iterations. Subsequently the effective segment length increases, and eventually it becomes
comparable to the radius of gyration of the initial chain when the coarse graining terminates. 
This can be understood so that  
initially, the effect of coarse graining is to suppress any abrupt short wave-length oscillation in the geometry; 
those sections of the chain with many twists and turns become more regular, in line with Figure~\ref{fig-5}.
This leads to an initial decrease in the segment length. 
Eventually, when the coarse graining progresses, since $s>1$ the effective chain length then starts increasing. 

%


\subsection{The observable}

We proceed to investigate how the  statistical ensemble average (\ref{density-m}) evolves during repeated coarse graining, 
in the SARW and collapsed phase of the homopolymer model.

\subsubsection{Homopolymer in the SARW phase} 

We evaluate the statistical average of the observable (\ref{density-m}) using the homopolymer in the SARW phase, with
chains that have $N=300$ and $N=1000$ initial vertices. 


We recall that for a RW chain the correlations between neighbouring vertices vanish; see  for example (\ref{vanish}) and Figure
\ref{fig-8} (a). But we have pointed out that in RW phase, coarse
graining  introduces correlations between neighbouring vertices. In Figure~\ref{fig-8} (a) we estimate that 
these  correlations have a finite extent in the RW phase, 
they appear to be effectively vanishing when vertices are 
a distance of $k \sim 10$ segments apart.   

In the SARW phase the correlation length can be expected to be longer,
there are native correlations between vertices along the entire
chain such as Pauli repulsion that ensures self-avoidance and acts between
any pair of vertices. We estimate
to what extent the {\it additional} correlations that are introduced by the coarse graining process, interfere with the
correlations that are native to  the SARW phase.

Figure~\ref{fig-11} shows our simulations 
results for the correlation length (\ref{corr-theta}) in the SARW phase homopolymer model, using various levels of coarse graining.
%
%
%
%
%
%
%
%
\begin{figure}
 \includegraphics[width=0.45\textwidth]{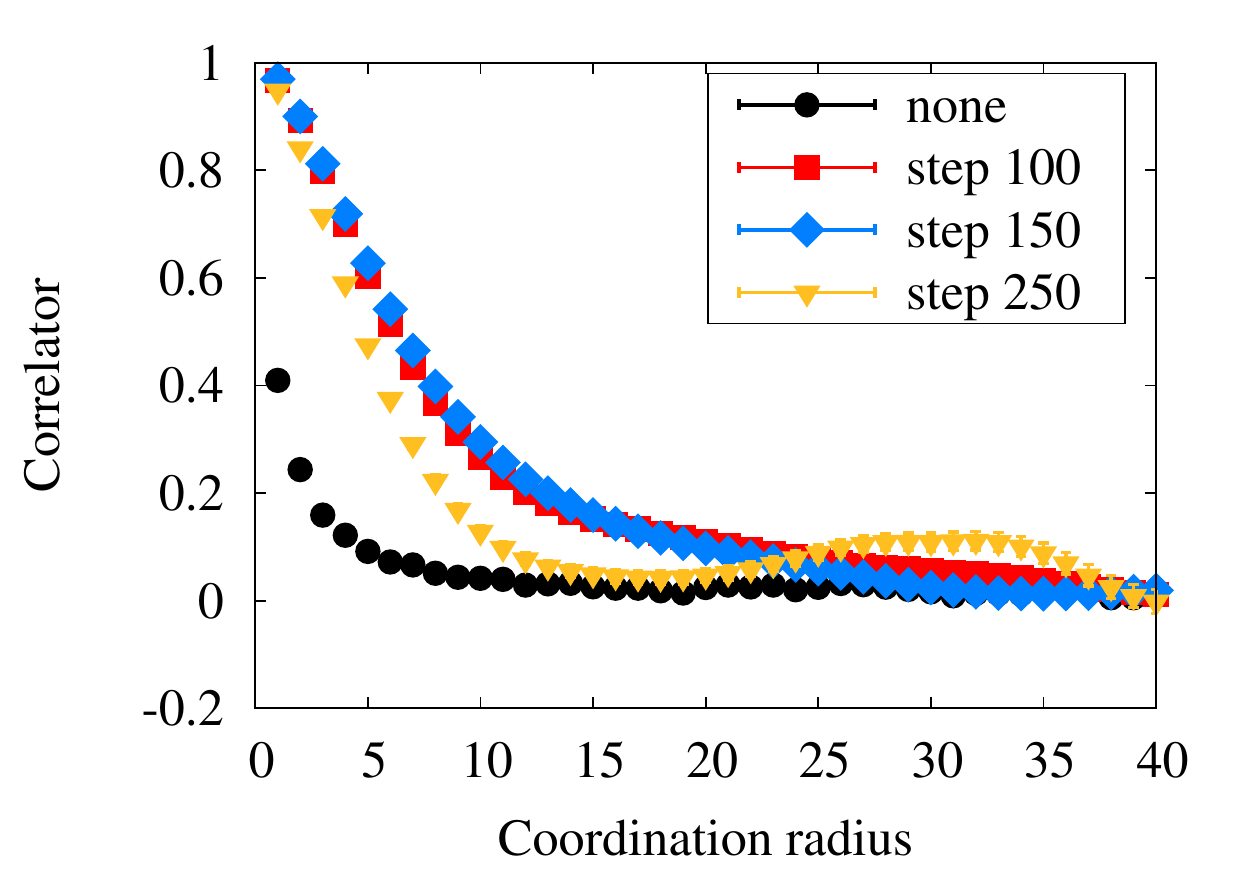}
 \vskip 0cm 
 \caption{(Color online) The dependence of the correlator (\ref{corr-theta}) on the coordination radius $k=|i-j|$ for different numbers of coarse graining steps in the SARW phase of a homopolymer.}
\label{fig-11}
\end{figure}
We observe that in line with the RW, in the SARW phase the coarse graining introduces short range correlations between 
vertices. But these correlations, together with the effect of Pauli repulsion,
seem to be observable only up to distances that are $k= |i-j| \sim 20$ segments apart from each other
along the chain, in our model. Moreover, already after $k\sim 10$ the influence becomes quite small.

In Figures  \ref{fig-12} we show simulation results for the flow of  observable (\ref{eq_cos_cutoff}) under the coarse graining, 
in  the SARW phase. In these Figures we can 
compare
the case $k=1$ where we  sum over all pairs in the observable (\ref{eq_cos_cutoff}), with the case $k=10$
where we only consider the contribution from those pairs where the vertices are a minimum segment distance 
$k=|i-j|\geqslant 10$ apart from each other along the chain. 
%
%
%
%
%
%
%
%
\begin{figure}
	\begin{subfigure}{0.45\textwidth}
 		\includegraphics[width=1\textwidth]{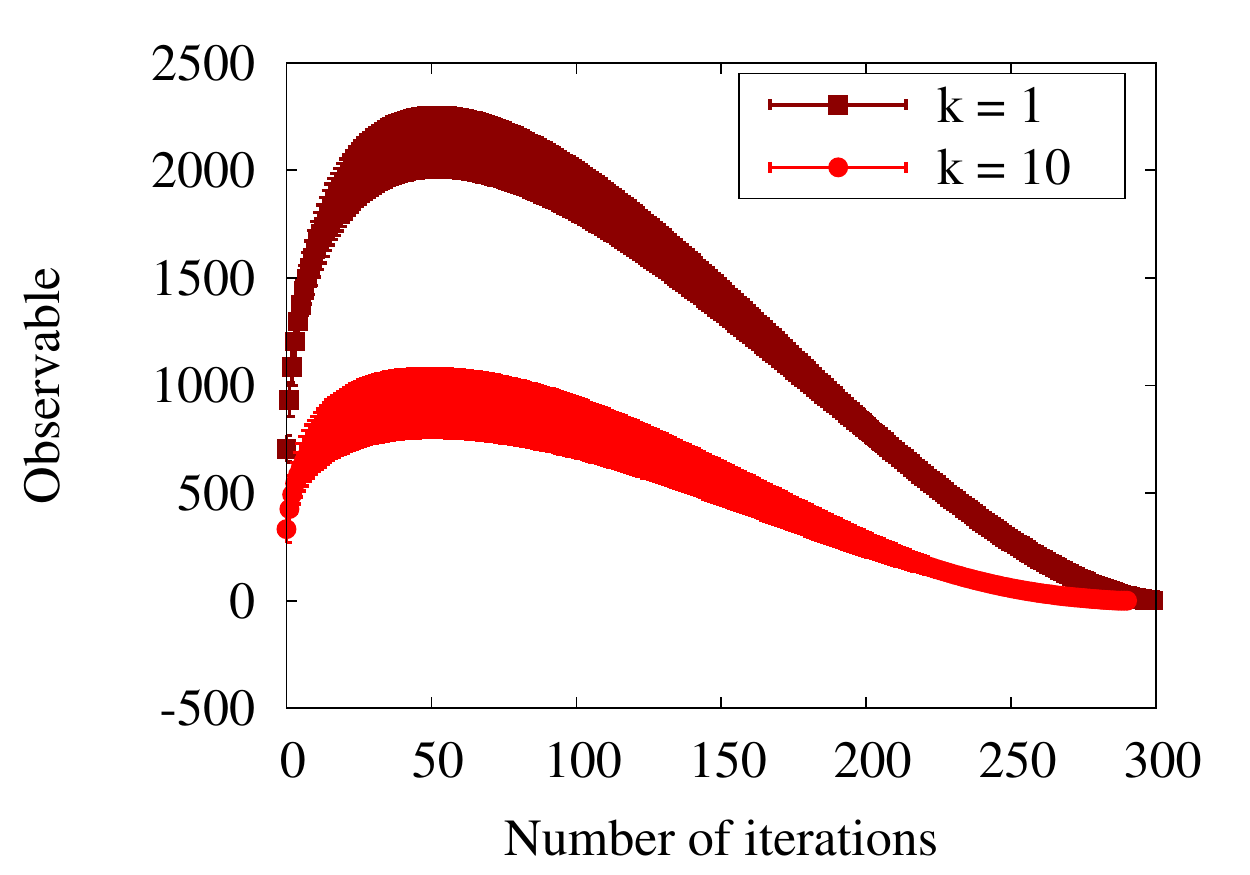}
 		\vskip 0cm 
  		\caption{$N=300$ initial vertices.}
 	\end{subfigure}
 	\hfill
 	\begin{subfigure}{0.45\textwidth}
 		\includegraphics[width=1\textwidth]{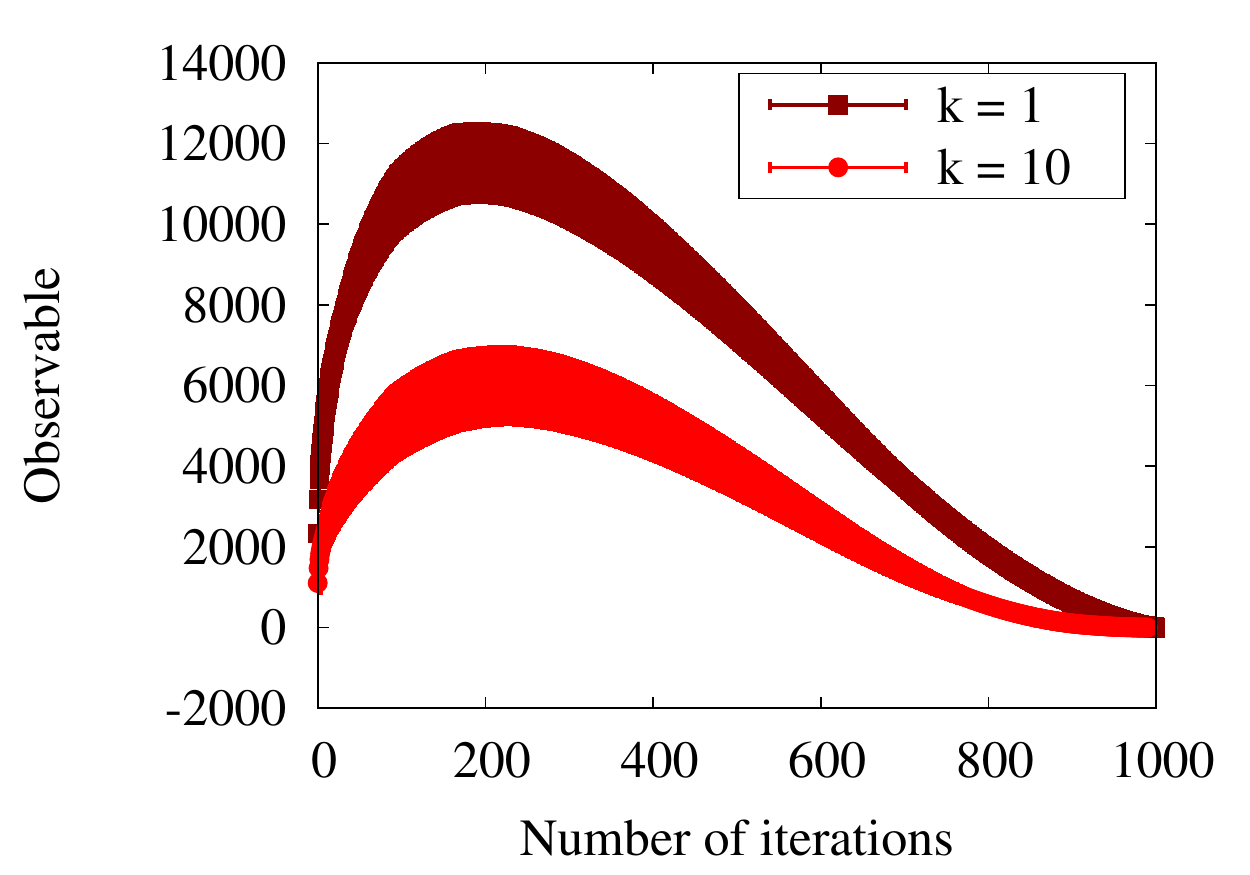}
 		\vskip 0cm 
  		\caption{$N=1000$ initial vertices.}
 	\end{subfigure}
 \caption{(Color online) Dependence of the observable (\ref{eq_cos_cutoff}) in the SARW phase on the number of scaling steps $n$ for  chains with various number of initial vertices, 
 and with different values of subtraction $k$. The average value is shown, together with the one standard deviation fluctuation regime.} 
 \label{fig-12}
\end{figure}
We observe that overall, the profiles in the Figures display self-similarity in their shape. The 
same conclusion persists  for larger values of $k$:
For a homopolymer in the SARW phase, 
the only visible finite size effect on the observable seems to be, that the height of the curve becomes lower as
$k$ increases. In particular, each of the curves in Figures (\ref{fig-12}) have initially a positive value, both 
display convergence towards vanishing values as the coarse graining proceeds. 

The qualitative behaviour shown in Figures  (\ref{fig-12}) 
is a characteristic of the SARW phase. In particular, the value of the observable is positive throughout, in line with 
Table \ref{table-1}.

In Figures  \ref{fig-13} we show the histograms for our statistical ensemble of  the $\cos\kappa_{ij}$ of
(\ref{eq_cos}) for $N=300$ initial vertices, 
in the SARW phase.
%
%
%
%
%
%
%
%
\begin{figure}
	\begin{subfigure}{0.45\textwidth}
		\includegraphics[width=1\textwidth]{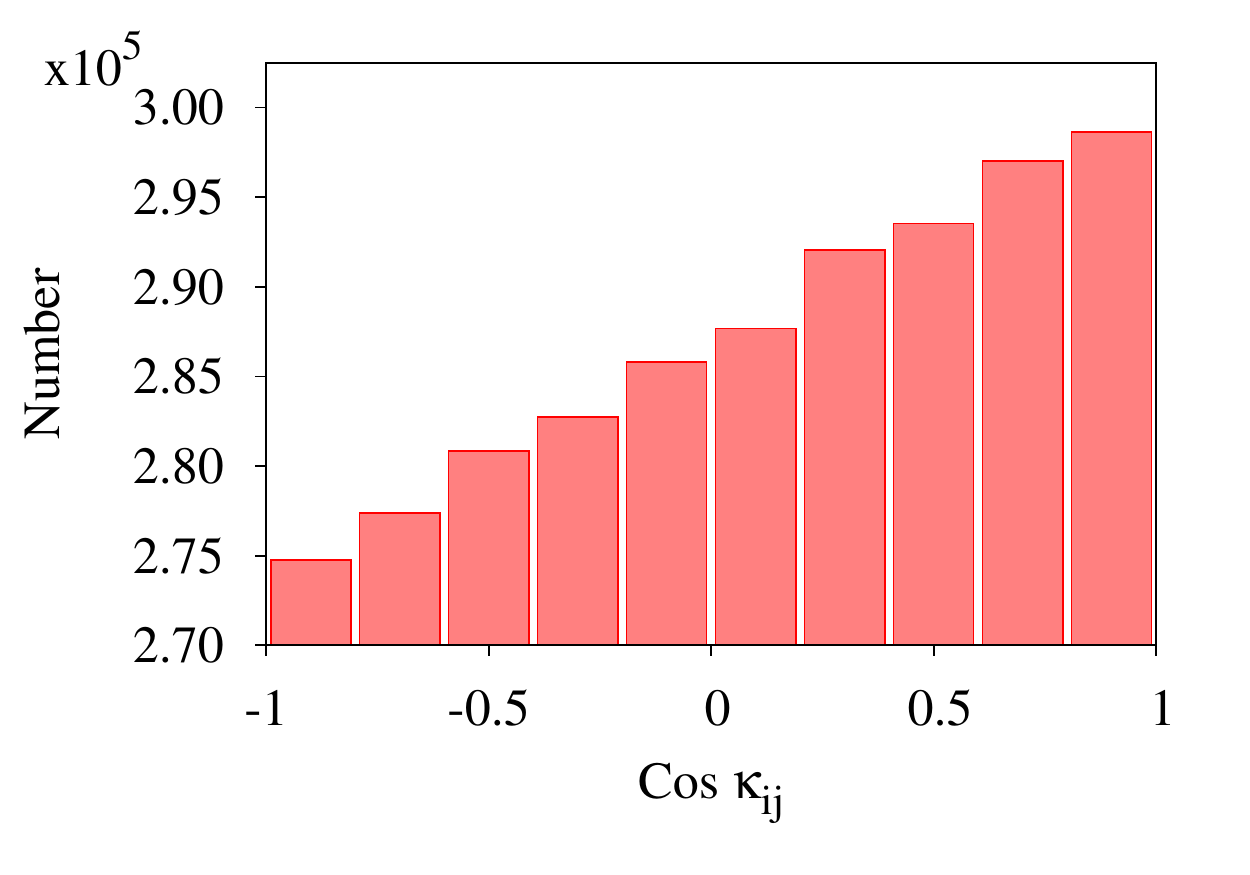}
		\caption{The initial distribution.}
	\end{subfigure}
	\hfill
	\begin{subfigure}{0.45\textwidth}
		\includegraphics[width=1\textwidth]{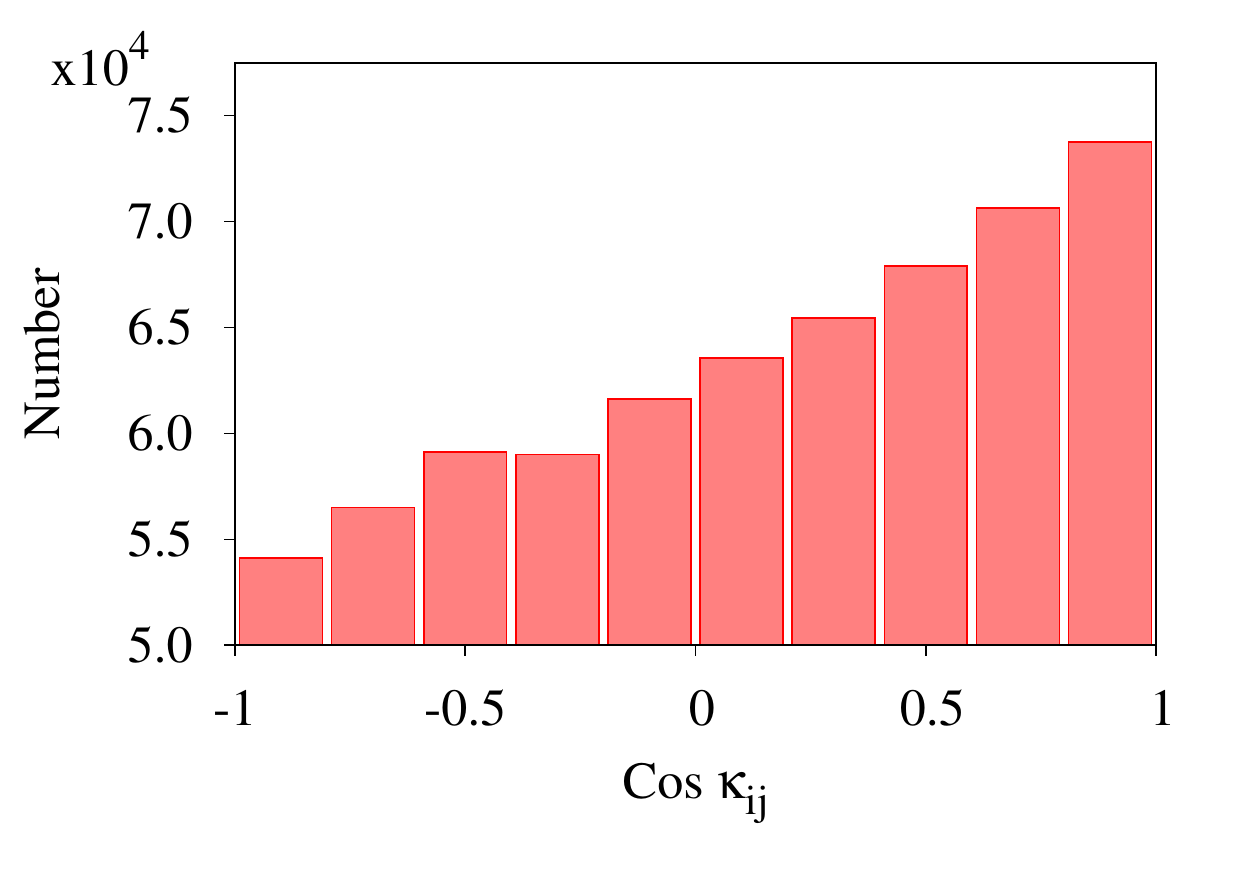}
		\caption{The distribution obtained after 150 coarse graining steps and with $k=10$ subtraction.}
	\end{subfigure}
\caption{(Color online)  Histograms for the values of $\cos\kappa_{ij}$ 
in the SARW phase for $N=300$ initial vertices.}
\label{fig-13}
\end{figure}
The Figure~\ref{fig-13} (a) shows the initial SARW distribution of the cosines in (\ref{eq_cos}) with no subtraction for the  
nearest neighbours, and the Figure~\ref{fig-13} (b)  shows the SARW distribution we obtain after we repeat the coarse graining
150 times and in addition introduce the nearest neighbour subtraction (\ref{sarw-as2}) with $k=10$ segments 
along the chain. The Figures
confirm the self-similarity, that we already observed in Figures \ref{fig-12}:
In the SARW phase, the histogram profile is stable under 
the coarse graining flow.

Finally, we inquire whether a relation akin to (\ref{sarw-as2}) can be introduced,  
%
%
%
%
%
%
%
%
\begin{figure}
\includegraphics[width=0.45\textwidth]{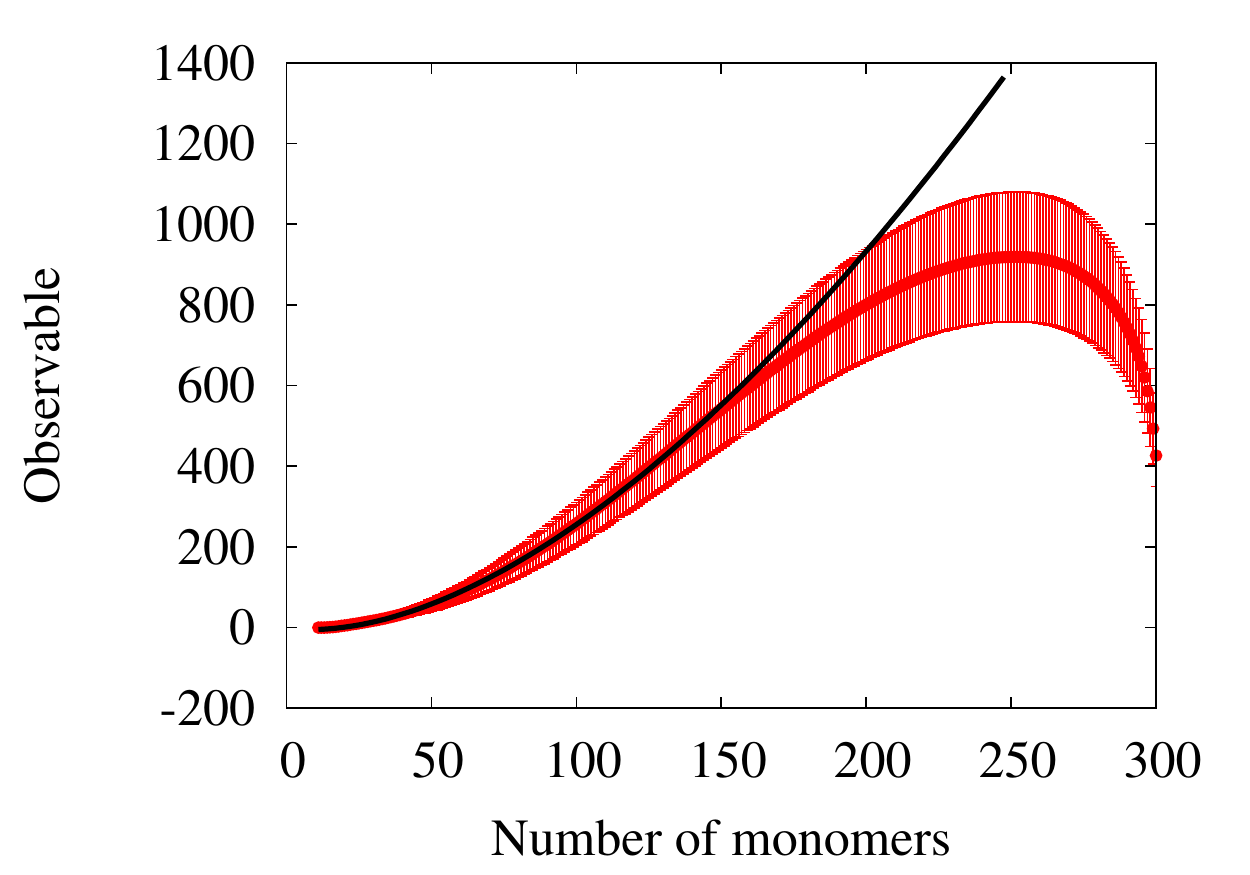}
\caption{(Color online) A fit of the form (\ref{fit-sarw}) to the observable in the SARW phase with $k=10$ subtraction. }
\label{fig-14}
\end{figure}
to model how our observable depends on the number of vertices. 
Instead of considering an ensemble of chains with an increasing number of vertices, which can be very CPU-time consuming,
we proceed as follows.
We have found that in SARW phase the observable
(\ref{eq_cos_cutoff}) displays self-similarity, under coarse graining. Thus,  we consider a statistical
pool of chains with $N=300$ vertices  
and inquire how a relation such as (\ref{sarw-as2}) can describe the flow of the observable during coarse graining:  
A large number of coarse graining iterations yields a chain with a small number of vertices. The relevant question
to address is then, how a relation like (\ref{sarw-as2}) models the coarse grained observable (\ref{eq_cos_cutoff}) when number
of coarse graining iterations increases. For this, let $r$ denote the number of vertices in the coarse grained chain. 
A small value of $r$ corresponds to a large number
 of coarse graining iterations, and when $r$ becomes large the number of coarse graining iterations becomes small. 
The relation (\ref{sarw-as2}) instructs us to inquire how the ensuing observable ${\mathcal P}_\Gamma (m)$ depends on $r$ as its
 value increases. For this we use an {\it Ansatz} of the form
 \begin{equation}
{\mathcal P}_\Gamma (r) \ \approx \ a r^b +  c r
\label{fit-sarw}
\end{equation}
We use the pool of chains shown in  Figure~\ref{fig-12} (b),   to get the result shown in Figure~\ref{fig-14}. We 
find that 
\begin{equation}
\begin{matrix} a & = & \ 0.24 &  \pm \ 0.03 \\ b & = &  \ 1.61 &  \pm \ 0.02 \\  c & = & \hspace{-0.2cm} \ -1.45 \ &  \pm \ 0.12
\end{matrix}
\label{fit-sarw-2}
\end{equation}
Here we use the Levenberg-Marquardt  nonlinear least-square algorithm for fitting, with one-sigma (standard deviation) errors.
Note the difference between Figures \ref{fig-12} and \ref{fig-14}. The former displays the observable when the number of coarse
graining steps increases. In the latter the observable is displayed in terms of increasing number of vertices during the coarse graining
flow. 

We make the following comment: We have arrived at the value $\sigma = 3/2$  in Table \ref{table-1} 
by assuming a chain with a very large number of vertices $N$, and the value $\sigma = 3/2$ is
very close to the value of $b$ we deduce in  (\ref{fit-sarw}), (\ref{fit-sarw-2}). 
The relation (\ref{sarw-as2}) is derived using the perturbation theory in the vicinity of the RW phase, and
our coarse grained chain reproduced this regime, with a
large number of iterations. This is because when the number of iterations increases the ensuing segment length also increases. Thus
the influence of the self-avoiding condition gradually disappears, with the observable approaching the 
a vanishing value of RW phase: When the number of iterations grows the perturbation theory works increasingly well.  
We comclude that  our coarse graining method is an efficient way to describe properties of chains with varying lengths, 
in terms of a pool of fixed length chains.

\subsubsection{The collapsed phase} 

In the case of RW we have investigated a statistical pool of chains that do not depend on the details of the 
homopolymer model. In SARW phase we have used  the homopolymer energy function 
(\ref{eq:A_energy}). However, as in the case of the RW phase we expect the results to be universal:   
The SARW phase describes the high temperature limit of the homopolymer model. In this limit   
the details of the energy function become irrelevant, as in this limit the temperature factor 
$\beta$ in the Gibbsian (\ref{gibbs}) vanishes; only the hard-core $r<R_0$ Pauli repulsion 
of (\ref{U}) survives, and the details of the repulsive interaction become increasingly irrelevant.
%
%
%
%
%
%
%
%
\begin{figure}
 \includegraphics[width=0.45\textwidth]{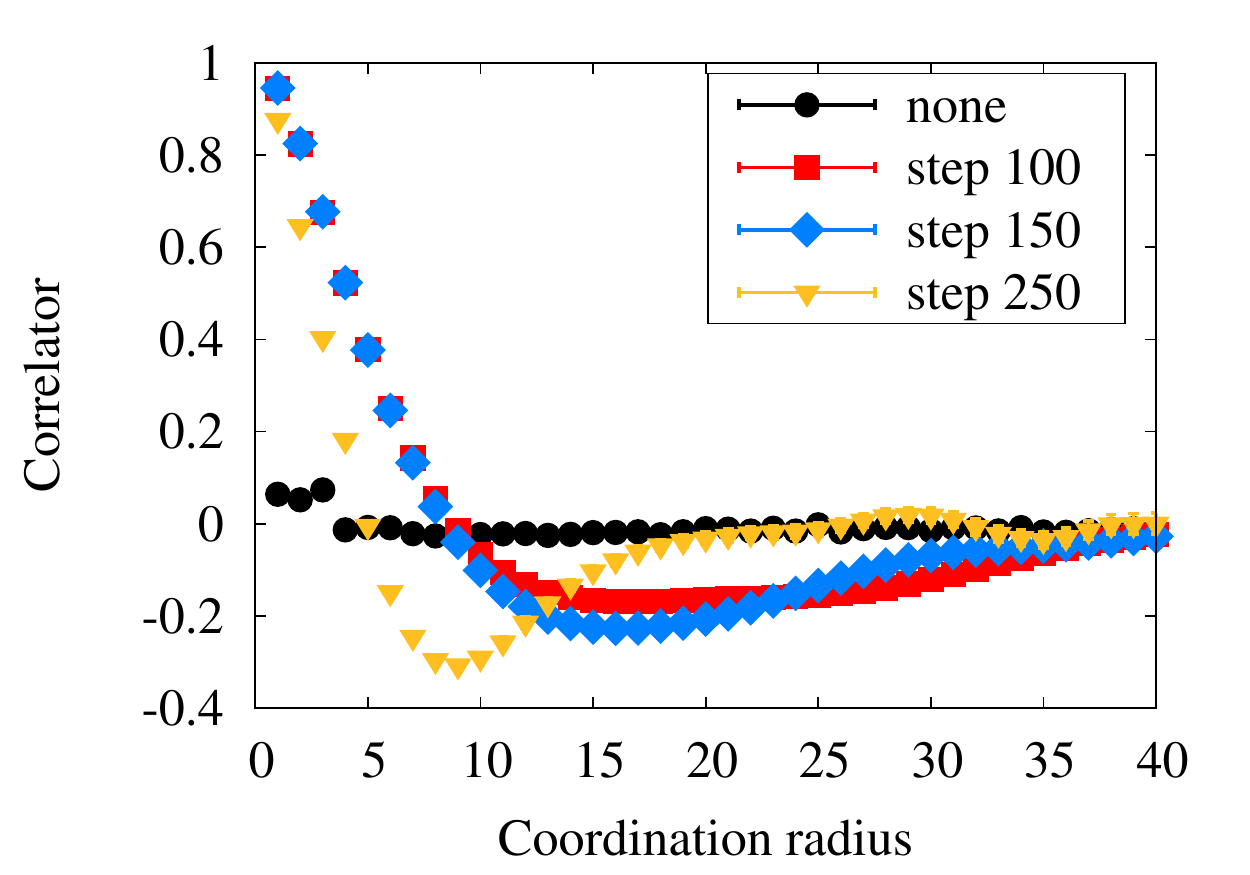}
 \vskip 0cm 
 \caption{(Color online) The dependence of the correlator (\ref{corr-theta}) on the coordination radius $k=|i-j|$ for different numbers of coarse graining steps in the collapsed phase of a homopolymer.}
 \label{fig-15}
\end{figure}

The situation is {\it very} different in the collapsed phase that occurs at low temperatures in the 
homopolymer model.  Now the temperature factor $\beta$ becomes large, and the  thermodynamics becomes
increasingly ruled by the energy function: Unlike in the case of RW and SARW phases
where universality is due to the apparent insensitivity of the phase on the details of the ensuing 
chain Hamiltonian, in the collapsed phase
the model specific details matter most. Indeed,
despite the asserted universality of (\ref{nuval}) we are not aware of any compelling 
argument why the low temperature phase properties should be insensitive to dynamical details. Quite to the
contrary: Discrete flows towards fractal attractors of all kinds are abundant in three dimensions.
Accordingly, we
scrutinize the collapsed phase of the homopolymer model (\ref{eq:A_energy}), with the parameter values in Table
\ref{table-2} and using the heat bath method described in \cite{Ann}.
%
%
%
%
%
%
%
%
\begin{figure}
 \includegraphics[width=0.45\textwidth]{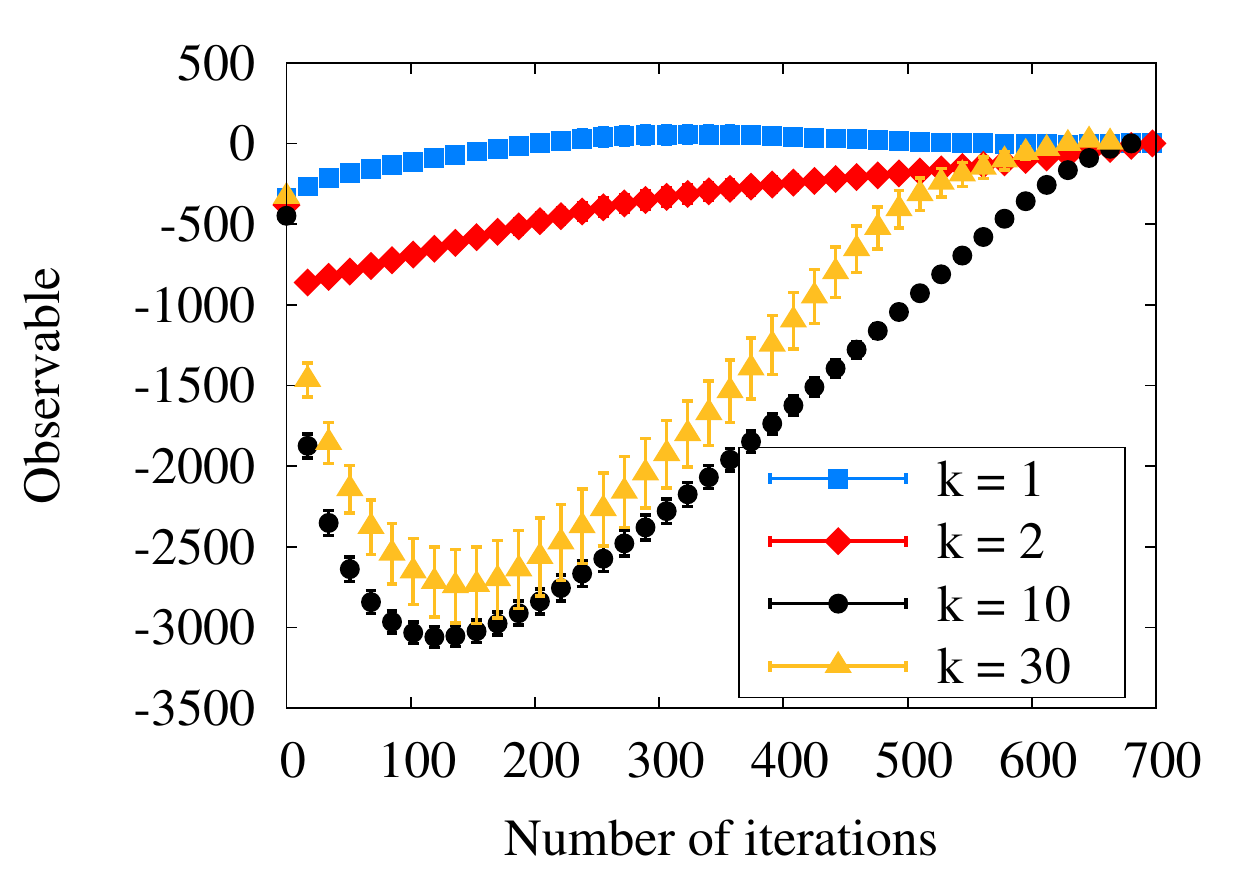}
 \vskip 0cm 
 \caption{(Color online) Dependence of the observable (\ref{eq_cos_cutoff}) on the number of scaling steps $n$ for chains with $N=700$  
 initial vertices, and with different values of subtraction $k=|i-j|$ in the collapsed phase. The average value is shown, together with the one 
 standard deviation fluctuation regime.} 
 \label{fig-16}
\end{figure}
The Figure~\ref{fig-15} shows the correlation length (\ref{corr-theta}) in the collapsed phase, for different values of the 
coarse graining steps. 

As in the SARW phase, there is a  hard core repulsion between all vertex pairs. There are also 
interactions that are due to the dynamical details of  (\ref{eq:A_energy}), including solitons that are absent in the
high temperature SARW phase.  Finally, we have the correlation between vertices due to 
the coarse graining process. But in line with the RW and SARW phases, we find that the 
effect of coarse graining extends only over a relatively short range in the segment distance $k=|i-j|$:
From Figure~\ref{fig-15} we deduce that the effects of coarse graining are 
largely unobservable when $k$ is greater than $k\sim 35$,
a somewhat longer segment distance than what we found in the RW and SARW phases. 


In Figure~\ref{fig-16} we show the evolution of the observable (\ref{eq_cos_cutoff}) during the coarse graining,
when we increase the number  $k = |i-j|$ of 
finite size subtractions (\ref{coll-as}); the initial chain has $N=700$ vertices. 

When there is no subtraction {\it i.e.} $k=1$, we find that the observable is  initially
negative, in line with our general arguments in Table \ref{table-1}.  
Then, after around 200 coarse graining steps the observable vanishes, the chain appears to reside in the RW phase.
However, this is an apparent short-range effects, due to correlations 
between nearest neighbour vertices with $k=1$:  
When we subtract the nearest neighbour contribution  {\it i.e.} we set 
$k=2$ in (\ref{coll-as}), the value of the observable is negative throughout the coarse scaling process, in line with the general arguments of
Table \ref{table-1}. 

In Figure~\ref{fig-16}   we 
show the result also with $k=10$ and with $k=30$. Comparison of the profiles proposes self-similarity, in line what we observed
previously in the SARW phase Figure~\ref{fig-12}. Note that for a chain with $N=700$ vertices, there can be additional finite length
effects when $k$ becomes much larger.
%
%
%
%
%
%
%
%
\begin{figure*}
	\begin{subfigure}{0.333\textwidth}
 		\includegraphics[width=1\textwidth]{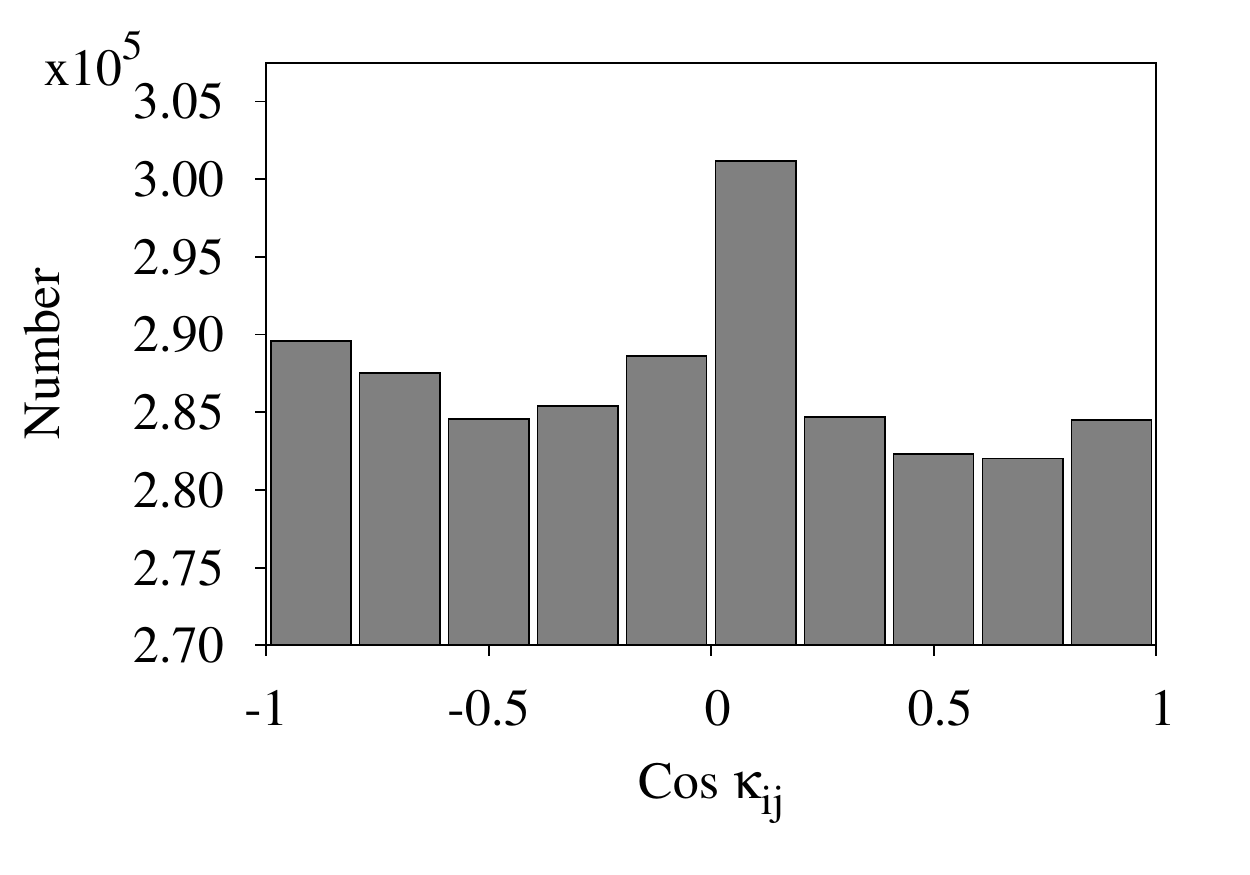}
 	\end{subfigure}
 	 \hskip -0.2cm 
 	\begin{subfigure}{0.333\textwidth}
 		\includegraphics[width=1\textwidth]{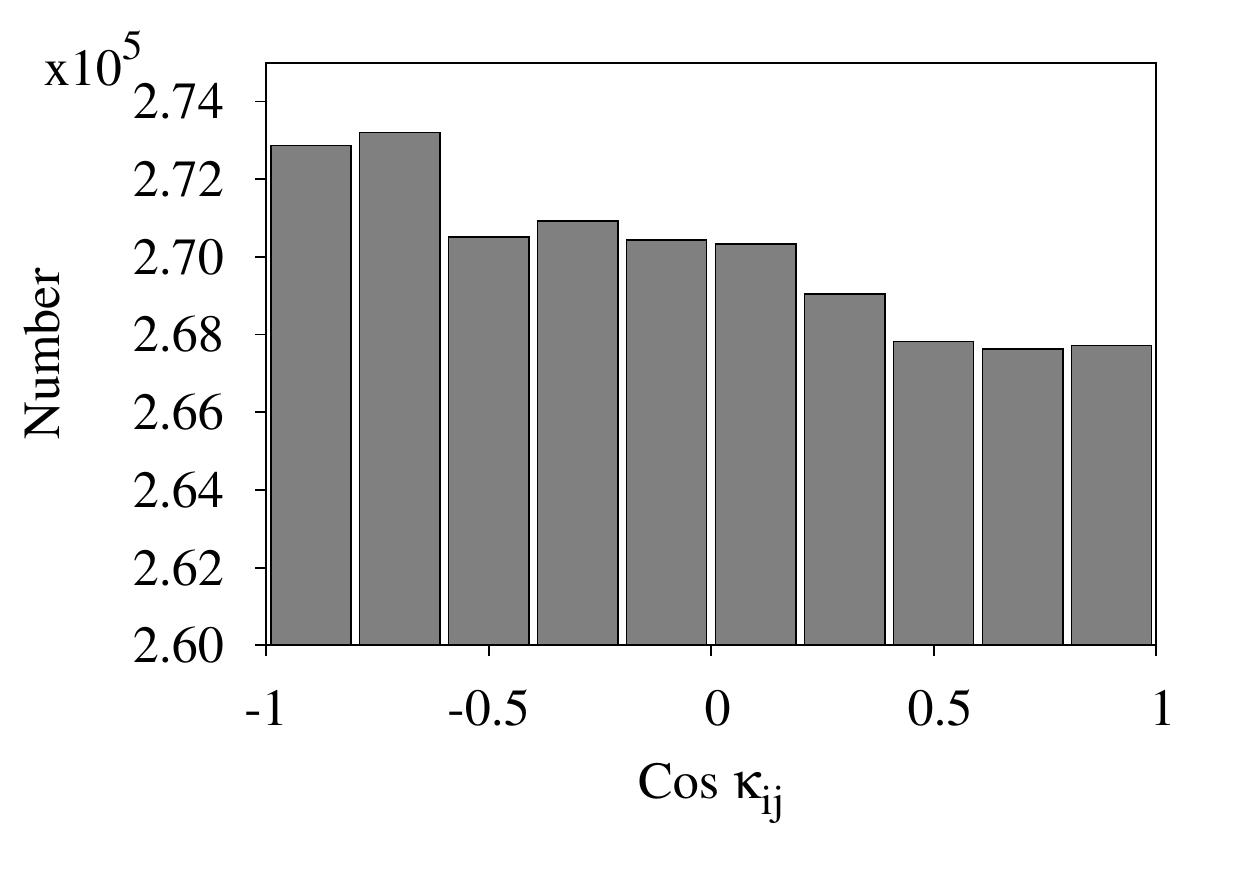}
 	\end{subfigure}
 	\hskip -0.2cm 
 	\begin{subfigure}{0.333\textwidth}
 		\includegraphics[width=1\textwidth]{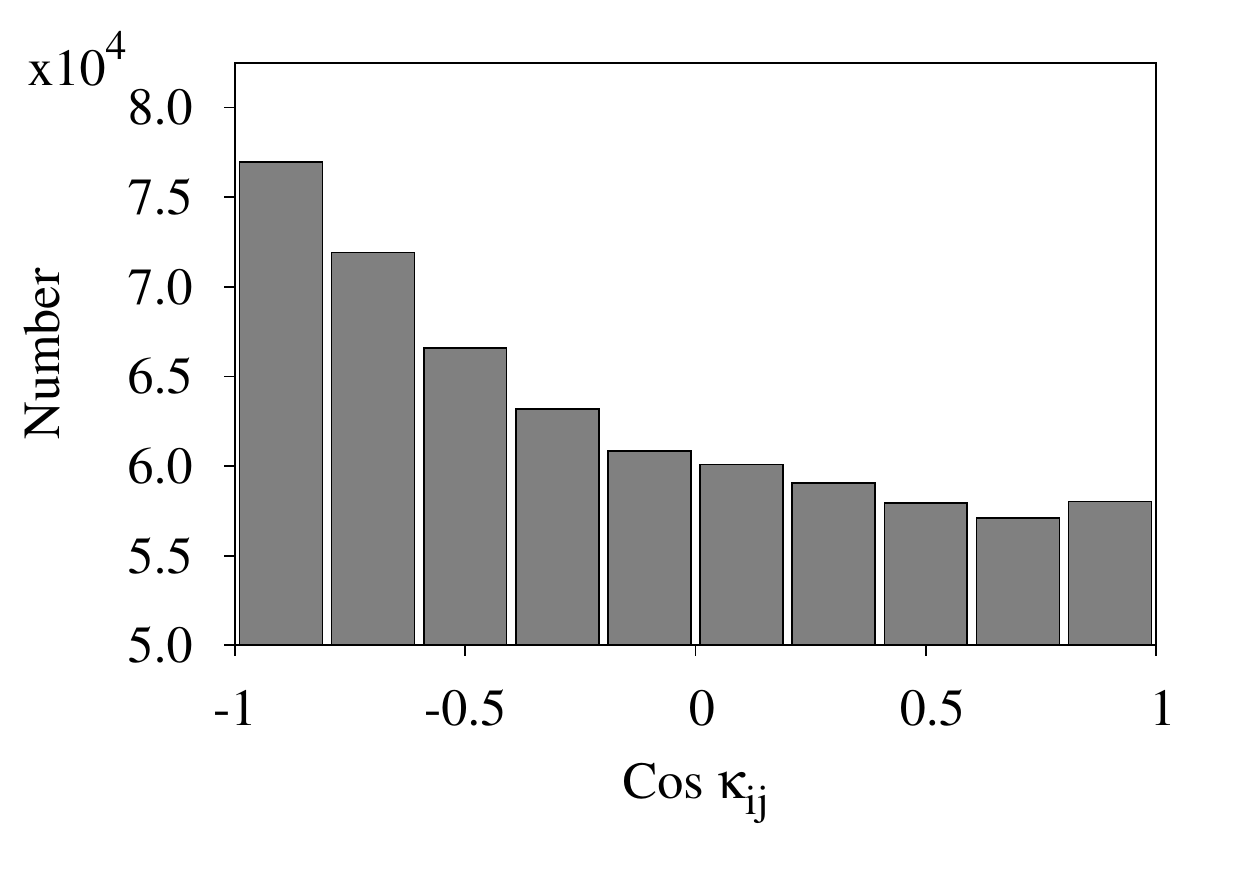}
 	\end{subfigure}
 \caption{(Color online) Histograms  for the values of $\cos\kappa_{ij}$ for a chain with $N=300$ initial vertices in the collapsed phase. (a) The full initial chain. (b) With $k=10$ nearest neighbour subtractions. (c) After $n=150$ coarse graining steps and with $k=10$ nearest neighbour subtractions.} 
 \label{fig-17}
\end{figure*}
In Figure~\ref{fig-17} we show three representative histograms of $\cos \kappa_{ij}$ in (\ref{eq_cos}), in the collapsed phase. 
The Figure~\ref{fig-17} (a)  is for the initial chain. Here,  we observe a clear accumulation of values between 
$0 < \cos \kappa_{ij} < 0.2$.  This reflects the effect of local minima for $\kappa$ angle in the Hamiltonian (\ref{eq:A_energy}). Since these minima are located at $\kappa=m=1.5\approx \pi/2$, the peaks appear due to values of $\cos \kappa_{ij}$ for $|i-j|=1$.

In the Figure~\ref{fig-17} (b)   we remove all nearest neighbour
contributions with  segment distances $k=|i-j| < 10$. Now, there is a clear excess of negative values. 
Finally, in  Figure~\ref{fig-17} (c)   we introduce $n=150$ coarse graining steps in the histograms of  
Figure~\ref{fig-17} (b).  Now we obtain a monotonic, decreasing distribution
of the $\cos \kappa_{ij}$ values. Note that the monotonic character of the distribution is  quite in line with that in Figures \ref{fig-13},
except for the sign.

Finally, in analogy with Figure~\ref{fig-14} of the SARW phase, in Figure~\ref{fig-18}
we introduce a fitting of the form (\ref{fit-sarw}) to the evolution of the collapsed state observable, during the coarse graining process. 
%
%
%
%
%
%
%
%
\begin{figure}
 \includegraphics[width=0.45\textwidth]{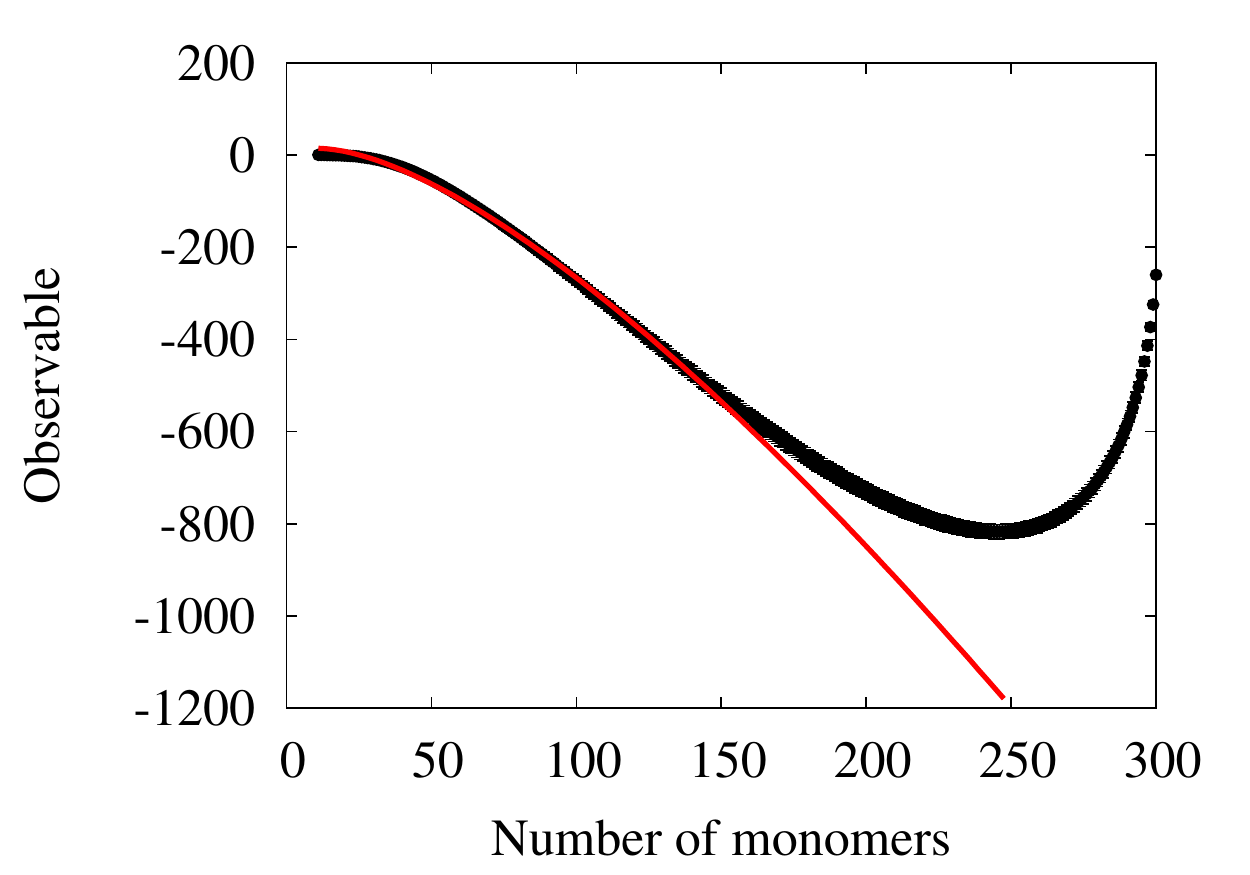}
 \vskip 0cm 
 \caption{(Color online) A fit of the form (\ref{fit-sarw}) to the observable in the collapsed phase with $k=10$ subtraction.} 
 \label{fig-18}
\end{figure}
We find 
\[
\begin{matrix} a & = & \ 6.22 &  \pm \ 2.05 \\ b & = &  \ 1.16 &  \pm \ 0.03 \\  c & = & \hspace{-0.2cm} \ -10.5 \ &  \pm \ 2.4
\end{matrix}
\]

\subsection{Summary of homopolymer simulations}

Our results show that in the case of a homopolymer,  the observable (\ref{eq_cos_cutoff}) flows in a self-similar
manner during repeated coarse graining.

\vskip 0.2cm
$\bullet$ In the RW the observable initially vanishes, in line with (\ref{vanish}). The coarse graining introduces 
correlations between neighbouring segments causing the observable to become positive valued. The observable
then flows asymptotically towards a vanishing value, as we proceed and iterate the coarse graining.  
The qualitative features of the flow are universal with the profile shown in Figure~\ref{fig-7}, and with an evenly and 
uniformly distributed histogram as shown in Figure~\ref{fig-8} (b).

\vskip 0.2cm
$\bullet$ In the SARW phase the observable is positive during the entire coarse graining process, 
with a self-similar profile as shown in Figure~\ref{fig-12}. The histogram profiles shown in Figure~\ref{fig-13}
are also qualitatively universal, for chains in this phase.

\vskip 0.2cm 
$\bullet$ We have simulated the observable in the collapsed phase of the homopolymer model (\ref{eq:A_energy}); unlike
in the case of universal RW and SARW phases, the results are now model dependent. 
The observable is negative and increases towards a vanishing value as the coarse graining proceeds. 
Once we remove the effect of very short distance repulsion between neighbouring segments, 
the profile of the flow becomes self-similar as shown in Figures  \ref{fig-16}; the histogram in Figure~\ref{fig-17} (c) is
also self-similar over a wide range of chains.

\vskip 0.2cm
We note that in all cases, the observable converges towards a vanishing value 
when the number of coarse graining iterations becomes large.
This can be understood as follows: When the coarse graining terminates, we are left with only three vertices and two segments.
In a statistical ensemble of long chains, the angle between these two remaining segments is randomly distributed  with a
vanishing average value.

\section{Applications to collapsed proteins}

We have proposed  that in the RW and SARW phases our results remain valid beyond the homopolymer model; these
two phases describe universality classes of chains. However, in the collapsed phase our results depend in an essential 
manner on the details of the energy function; {\it a priori} the results are model dependent and 
we have no reasons to expect that in the collapsed phase the profile 
shown in Figure~\ref{fig-16} for $k=10$  and $k=30$   persists beyond a homopolymer, {\it as such:} We remind that 
there are many examples of discrete
3D curves that describe deterministic evolution towards  all  kind of chaotic attractors.  
We proceed to analyse the coarse graining 
flow of the observable in the case of collapsed heteropolymers, using  crystallographic PDB protein structures 
as examples.

\subsection{Myoglobin}

The first example we consider is myoglobin. There are 154 amino acids, and we use the crystallographic structure with 
PDB code 1ABS. In Figure~\ref{fig-19} (a)
we show how the observable (\ref{eq_cos_cutoff}) flows during the coarse graining, for the entire chain with $k=1$. In Figure~\ref{fig-19} (b)
we introduce a short distance subtraction with $k=2$ and in Figure~\ref{fig-19} (c) we increase the subtraction distance 
to $k=10$. 
%
%
%
%
%
%
%
%
\begin{figure}
	\begin{subfigure}{0.45\textwidth}
 		\includegraphics[width=1\textwidth]{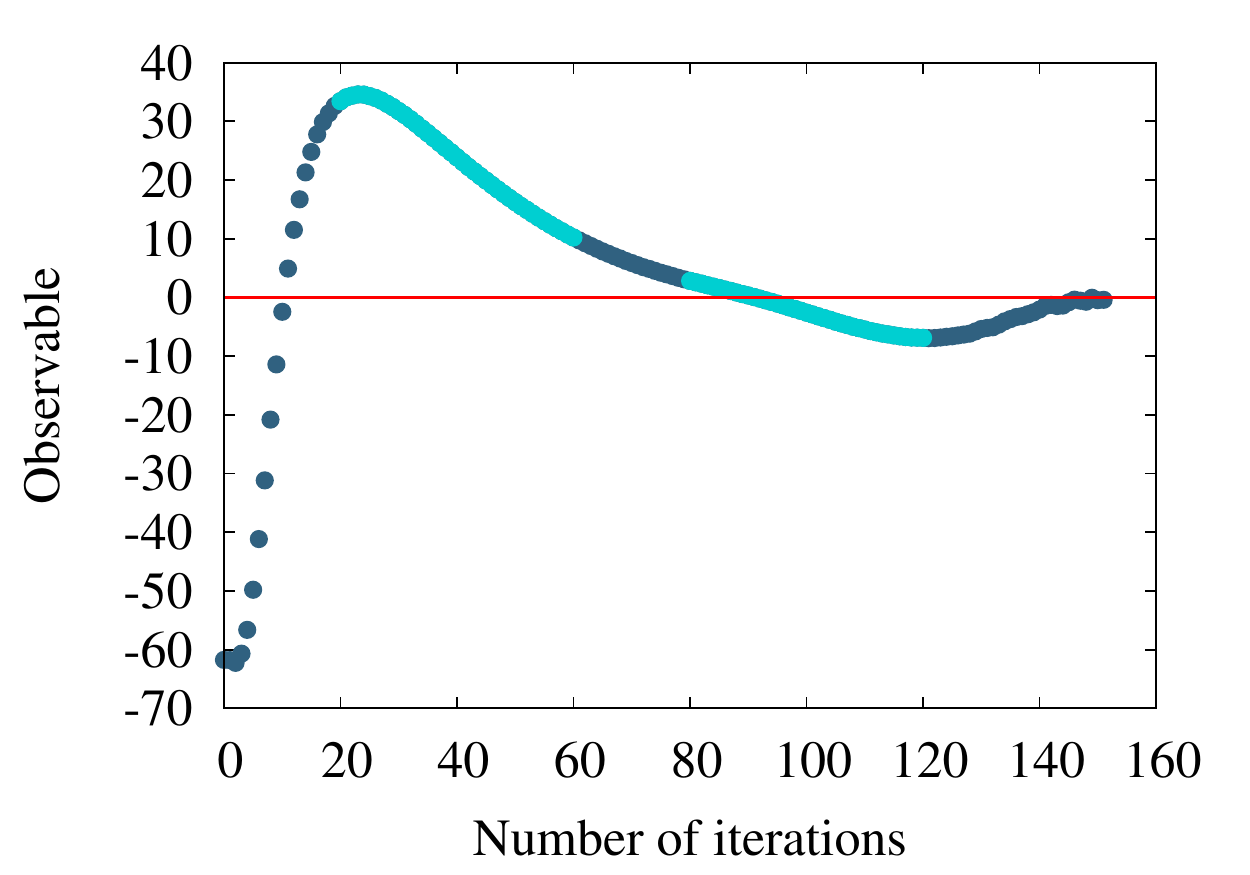}
 		\vskip 0cm 
 		\caption{For the entire chain \textit{i.e.} $k=1$.}
 	\end{subfigure}
 	\hfill
 	\begin{subfigure}{0.45\textwidth}
 		\includegraphics[width=1\textwidth]{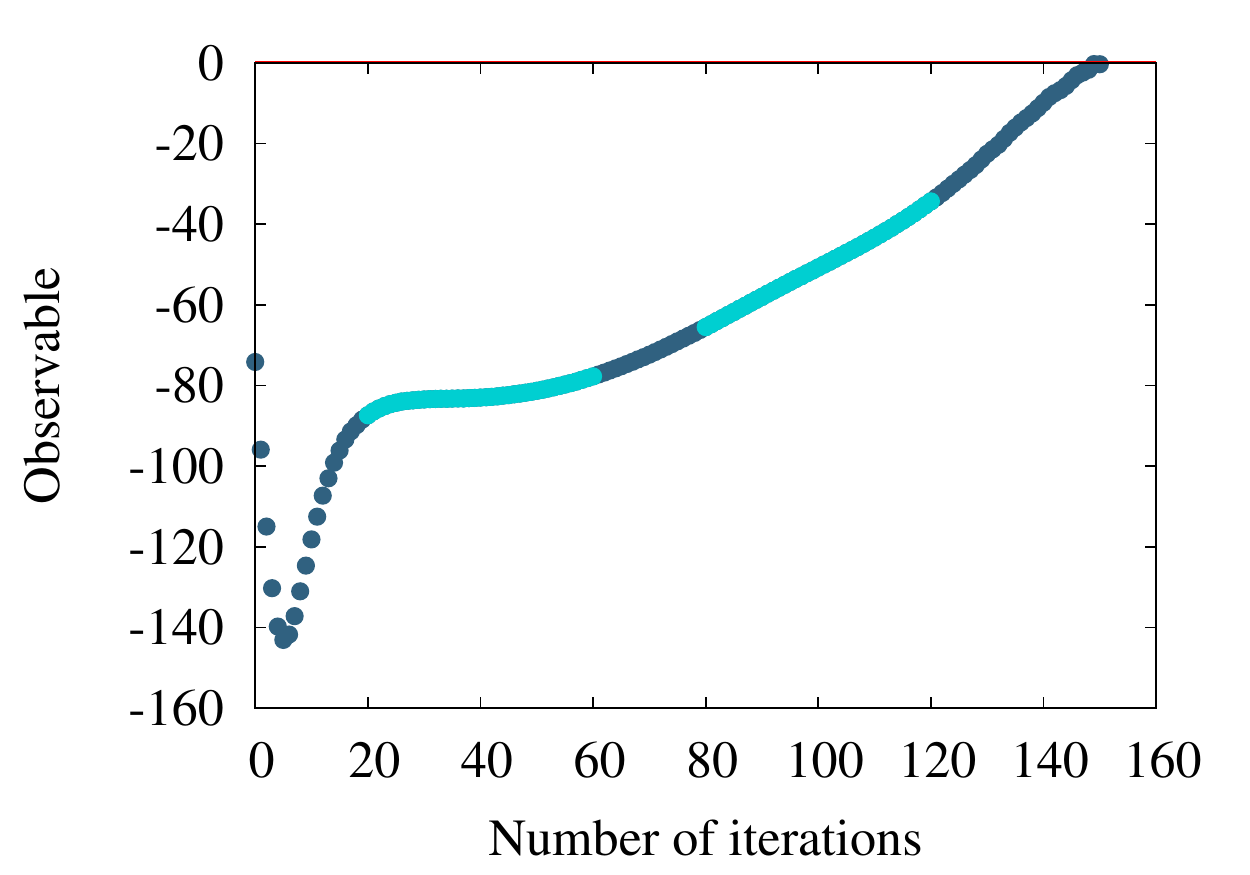}
 		\vskip 0cm 
 		\caption{For $k=2$ nearest neighbour subtractions.}
 	\end{subfigure}
 	\hfill
 	\begin{subfigure}{0.45\textwidth}
 		\includegraphics[width=1\textwidth]{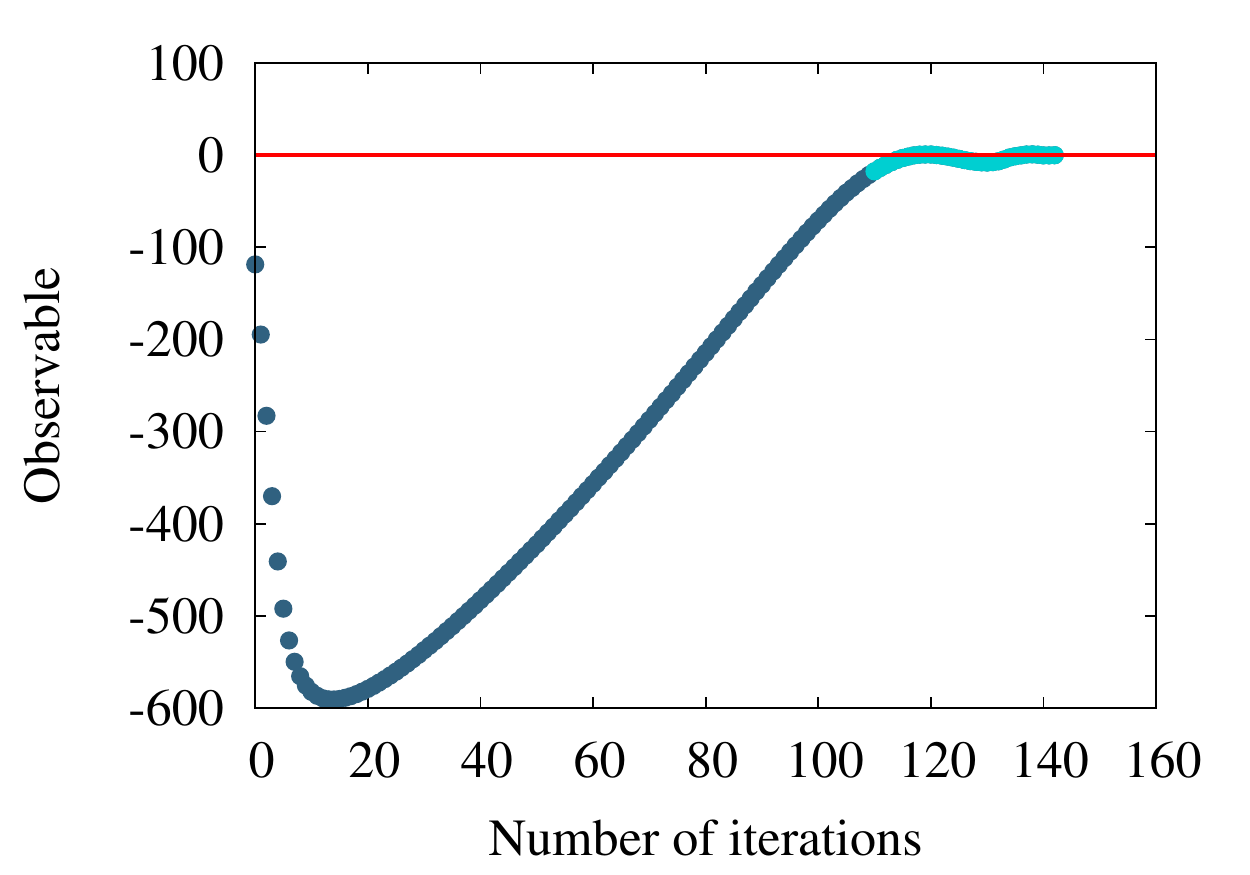}
 		\vskip 0cm 
 		\caption{For $k=10$ nearest neighbour subtractions.}
 	\end{subfigure}
 \caption{
 (Color online) Coarse graining flow of the observable (\ref{eq_cos_cutoff}), for myoglobin 1ABS.} 
 \label{fig-19}
\end{figure}
%
%
%
%
%
%
%
%
The flows in these Figures are remarkably similar to the flows in the corresponding curves of Figure~\ref{fig-16}   of the homopolymer model.
The only (slight) differences are in the non-uniformities pointed out by the coloring in Figures \ref{fig-19}, and that in Figure~\ref{fig-19}
(a)  the observable becomes negative between coarse graining iterations 100-140: In the case of a homopolymer there is uniformity in
the length scale, thus the flow profile is also uniform.  The non-uniformity in the case of myoglobin implies that
there are additional length scales, these scales affect the profile. 

\subsection{$\beta$-barrel}
%
%
%
%
%
%
%
%
\begin{figure*}
	\begin{subfigure}{0.45\textwidth}
 		\includegraphics[width=1\textwidth]{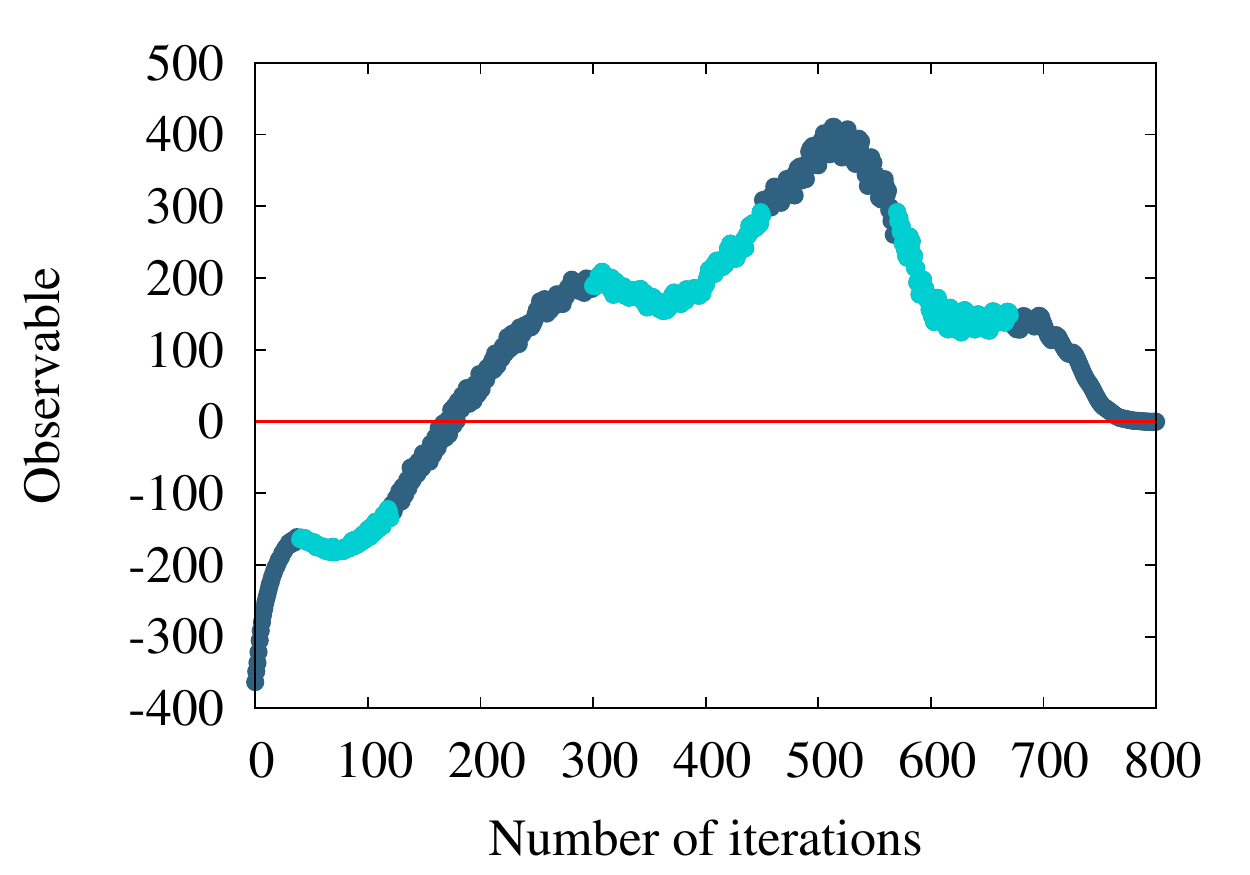}
 		\vskip 0cm 
 		\caption{For the entire chain \textit{i.e.} $k=1$.}
 	\end{subfigure}
 	\begin{subfigure}{0.45\textwidth}
 		\includegraphics[width=1\textwidth]{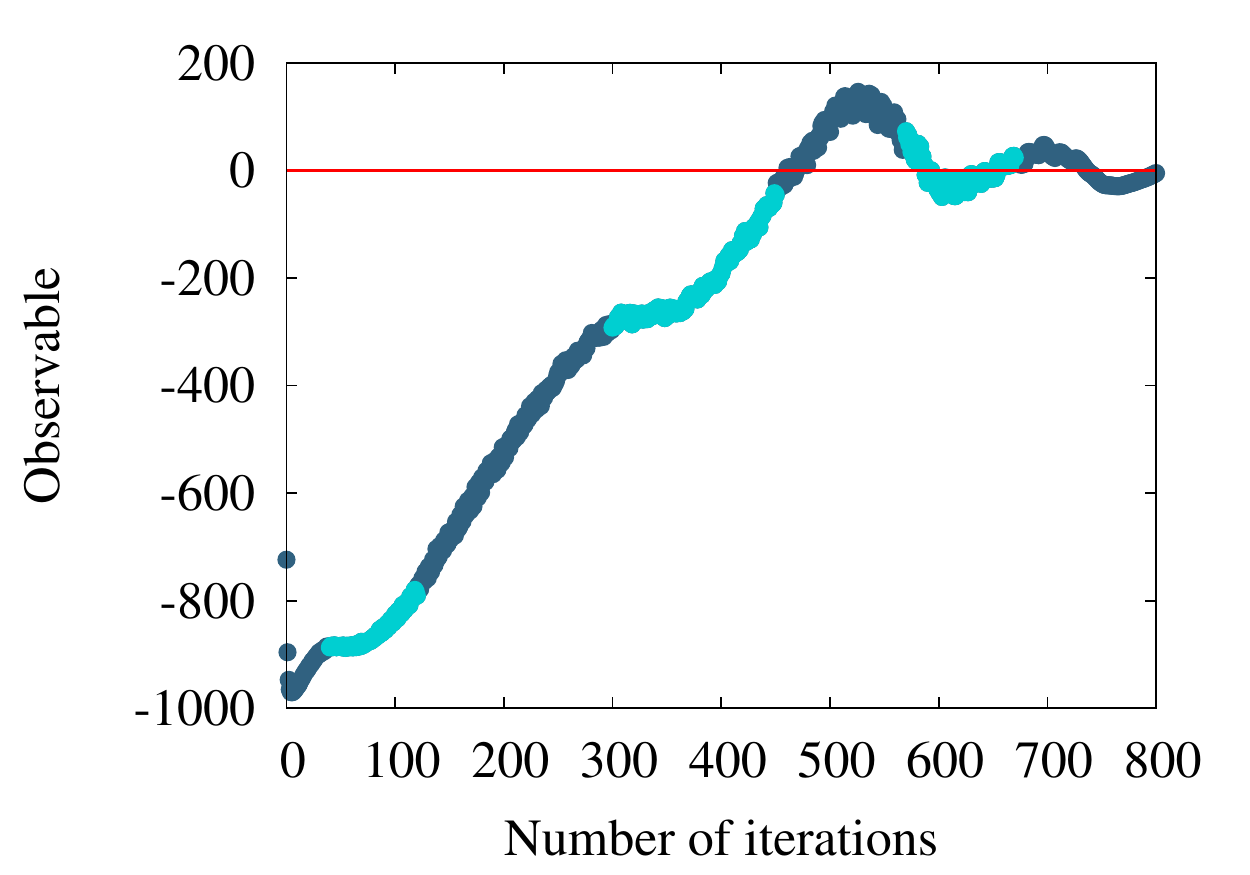}
 		\vskip 0cm 
 		\caption{For with $k=2$ nearest neighbour subtractions.}
 	\end{subfigure}
 	\hfill
 	\begin{subfigure}{0.45\textwidth}
 		\includegraphics[width=1\textwidth]{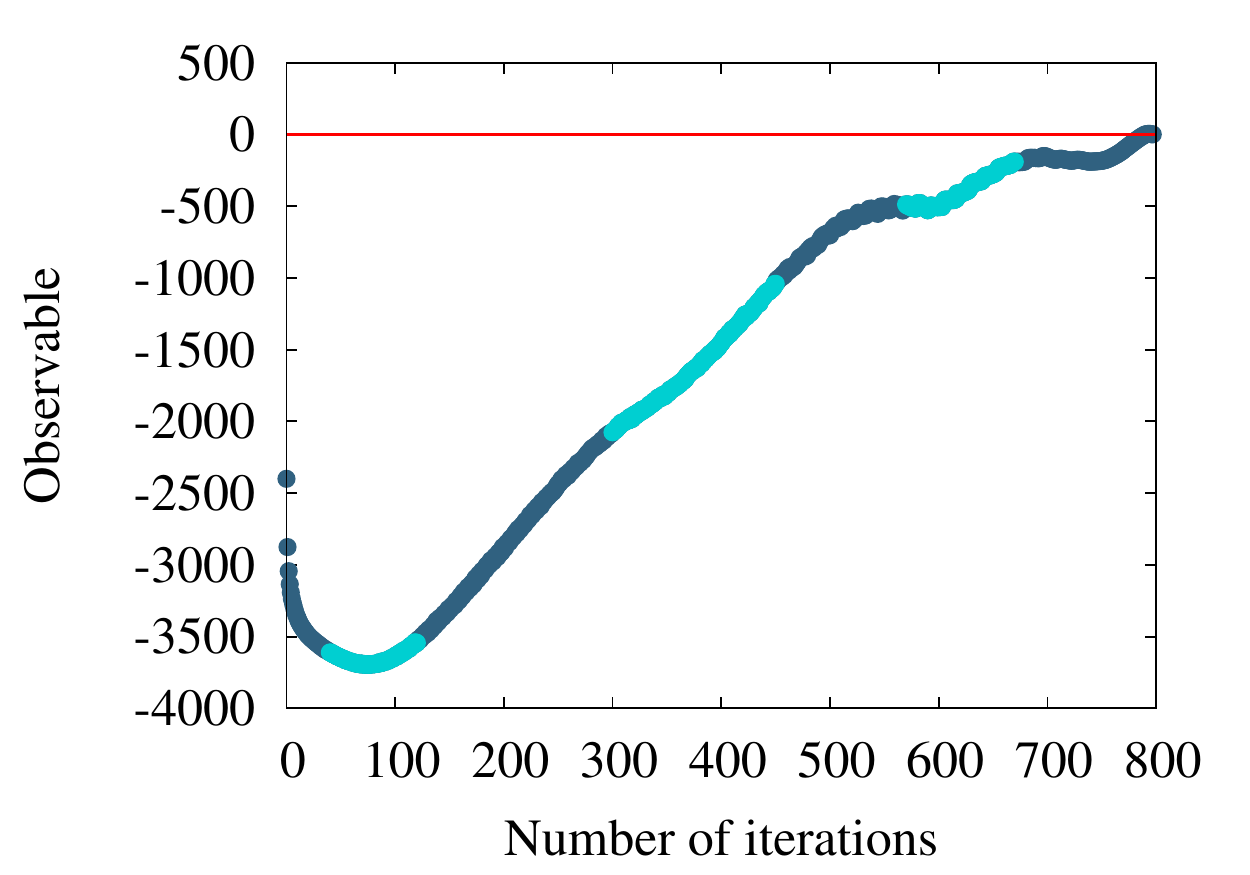}
 		\vskip 0cm 
 		\caption{For $k=10$ nearest neighbour subtractions.}
 	\end{subfigure}
 	\begin{subfigure}{0.45\textwidth}
 		\includegraphics[width=1\textwidth]{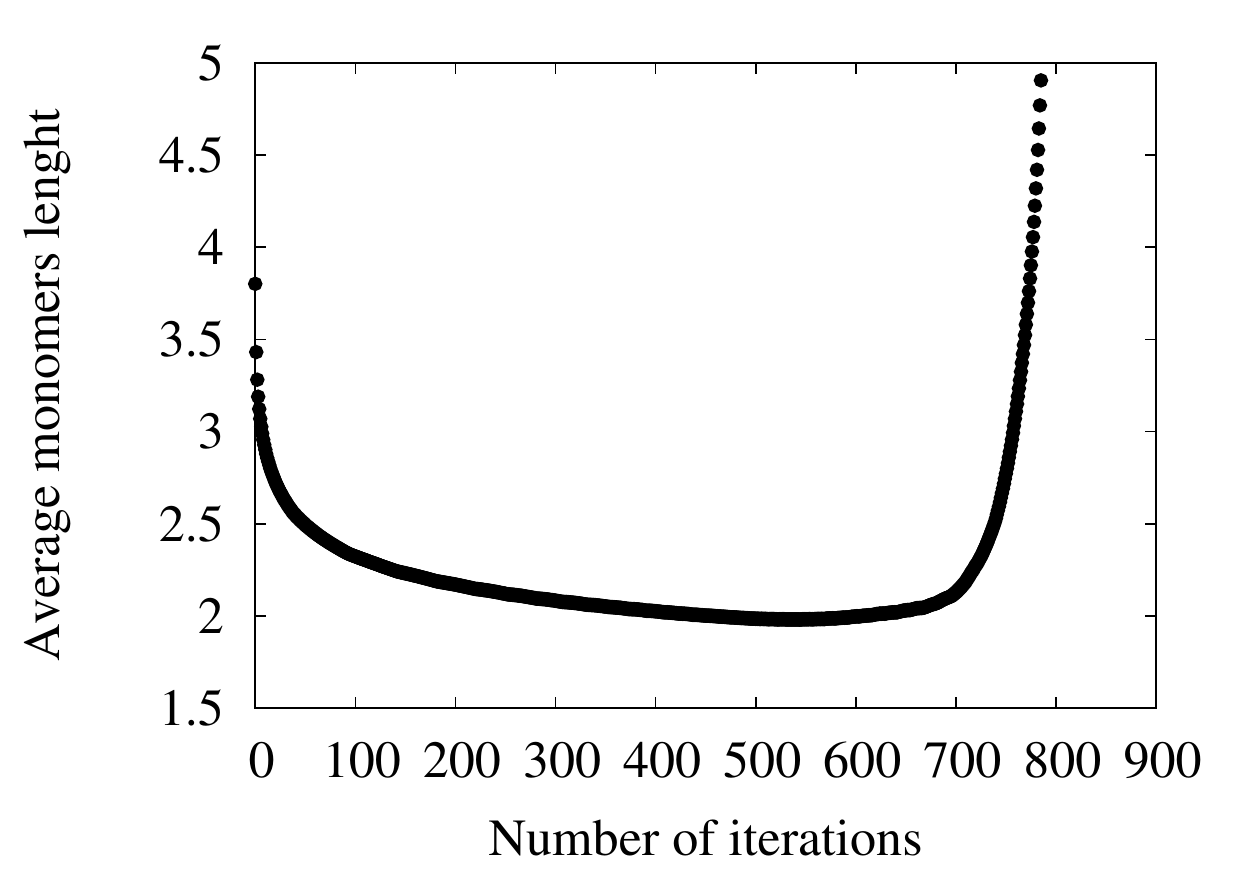}
 		\vskip 0cm 
 		\caption{The evolution of effective Kuhn  length.}
 	\end{subfigure}
 \vskip 0cm 
 \caption{(Color online) Coarse graining flow of the observable (\ref{eq_cos_cutoff}), for the $\beta$-barrel structure 4ZGV. The Figure  (d) shows the evolution of the effective Kuhn  length during coarse graining. } 
 \label{fig-20}
\end{figure*}
$\beta$-barrel is a structural motif which is 
prevalent in many transmembrane proteins. An example is the PDB entry 4GZV, with 868 amino acids.
In Figures \ref{fig-20} (a)-(c)   we have the coarse graining flow of the ensuing observable (\ref{eq_cos_cutoff}), with 
$k=1$, $k=2$ and $k=10$ respectively. 

Qualitatively, the overall pattern of the flow is very similar to Figures \ref{fig-16} and \ref{fig-19}, in the case of homopolymer and myoglobin.
In Figure~\ref{fig-20} (d)   we show the evolution of the effective segment length in 4ZGV during coarse graining. We find that
it decreases from the initial value 3.8 \AA~  which is the distance between two neighbouring C$\alpha$ atoms
along the 4ZGV backbone, to a minimum value around 2.0 \AA,  only after a quite large $n \sim 700$ number of
coarse graining steps. Then, there
is a rapid increase towards the end of the coarse graining process. This evolution parallels that we have previously recorded
in Figure~\ref{fig-10}, in the case of the homopolymer model. Note that in that case the parameter value $m$ in (\ref{eq:A_energy})
corresponds to $\alpha$-helical structures while the 4ZGV backbone is dominated by $\beta$-sheets.

\subsection{$\alpha$-helical protein} 

As a third example, we consider the structure 2PO4 in PDB. This is an $\alpha$-helical protein with 1104 amino acids.
In Figures \ref{fig-21} we show the flow of the observable (\ref{eq_cos_cutoff}) in the cases $k=1$, $k=2$ and $k=10$.
%
%
%
%
%
%
%
%
\begin{figure}
 \begin{subfigure}{0.45\textwidth}
 		\includegraphics[width=1\textwidth]{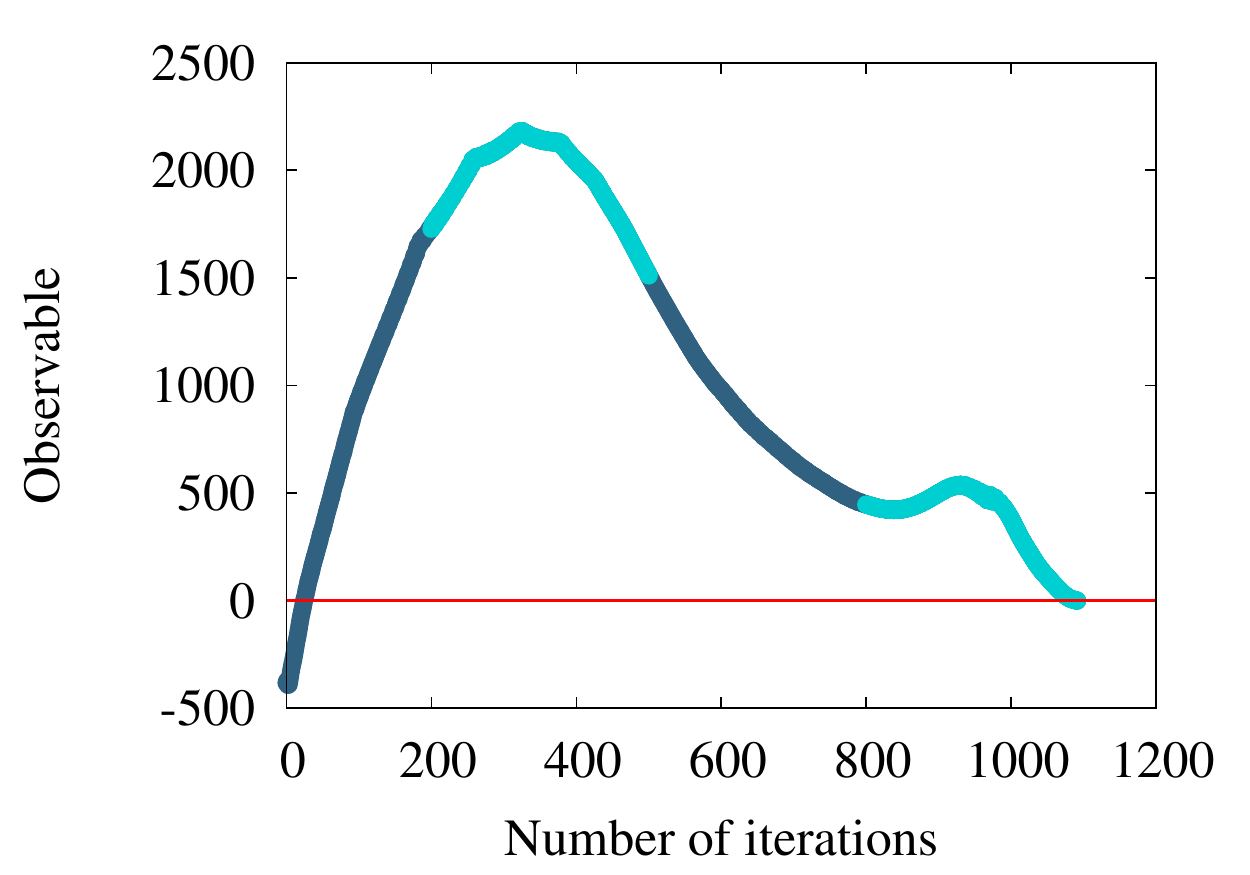}
 		\vskip 0cm 
 		\caption{For the entire chain \textit{i.e.} $k=1$.}
 	\end{subfigure}
 	\hfill
 	\begin{subfigure}{0.45\textwidth}
 		\includegraphics[width=1\textwidth]{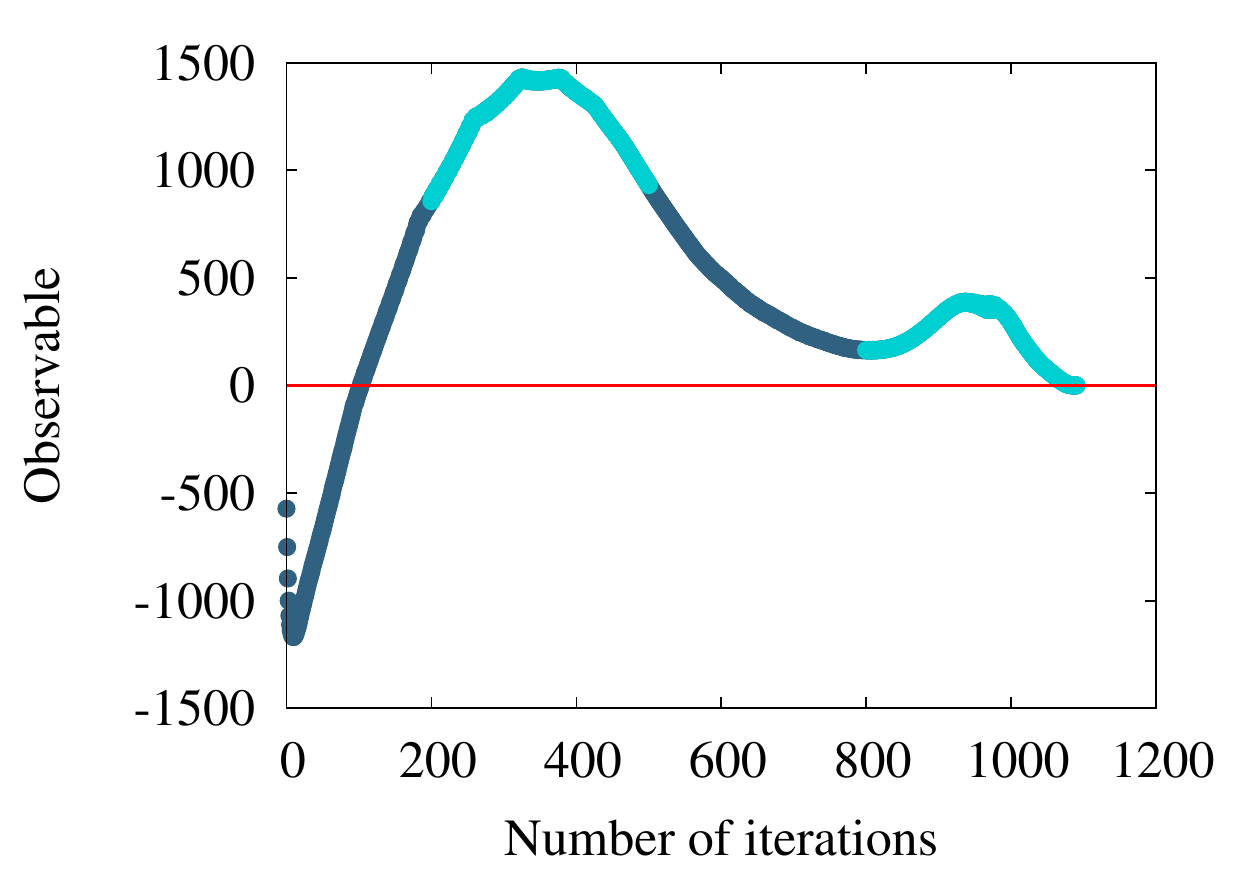}
 		\vskip 0cm 
 		\caption{For $k=2$ nearest neighbour subtractions.}
 	\end{subfigure}
 	\hfill
 	\begin{subfigure}{0.45\textwidth}
 		\includegraphics[width=1\textwidth]{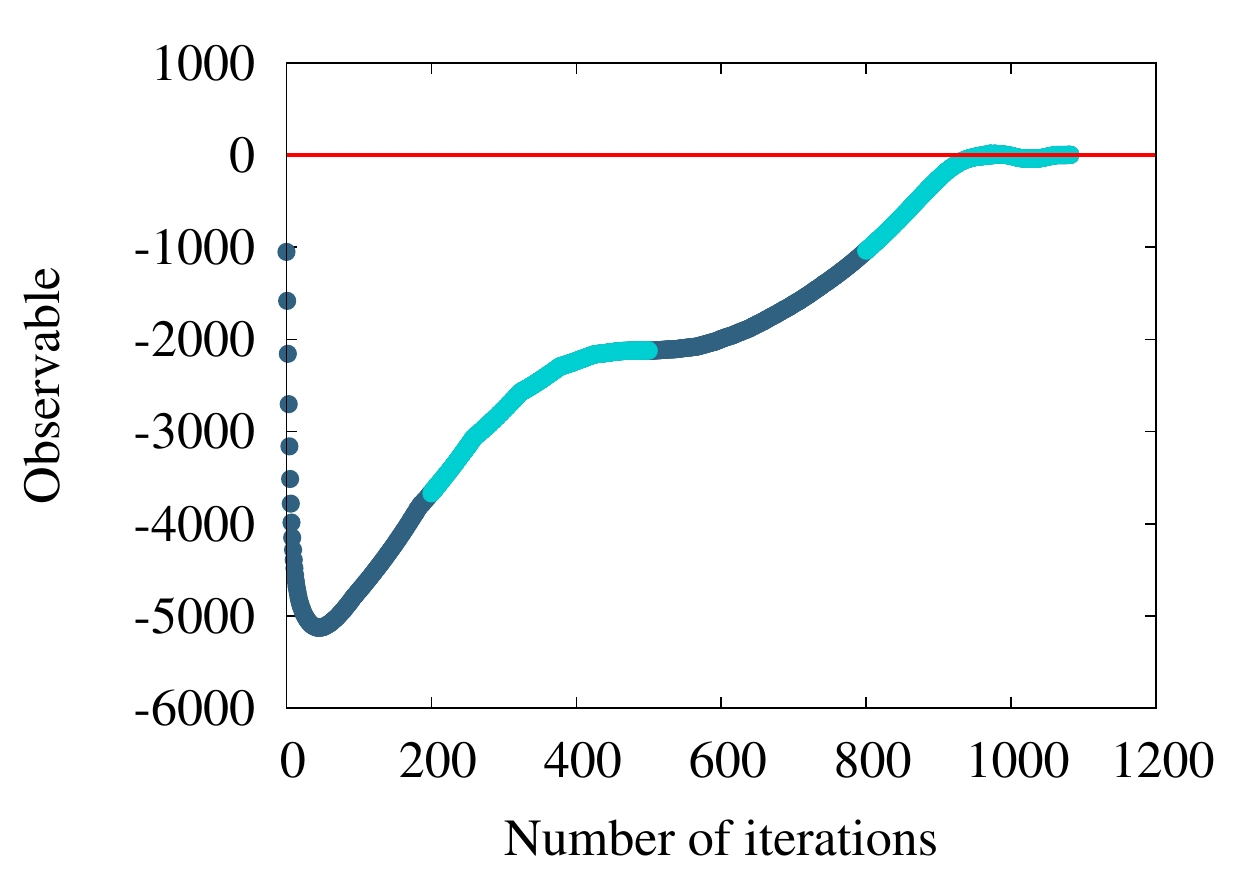}
 		\vskip 0cm 
 		\caption{For $k=10$ nearest neighbour subtractions.}
 	\end{subfigure}
 \caption{
 (Color online) Coarse graining flow of the observable (\ref{eq_cos_cutoff}), for the $\alpha$-helical structure 2PO4.} 
 \label{fig-21}
\end{figure}
We find that the qualitative features we have observed in the case of homopolymer, myoglobin and the protein 4ZGV persist,
except that now the evolution from the profile in Figure~\ref{fig-21} (a) to that in Figure~\ref{fig-21} (c) proceeds more slowly in $k$:
The structure of  2PO4 is highly helical, and for 
$\alpha$-helix we have the nearest neighbour $\kappa_{i,i+1} \approx \pi/2$, thus the effect of $k=1$ 
subtraction is small.

Since the chain 2PO4 is much longer than the  coarse graining correlation length $\sim 35$ segments according to Figure~
\ref{fig-15}, we may safely consider larger $k$ values; {\it e.g.} in the case of myoglobin the relatively small 
number of  residues might introduce hard-to-resolve effects for such a large values of correlation length. 
In Figure~\ref{fig-22}
%
%
%
%
%
%
%
%
\begin{figure}
 \begin{subfigure}{0.45\textwidth}
 		\includegraphics[width=1\textwidth]{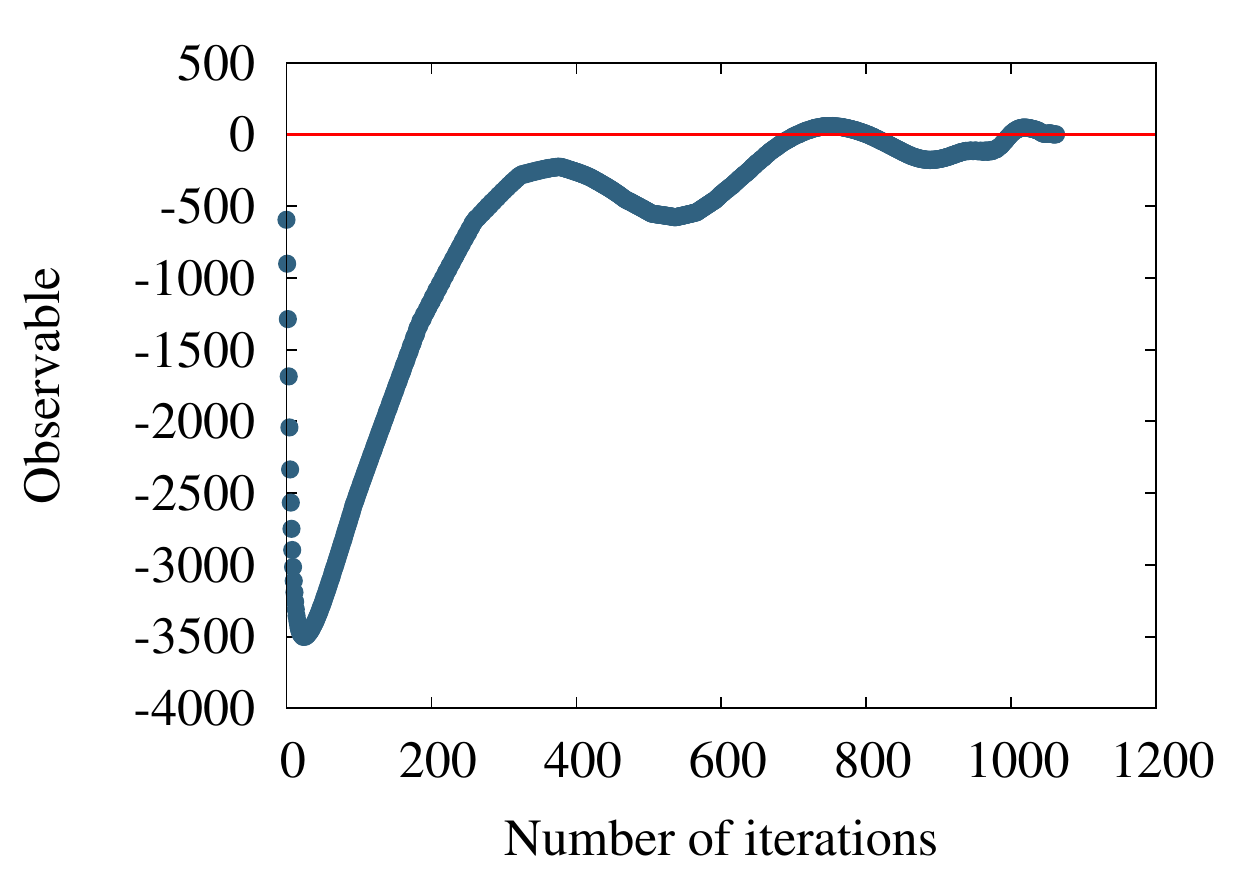}
 		\vskip 0cm 
 		\caption{For $k=30$ nearest neighbour subtractions.}
 	\end{subfigure}
 	\hfill
 	\begin{subfigure}{0.45\textwidth}
 		\includegraphics[width=1\textwidth]{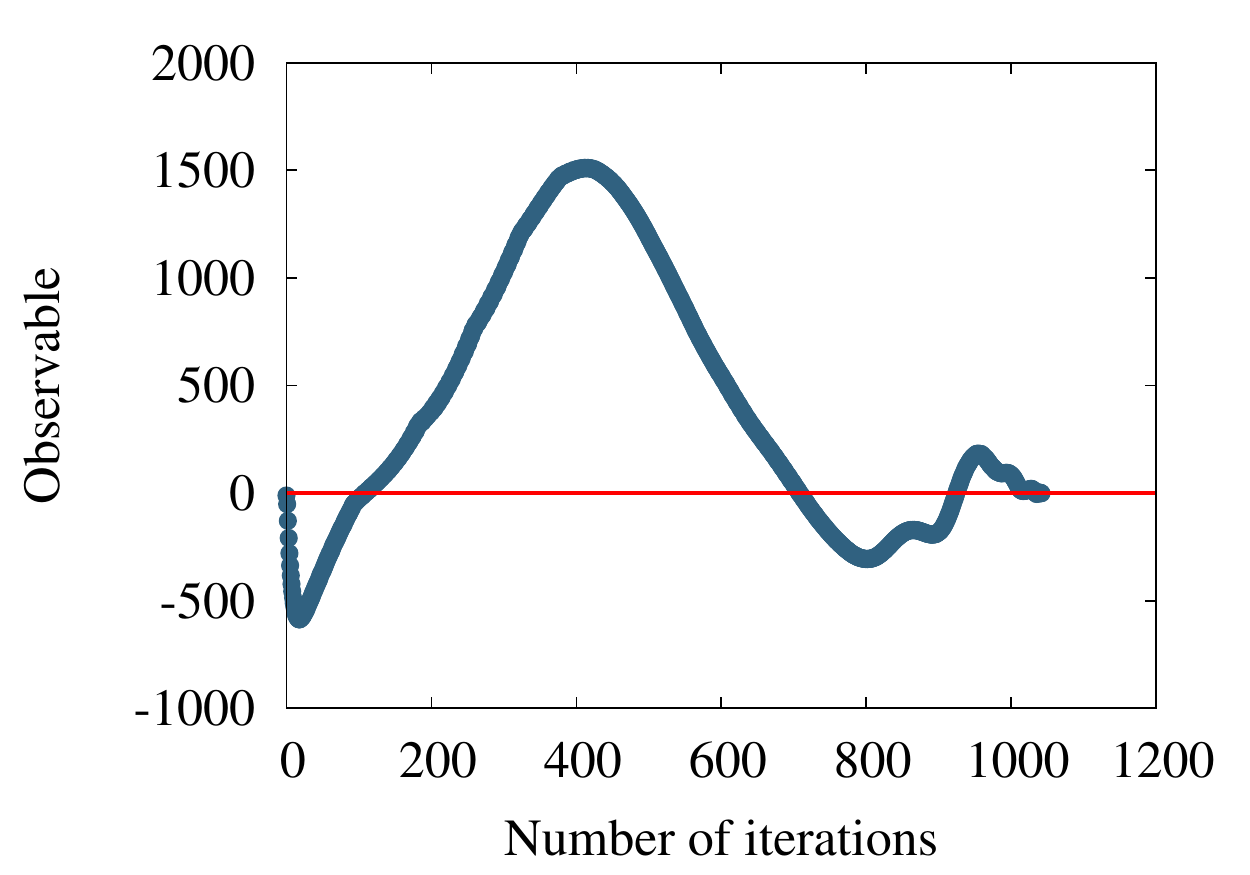}
 		\vskip 0cm 
 		\caption{For $k=50$ nearest neighbour subtractions.}
 	\end{subfigure}
 	\hfill
 	\begin{subfigure}{0.45\textwidth}
 		\includegraphics[width=1\textwidth]{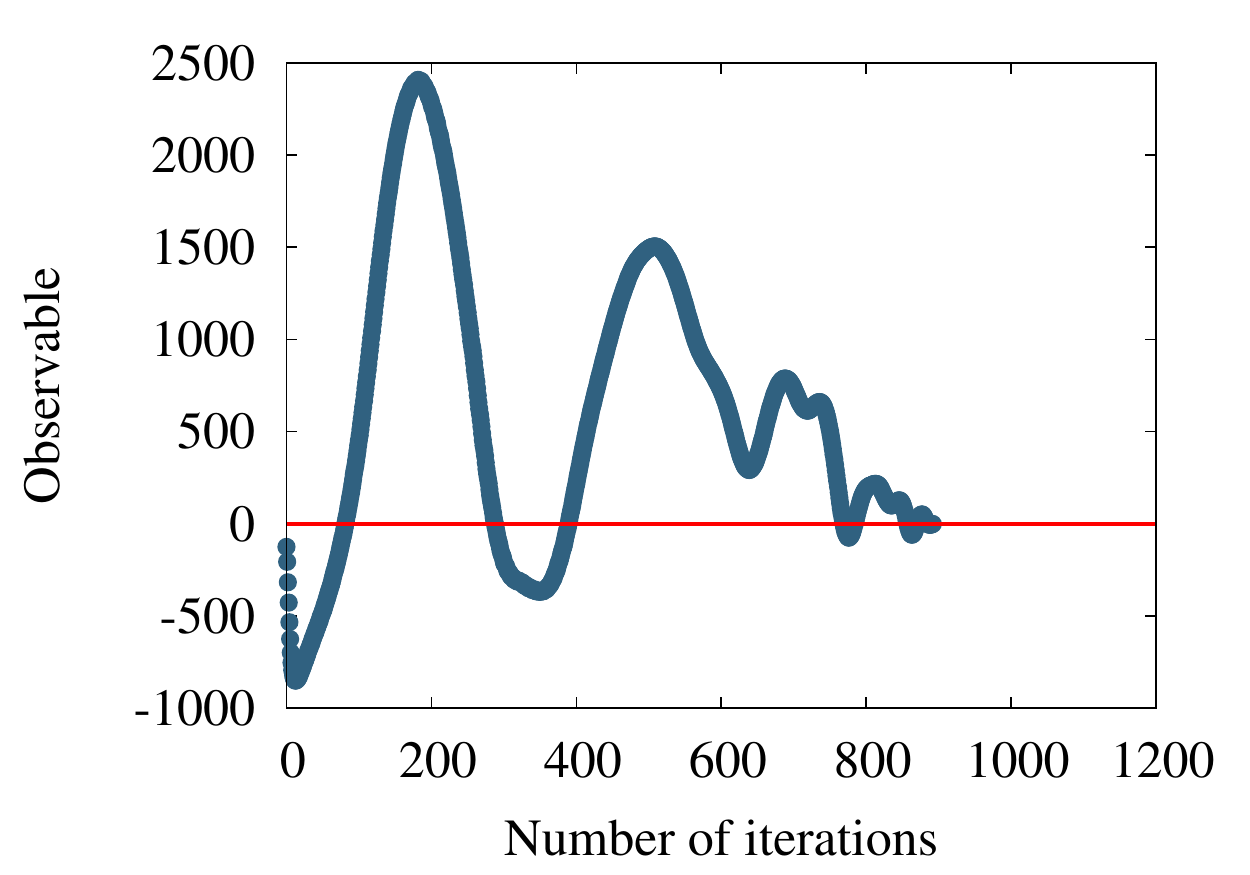}
 		\vskip 0cm 
 		\caption{For $k=200$ nearest neighbour subtractions.}
 	\end{subfigure}
 \caption{
 (Color online) Coarse graining flow of the observable (\ref{eq_cos_cutoff}), for the $\alpha$-helical 2PO4.} 
 \label{fig-22}
\end{figure}
we observe that there is an increase in oscillatory behaviour, when the value of $k$ increases. This  proposes that additional
scales become excited when we proceed and increase the resolution at which the chain is inspected: 
In Figure~\ref{fig-22} (c)  there are already several local maxima and minima, along the chain.

\subsection{Multiple scales and co-existing phases in heteropolymers?}

We compare the profiles in Figure~\ref{fig-16} with those in Figures \ref{fig-19}-\ref{fig-22}. We observe that there is an 
overall qualitative similarity.  However, there is also a difference: The flows of the observable in the homopolymer case 
shown in Figures \ref{fig-16} are monotonic, with no apparent sign of oscillatory behaviour beyond statistical fluctuations. 
On the other hand,  
in the protein examples of Figures  \ref{fig-19}-\ref{fig-22} we observe an increasingly oscillatory behaviour in the flows,
as the value of $k$ increases.

In the case of a homopolymer the parameter values in (\ref{eq:A_energy}) are uniform.  Thus there should be no intrinsic 
structural inhomogeneity along the chain to cause oscillatory behaviour on the flow of the observable, 
in a large statistical sample. 
On the other hand, in the case of a protein the homogeneity along the chain is  broken by the  
amino acid structure: Globular proteins have been found to have a modular structure \cite{scop,cath}. They are built from
super-secondary structures that can be modelled in terms of the DNLS solitons \cite{cherno,nora},  interpolating between regular segments
of  $\alpha$-helices, $\beta$-strands {\it etc.}  and each with its characteristic length. Furthermore, in a long protein chain
these super-secondary structures can form clusters of solitons (protein loops) of various sort and size.  
This kind of structure formation introduces
and engages miscellaneous scales.  We propose that these emergent scales are the source of of the oscillations we observe, they
become visible when we scrutinize the backbone at different scales {\it i.e.} at different stages of coarse graining.  

The presence of multiple scales along a chain should have an influence on its 
phase structure: In the case of point-like particles we have
the Gibbs phase rule (\ref{phase-rule}) that estimates the upper bound for the number of co-existing phase in terms of 
dimensional parameters that characterise the system. In the case of heteropolymers, no such result is known. However, we learn from 
Figures \ref{fig-19}-\ref{fig-22} that our observable oscillates between different values when 
we scrutinise the chain at different length scales. Accordingly we propose that each of these
emergent scales can give rise  to its own phase, and the phase-state of the entire chain is then a mixture of these various phases.

Indeed, globular protein are assumed to reside in a collapsed phase. At the same time, at distance scales that are short in 
comparison to the radius of gyration, the geometry is often dominated by straight rod phase 
structures such as $\alpha$-helices and $\beta$-strands. The phenomenon of  phase co-existence
can then be present in a single linearly conjugated heteropolymer such as a protein.  
A protein backbone such as 2PO4, with a wildly oscillating 
observable as shown in Figure~\ref{fig-22} (c), appears to display the characteristics of several different phases.

The relation between the oscillatory pattern of the observable and the folding pathway of the protein deserves to be
investigated, in more detail. For example, a myoglobin is not a two-stage folder but possesses a molten globule folding intermediate
\cite{ptitsyn-1,ptitsyn-2}. Indeed, in Figures \ref{fig-19} we observe evidence for 
at least one additional oscillatory transition, on top of the monotonous 
homopolymer profile. It remains to be understood, how this variation from the monotonous homopolymer
profile relates to the emergence of a 
molten globule folding intermediate. Proteins such as 4ZGV and 2PO4 with a wildly oscillating observable 
could then have several``molten globule" intermediates along their folding pathway.

Multiple length and time scales are a prerequisite for the emergence of 
the kind of complex structures and structural self-organisation that takes place in proteins of
live matter. However, our understanding of the relevance and physical origin of the diverse 
scales that are observed in larger globular proteins, remains incomplete. 
Here we  have introduced a new observable,  in combination with a new approach to
coarse grain the protein backbone, to inspect the length scales that can reside  
along a discrete piecewise linear chain such as the C$\alpha$ 
backbone of a protein. We have analysed the generic properties of our 
observable, in particular how its numerical value flows under coarse graining of a chain. 
We have confirmed these properties, and we have revealed additional ones, by numerical simulations 
of a homopolymer in thermal equilibrium. In particular, we have shown how our observable 
recuperates the finite temperature phase diagram. We have extended the methodology to 
analyse PDB proteins, and we have found that our observable can indeed detect the presence
of multiple scales in a heteropolymer such as a globular protein. We have noted that in terms of our observable,
a complex globular protein can exhibit many different phase characteristics, when we inspect it at 
different length scales. A complex protein and more generally a heteropolymer 
chain could exemplify the phenomenon of phase coexistence. 

\vskip 0.3cm
\section{Code}
The code which we used for coarse graining and calculating the observable for polymer chains is accessible online \cite{code}.

\section{Acknowledgements}

AS and JN acknowledge funding from the Knut and Alice Wallenberg Foundation and  Vetenskapsr{\aa}det.
The work of AJN is supported by a grant from Vetenskapsr\aa det and by a Qian Ren grant.
The work of MU is supported by  the DFG Grant BU 2626/2-1.


\begin{thebibliography}{10}

\bibitem{huggins} M. L. Huggins Journ. Chem. Phys. {\bf 9} 440 (1941)
 
\bibitem{flory1} P. J. Flory, Journ. Chem. Phys. {\bf 9} 660 (1941)

\bibitem{flory2} P. J. Flory, {\it Principles of Polymer Chemistry} (Cornell  University Press, Ithaca, 1953)

\bibitem{flory3} P. J. Flory, Journ. Chem. Phys. {\bf 17} 303 (1949)

\bibitem{degen} P. G. de Gennes, J. Chem. Phys. {\bf 55} 572 (1971)

\bibitem{degen1} P. G. de Gennes, Phys. Lett. {\bf 38A} 339 (1972)

\bibitem{desc} J. Des Cloizeaux, J. Phys. (Paris) {\bf 36} 281 (1975) 
 
\bibitem{edw1} S. F. Edwards, Polymer {\bf 6} 143 (1977)

\bibitem{degen2} P. G. de Gennes,  {\it Scaling Concepts in Polymer Physics} (Cornell  University Press, Ithaca, 1979)

\bibitem{edw2} M. Doi, S. F. Edwards, {\it The Theory of Polymer Dynamics}  (Oxford University Press, New York, 1986)

\bibitem{grosberg-1994} A. Yu. Grosberg, A.R. Khokhlov  {\it Statistical physics of macromolecules}
(AIP Series in Polymers and Complex Materials, Woodbury, 1994)

\bibitem{naka} T. Nakayama, Y. Kousuke, R.L. Orbach
Rev. Mod. Phys. {\bf 66} 381 (1994)

\bibitem{sokal} B. Li, N. Madras, A. Sokal, Journ.  
Stat. Phys. {\bf 80}  661 (1995).

\bibitem{schafer} L. Sch\"afer, {\it Excluded Volume Effects in Polymer Solutions, as Explained by the Renormalization Group}
(Springer Verlag, Berlin, 1999)

\bibitem{pdb} H. M. Berman, J. Westbrook, Z. Feng, G. Gilliland, T. N. Bhat, H. Weissig, I.N. Shindyalov, P.E. Bourne
Nucl. Acids Res. 28, 235 (2000).

\bibitem{dewey} T. G. Dewey, Journ. Chem. Phys. {\bf 98} 2250 (1993)

\bibitem{hong} L. Hong, J. Lei, Polym. Sci. {\bf B47} 207 (2009)

\bibitem{huang} J. Lei, K. Huang, EPL {\bf 88} 68004 (2009)

\bibitem{jcp}  N. Rawat, P. Biswas, Journ. Chem. Phys. {\bf 131} 065104 (2009)

\bibitem{kada} L.P. Kadanoff, Physics {\bf 2} 263 (1966)

\bibitem{wilson} K. Wilson, Phys. Rev. {\bf D4} 3174 (1971)

\bibitem{fisher} M. E. Fisher, Rev. Mod. Phys. {\bf 46} 597 (1974)

\bibitem{golden} N. Goldenfeld, {\it Lectures on phase transitions and the
renormalization group} (Addison-Wesley, Reading, 1992)

\bibitem{stam} A. Krokhotin, S. Nicolis, A. J. Niemi,  Journ. Chem. Phys. {\bf 140} 095103  (2014) 

\bibitem{frenet} S. Hu, M. Lundgren, A.J. Niemi, Phys. Rev. {\bf E83} 061908 (2011)

\bibitem{oma-old} A. J. Niemi, Phys. Rev. {\bf D67}  106004  (2003)

\bibitem{ulf} U. Danielsson, M. Lundgren, A. J. Niemi, Phys. Rev. {\bf E82} 021910 (2010)

\bibitem{hu-13} S. Hu, Y. Jiang, A. J. Niemi, Phys. Rev. {\bf D87} 105011 (2013)

\bibitem{theo-2014}  T. Ioannidou, Y. Jiang, A. J. Niemi, Phys. Rev. {\bf D90}
025012 (2014)

\bibitem{theo-2015} T. Ioannidou, A. J. Niemi, Phys. Lett. {\bf A380}  333 (2015)

\bibitem{ivan} I. Gordeliy, D. Melnikov, A. J. Niemi, A. Sedrakyan, Phys.  Rev. {\bf D94} 021701(R) (2016)

\bibitem{Ann} A. Sinelnikova, A. J. Niemi, M. Ulybyshev, Phys. Rev. {\bf E92} 032602 (2015)

\bibitem{priva-1}  P. L. Primalov, Adv. Protein Chem. {\bf 33} 167 (1979)

\bibitem{priva-2} P. L. Primalov, Ann. Rev. Biophys. Biophys. Chem. {\bf 18} 47 (1989)

\bibitem{priva-3} E. Shakhnovich, A. Finkelstein Biopolymers {\bf 28} 1667 (1989)

\bibitem{cherno} M. Chernodub, M. Lundgren, A. J. Niemi, Phys. Rev.  {\bf E 83} (2011) 011126 

\bibitem{nora} N. Molkenthin, S. Hu, A. J. Niemi, Phys.  Rev. Lett.  {\bf 106} (2011)  078102

\bibitem{DeGennes-book} P. G. de Gennes, J. Prost, {\it The Physics of Liquid Crystals} (Clarendon Press, Oxford, 1995)

\bibitem{ptitsyn-1} O. B. Ptitsyn, J. Protein Chem. {\bf 6} 273 (1987)

\bibitem{ptitsyn-2} O. B. Ptitsyn,  Curr. Opin. Struct. Biol. {\bf 5} 74 (1995)

\bibitem{scop} A. G. Murzin, S. E. Brenner, T. Hubbard, C. Chothia, J. Mol. Biol. {247} 536 (1995)

\bibitem{cath} L. H.  Greene {\it et.al} Nucl. Acids Res. {\bf 35} D291(2007)

\bibitem{code} A. Sinelnikova (2017), URL \url{http://doi.org/10.5281/zenodo.581166}

\end{thebibliography}
\end{document}